\documentclass[mnsc,nonblindrev]{informs3}

\OneAndAHalfSpacedXI 

\usepackage{natbib}
 \bibpunct[, ]{(}{)}{,}{a}{}{,}%
 \def\bibfont{\footnotesize}%
\TheoremsNumberedThrough     
\ECRepeatTheorems

\EquationsNumberedThrough    

\usepackage{dsfont}
\usepackage{mathtools}
\usepackage{bm}
\usepackage{booktabs}
\usepackage[official]{eurosym}
\usepackage{microtype}
\usepackage{lmodern}
\usepackage{comment}
\usepackage{mathrsfs}
\usepackage{hyperref}
\usepackage{nicefrac}
\usepackage{algorithmic,algorithm}

\newcommand{\av}[1]{{\leavevmode\color{black}{#1}}}

\newcommand{\tS}{S^+}

\newcommand{\balpha}{\bm{\alpha}}
\newcommand{\balphak}[1]{\bm{\alpha}^{(#1)}}

\newcommand{\gsp}{\text{GSP}}

\newcommand{\Cscr}{[n]}
\newcommand{\bM}{\bm{M}}
\newcommand{\Tcal}{\mathcal{T}}
\newcommand{\Pcal}{\mathcal{P}}

\newcommand{\Mcal}{\mathcal{M}}

\newcommand{\gspk}[1]{\Omega_{(#1)}}
\newcommand{\gsptypes}{\Omega}

\newcommand{\Exp}{\mathbb{E}}
\newcommand{\Real}{\mathbb{R}}
\newcommand{\Scal}{\mathcal{S}}
\newcommand{\rank}{\text{pos}}

\newcommand{\mus}[1]{\mu^{(#1)}}
\newcommand{\Lscr}{\mathscr{L}}
\newcommand{\bell}{\bm{\ell}}
\newcommand{\np}{{\sf np}}
\newcommand{\posvec}{\overrightarrow{\rank}}
\newcommand{\loss}{{\sf MAPE}}
\newcommand{\mpmnl}{{\sf MP-MNL}}

\newcommand{\indicator}[1]{\ensuremath{\mathds{1}\bigg[#1\bigg]}}

\newcommand{\abs}[1]{\left\vert#1\right\vert}
\newcommand{\set}[1]{\left\{#1\right\}}

\newcommand{\logll}{\mathcal{L}}
\newcommand{\LL}{\mathcal{LL}}

\newcommand{\vveciter}[1]{\bm{v}^{(#1)}}
\newcommand{\piter}[1]{\lambda^{(#1)}}
\newcommand{\osminus}[2]{{#1}^{-#2}}
\newcommand{\prefvec}{\bm{v}}

\newcommand{\blambda}{\bm{\lambda}}
\newcommand{\rat}{\sf R}
\newcommand{\nonrat}{\sf NR}
\newcommand{\itertypes}[1]{\Omega^{(#1)}}
\newcommand{\rattypes}{\Omega_{\rat}}
\newcommand{\nonrattypes}{\Omega_{\nonrat}}

\newcommand{\lambdak}[1]{\lambda^{(#1)}}
\newcommand{\tlambdak}[1]{\tilde{\lambda}^{(#1)}}
\newcommand{\tlambda}{\tilde{\lambda}}

\newcommand{\cond}{\; \vert \;} 
\newcommand{\ukvec}[1]{\bm{U}^{(#1)}}
\newcommand{\uk}[1]{U^{(#1)}}
\newcommand{\gvec}{\bm{g}}
\newcommand{\Gscr}{\mathcal{G}}
\newcommand{\hPcal}{\hat{\Pcal}}
\MANUSCRIPTNO{MS-0001-1922.65}

\begin{document}


\RUNAUTHOR{Berbeglia and Venkataraman}

\RUNTITLE{The Generalized Stochastic Preference Choice Model}

\TITLE{The Generalized Stochastic Preference Choice Model}

\ARTICLEAUTHORS{%
\AUTHOR{Gerardo Berbeglia}
\AFF{Melbourne Business School, University of Melbourne, \EMAIL{g.berbeglia@mbs.edu}} 
\AUTHOR{Ashwin Venkataraman}
\AFF{Jindal School of Management, University of Texas at Dallas, \EMAIL{ashwin.venkataraman@utdallas.edu}}
} 

\ABSTRACT{%
We propose a new discrete choice model, called the generalized stochastic preference (GSP) model, that incorporates non-rationality into the stochastic preference (SP) choice model, also known as the rank-based model. Our model can capture several context-dependent choice behaviors that cannot be represented by any SP model, such as the well-documented compromise and attraction effects, while still including the SP model as a special case. The GSP model is defined as a distribution over consumer types, where each type extends the choice behavior of rational types in the SP model. We build on existing methods for estimating the SP model and propose an iterative estimation algorithm for the GSP model that finds new types by solving an integer linear program in each iteration. We further show that our proposed notion of non-rationality can be incorporated into other choice models, like the random utility maximization (RUM) model class as well as any of its subclasses. As a concrete example, we introduce the non-rational extension of the classical MNL model, which we term the generalized MNL (GMNL) model and present an efficient expectation-maximization (EM) algorithm for estimating it. \av{For the GSP model, we demonstrate that the worst-case performance guarantee of revenue-ordered assortments is significantly worse than for the SP model. For the GMNL model, we establish that assortment optimization with totally unimodular constraints is NP-hard to approximate to within a factor of $O(n^{1-\epsilon})$ for any $\epsilon > 0$, where $n$ is the number of products.} Finally, numerical evaluation on synthetic and real choice data shows that the GMNL and GSP models can outperform their rational counterparts in out-of-sample prediction accuracy.
}%


\KEYWORDS{non-rational choice, nonparametric model, regularity violations, context-dependent effects} 

\maketitle

%

\vspace{-2em}
\section{Introduction}
Discrete choice models are commonly used to predict how consumers choose from a finite set of alternatives and have been studied extensively in multiple domains. 
For instance, they are used to predict which travel mode consumers prefer from available options such as car, bus, train, etc. in the transportation literature. In the economics literature, they are used to elicit consumer preferences from observed choices such as buying a house or joining the workforce. In the marketing and operations literature, they are used to predict which product a consumer purchases from the offered set of products and consequently, estimate the demand for different products. A key feature of such models is their ability to capture substitution behavior in consumer choices, that is, if a consumer's preferred product is not available, they can switch to 
 a (close) substitute that is available. 

A natural way to model consumer preferences is to associate a list or ranking of the products\footnote{We use product and alternative interchangeably throughout the paper.} with each consumer, with the top-ranked product being the most preferred. When making a choice, the consumer goes down her preference list and picks the first product that is available. This is precisely what the stochastic preference (SP) model~\citep{block1960random} prescribes, which is also referred to as the rank-based model in the operations literature, see, e.g.,~\cite{van2017expectation}. In particular, the SP model associates a probability distribution over all possible preference lists over the product universe---for $n$ products, the maximum support, therefore, is of size $n!$ ($n$ factorial). Assuming the set of alternatives is fixed during the choice process, the SP model can be shown to represent any model in the random utility maximization (RUM) model class, see, e.g,~\cite{block1960random,jagabathula2019limit}, which is by far the most widely studied class of choice models in the literature. The RUM model associates a joint distribution over product utilities from which an individual consumer samples a utility vector when making a choice; the consumer is assumed to choose the available product that has the largest utility. This utility maximizing behavior is consistent with the notion of {\em rationality} developed in the economics literature. 

The RUM model can capture a large array of consumer substitution patterns encountered in practice by varying the form of the joint utility distribution (see~\cite{feng2022consumer} for a recent review of different choice models in the RUM class). The SP model inherits this ability but has the added benefit of being a {\em nonparametric} choice model, because it does not make any parametric assumptions and allows the support of the distribution over preference rankings to grow as the number of consumers and/or choices increase. Moreover, the last decade has witnessed great interest in designing scalable estimation methods for the SP model from consumer choice data, which is critical because of the large support size mentioned earlier. 
 The first robust method to estimate the SP model was proposed by \citet{farias2013nonparametric} using LP and constraint sampling techniques. More recent methods include the column generation approaches of \citet{van2014market} and \citet{mivsic2016data}, and the tree-growing procedure by \citet{jena2020partially}. 
Despite the generality of the SP model and the availability of scalable estimation methods, it still suffers from a major limitation. It is easy to check that the SP model must satisfy a property known as {\em regularity}, which states that the probability of choosing an alternative cannot increase if the set of available options is enlarged~\citep{block1960random}. Although regularity seems natural, experiments (e.g., \citet{huber1982adding,simonson1992choice,ariely03}) have shown that it can be violated. In these cases, the SP model may fit poorly regardless of estimation procedure. 


While strong regularity violations have been observed in controlled experiments, one may question whether these violations also appear in more operations relevant contexts such as retail sales. To answer this question, \citet{jagabathula2019limit} developed an efficient procedure to quantify the \emph{limit of rationality} (LoR), which the authors defined as the cost of approximating aggregate sales transactions data using an SP model. Their analysis on the IRI Academic Dataset~\citep{bronnenberg2008database}, a real-world dataset consisting of weekly sales transactions for consumer packaged goods across different grocery and drug store chains, reveals that for some categories of products such as coffee and yogurt, the LoR can be quite high. In other words, the SP choice model, and consequently any model in the RUM class, cannot reasonably approximate the observed sales data.

\av{Following the work of~\cite{jagabathula2019limit}, there has been a growing literature dedicated to developing models that can capture regularity violations and therefore, lie outside the RUM class. 
On the one hand, there are several parametric models such as the general attraction model (GAM) \citep{gallego2014general}\footnote{The GAM can only fall outside RUM when the ``shadow'' attractions are larger than the natural attractions.}, the general Luce model \citep{echenique2019general}, the perception adjusted Luce model (PALM) \citep{echenique2018perception}, the focal Luce model~\citep{kovach2022focal}, the Halo-MNL model~\citep{maragheh2018customer,yousefi2020choice} and many others. The problem with using any of these models is that they can potentially fit the data worse than the SP model, because they do not contain the RUM class. On the other hand, nonparametric extensions that do subsume the RUM class such as~\cite{chen2022decision,chen2021estimating} have the flexibility to represent {\em all} possible discrete choice models. This universality property  makes them prone to strong overfitting issues, especially when trained with limited data.}

\av{In this paper, we address this gap in the literature by introducing the \emph{generalized stochastic preference} (GSP) model—a new model that strictly extends the standard stochastic preference (SP) framework and can capture a variety of choice phenomena that violate regularity, such as the \emph{attraction (or decoy) effect} and the \emph{compromise effect}~\citep{simonson1989choice}. The attraction effect describes situations where the introduction of a clearly dominated, never-chosen option (a decoy) increases the probability of selecting a particular target alternative; see Examples~\ref{ariely_solution} and~\ref{herne_solution} in Appendix~\ref{sec:more_regularity_violations}. The compromise effect refers to the tendency of consumers to prefer an option that appears as a middle-ground among the available alternatives, even if that same option would not have been chosen (or chosen with a lower frequency) when offered within a smaller subset of the choice set; refer to Section~\ref{sec:gsp_power} for a concrete example.

As the name suggests, the GSP model is a natural extension of the SP model consisting of ``standard" consumer types who follow a preference list and additional ``non-standard'' consumer types; the latter allow the GSP model to capture rich classes of consumer choice behaviors. Given the relatively simple and structured nature of the choice process for non-standard consumer types, the GSP model provides a key advantage: it helps avoid overfitting and allows a more interpretable estimation. As a direct consequence, the GSP model does not capture all possible consumer choice patterns like the aforementioned universal models.}



{\bf Summary of Contributions.} We make the following contributions in this work:
\begin{enumerate}
    \item {\em Nonparametric extension of SP model that admits regularity violations.} We introduce the generalized stochastic preference (GSP) choice model that can capture regularity violations in consumer choices. Unlike existing parametric non-rational choice models, the GSP model is nonparametric and therefore, has the ability to flexibly explain (fit) complex consumer choice behaviors (data). The model is simple to describe---it is defined as a distribution over consumer types, where each type is either {\em standard}, in which case its choice behavior is the same as in the SP model, or {\em non-standard}, which extends the choice process of a standard type; 
 refer to Section~\ref{sec:model} for a formal description of the model. As a result, the GSP model naturally subsumes the SP model. \av{Moreover, we show that the non-standard types in our model have a natural interpretation based on behavioral considerations, such as cognitive bias (e.g., compromise effect), limited attention, or choice under uncertainty}, all mechanisms that have been employed to explain non-rational choices in existing literature; see the discussion in Section~\ref{sec:gsp_interpretation}. \av{We also study the representation power of GSP models, specifically in relation to the broader class of regular choice models and provide sufficient conditions under which the GSP model satisfies regularity; see Propositions~\ref{prop:regular_but_not_gsp} and~\ref{prop:when_is_gsp_regular}.}
\item {\em General notion of non-rationality applicable to any RUM choice model.} We show that our proposed notion of non-rationality is general and can be incorporated within the RUM model class as well as any of its subfamilies. We illustrate the idea by introducing the {\em generalized MNL (GMNL) model}, the non-rational extension of the classical MNL model. Because the MNL model (and other parametric models in the RUM class) can account for product features such as price, this approach significantly broadens the scope of non-rational models available to researchers and practitioners. Refer to Section~\ref{sec:general_RUM_models} for the details. 

    \item {\em Estimation methods for the GSP and GMNL models.} The similarity in construction between the GSP and SP models allows us to build upon the nonparametric estimation framework for the SP model proposed in \citet{van2014market} and \citet{jagabathula2019limit}. In particular, we propose an iterative algorithm for maximum likelihood estimation (MLE) of the GSP model which, starting from an initial set of types, finds new consumer types in each iteration by solving an integer linear program. Our algorithm also constrains the complexity and proportion of non-standard types recovered in the solution as a means to reduce overfitting; the detailed approach is discussed in Section~\ref{sec:gsp_est}. Moreover, in Section~\ref{sec:kmnl_est}, we present an expectation-maximization (EM) algorithm for MLE of the GMNL model that iteratively fits a single MNL model on a transformed version of the choice data, resulting in a fast and efficient method in practice. 
    \item \av{{\em Assortment optimization.} Because GSP subsumes the SP model, existing hardness results~\citep{aouad2015approximability} apply. In terms of heuristics for the problem, we show that the worst-case performance guarantee of the well-known revenue-ordered assortments strategy~\citep{talluri2004revenue,berbeglia2016assortment} can be significantly worse compared to the SP model; see Proposition~\ref{prop:gsp_assort}. Perhaps more surprisingly, we show that assortment optimization with totally unimodular (TU) constraints under the GMNL model is NP-hard to approximate; see Proposition~\ref{prop:gmnl_assortment} for the formal result. We derive this result by constructing a reduction from the recently proposed multi-purchase MNL (MP-MNL) model of~\cite{bai2023assortment}, and to the best of our knowledge, is one of the first formal links between single-purchase and multi-purchase models. Our result reveals a strong separation from the MNL model for which TU-constrained assortment optimization can be solved using a linear program~\citep{sumida2021revenue}. On the positive side, we demonstrate that revenue-ordered assortments perform robustly in practice, achieving near-optimal average-case performance (exceeding 97\% of the optimal revenue) and guaranteeing at least 60\% of the optimal revenue in the worst case across a range of ground-truth problem instances.}
    \item {\em Numerical evaluation on real data.} We test the performance of our proposed models using real choice data. Using data collected from two incentive-aligned experiments in the economics literature, we show further evidence of non-rational consumer preferences, \av{resulting in poor fit from the SP model---mean absolute percentage error (MAPE) of $2\%$ and $7\%$ between predicted and observed choice probabilities.} The GSP model, on the other hand, is able to perfectly explain the observed choices while utilizing at most 10-15\% proportion of non-standard types. \av{In contrast, the Halo-MNL model~\citep{maragheh2018customer, yousefi2020choice}---a parametric model capable of capturing regularity violations—achieves MAPE values of $5\%$ and $10\%$, highlighting the advantage of a nonparametric framework like GSP, which subsumes the SP model.
} See Section~\ref{sec:explain} for the details.
    
    We also showcase the predictive performance of our proposed models on three datasets reporting travel mode choices. Our evaluation, described in Section~\ref{sec:pred_study}, reveals that the non-rational GSP and GMNL models can outperform their rational counterparts by \av{up to} $40\%$ in out-of-sample log-likelihood. The GMNL model, in particular, emerges as promising candidate for practical use, providing better predictions than the SP model but with estimation complexity comparable to fitting the standard MNL model. 
    \end{enumerate}
\subsection{Related literature}
\label{sec:related_work}
\av{This paper introduces a new choice model designed to capture regularity violations in consumer behavior. To position our contribution, we begin by summarizing the existing literature on such models. In particular, we distinguish between parametric and nonparametric approaches to modeling regularity violations. We then review estimation procedures for choice models using choice or sales transaction data, as our work also develops tractable algorithms to estimate the parameters of the proposed models. 

{\bf Parametric choice models capturing regularity violations.}
About fifty years ago, economists and psychologists working in the field of behavioral decision theory began documenting empirical evidence that individuals frequently violate the principle of regularity (see, e.g., \cite{tversky1974judgment,tversky1981framing}). These violations are typically attributed to context effects, reference dependence, or cognitive costs incurred during the choice process \citep{McFadden2000}. In response, a number of choice models—primarily in the economics literature—have been proposed to account for such behavioral phenomena and explain violations of regularity (e.g., \cite{wernerfelt1995rational,kamenica2008contextual,bordalo2013salience,echenique2018perception,ahumada2018luce}). 


Many of the models proposed in the literature extend the classical MNL model~\citep{luce1959}. \citet{wang2018impact} show that incorporating search costs into the MNL model can lead to violations of regularity. \cite{echenique2018perception} proposed the {\em perception-adjusted Luce model} (PALM), in which consumers perceive the alternatives in a sequential fashion over $T \geq 2$ ``levels'', and in each level choose from amongst the available products according to an MNL model. The authors provided an axiomatic characterization for the PALM and showed that it allows regularity violations. \cite{flores2019assortment} proved that a simple algorithm based on revenue-ordered assortments by level solves the assortment optimization problem under the PALM when $T=2$. Along similar lines,~\cite{kovach2022focal} propose the focal Luce model (FLM) in which the utilities of certain products in the choice set, termed the focal set, are amplified according to a choice set-specific distortion function. The FLM framework offers the flexibility to capture well-studied context effects—such as the compromise effect~\citep{simonson1989choice} and choice overload~\citep{iyengar2000choice}—by appropriately specifying the combination of the focal set and the distortion function.

\cite{echenique2019general} and \cite{ahumada2018luce} independently developed another generalization of the MNL model, termed the {\em General Luce model} (GLM), which incorporates consideration sets~\citep{hauser2014consideration} into the choice process such that products not included in the consideration set have choice probabilities equal to zero; the probability of choosing products in the consideration set follows an MNL model. The authors demonstrate that this two-step procedure can admit regularity violations.
\cite{flores2017assortment} and~\cite{wang2022threshold} study assortment and pricing problems under a special case of the GLM, called the {\em two-stage Luce model}, which defines consideration sets based on a dominance relationship (strict partial order) among alternatives.


In the operations literature, the {\em general attraction model} (GAM)~\citep{gallego2014general} modifies the choice probabilities under the MNL model
to depend on {\em all} products, including those that are not offered. In particular, the model associates a ``shadow'' attraction with each product---in addition to the usual attraction in the MNL model---that appears in the
choice probability expression when the product is not part of the offer set. If the shadow attraction exceeds the usual attraction for any product, the GAM lies outside the RUM class; see~\citet[Appendix E.4]{jagabathula2019limit}.
More recently, the HALO-MNL model \citep{maragheh2018customer,yousefi2020choice} has been proposed to approximate context effects by introducing offer set-dependent utilities within the MNL framework. Specifically, the utility of each product includes a pairwise interaction term that captures the contextual influence with respect to every other product in the offer set. \citet{lo2019assortment} study the assortment optimization problem under a related model in which the product attraction values, instead of the utilities, are a function of the other products in the offer set.


Other work incorporates regularity violations by modeling consumer preferences as a ranked list of products. \citet{brady2016menu} introduce a framework in which the consideration set formation process is probabilistic: for any given offer set, the distribution over consideration sets follows an MNL structure, termed ``logit attention'' in the economics literature. The consumer then selects the highest-ranked product within her realized consideration set.
In the computer science literature, \citet{kleinberg2017comparison} propose the \emph{comparison-based choice} (CBC) model where each consumer ranks the alternatives according to some latent one-dimensional scale, and then selects the $k^{th}$ ranked item from the offered set of options, for some $k > 1$. This “position-based” rule naturally captures context effects where the presence of extreme options shifts preference toward intermediate (compromise) choices. Both of these models assume homogeneous consumer preferences, with all customers sharing a common ranked list over products. In contrast, as discussed in Section~\ref{sec:gsp_interpretation}, the proposed GSP model extends the CBC model by allowing for different product orderings across consumers to capture preference heterogeneity.
Due to their reliance on parametric and/or homogeneity assumptions, none of the models discussed above fully subsume the RUM class. Next, we discuss models that, like the proposed GSP model, allow for regularity violations while encompassing the RUM class as a special case.


\paragraph{\bf Nonparametric choice models capturing regularity violations.}
\citet{cattaneo2017random} introduce the \emph{random attention model (RAM)}, which generalizes the framework of \citet{brady2016menu} by allowing for a nonparametric attention (consideration set formation) process, subject to a natural identification restriction. This generalization enables the RAM to fully subsume the RUM class, despite assuming a common preference ranking across all consumers.\footnote{In fact, the RAM also subsumes the larger class of choice models that exhibit regularity.} However, the nonparametric nature of the attention rules requires specifying a probability distribution over all possible subsets for each distinct choice set, subject to monotonicity constraints. Consequently, the RAM is less amenable to sparse representations which can lead to difficulty in interpreting the customer choice behavior. In contrast, we propose an iterative estimation procedure for the GSP model that incrementally identifies consumer types, thereby offering a principled approach to controlling model complexity by limiting the number of iterations. Moreover, as shown in Appendix~\ref{app:ram_and_gsp}, the GSP and RAM models are not nested: neither model strictly generalizes the other. Recent extensions of RAM~\citep{kashaev2022random,aguiar2023random} have incorporated preference heterogeneity, allowing consumers to have different preference rankings. 
\cite{feng2017relation,feng2018substitutability} propose a welfare-based class of choice models which strictly subsumes the RUM class (as long as $n \geq 3$) and can capture certain kinds of regularity violations such as the halo effect~\citep{thorndike1920constant}. However, unlike the GSP model, the authors do not provide any methods to estimate the model from choice data.

Complementary to the models discussed above, a separate line of work focuses on the development of \emph{universal} choice models---models which can represent \emph{any} stochastic choice rule. 
In the economics literature, \cite{li2017every} proposed a randomized game tree model and showed that every discrete choice model can be represented by it.
Similarly, \citet{dogan2023every} introduce a universal model based on two disjoint sets of rankings—termed \emph{pro} and \emph{con} rankings—along with a weight function that assigns positive (negative) weights to pro (con) rankings. 
While these models are theoretically expressive and conceptually elegant, their generality and combinatorial complexity raise practical estimation challenges which limit their applicability in operational settings.


In the operations literature, \citet{chen2022decision} and~\citet{chen2021estimating} independently proposed a universal choice model where each consumer makes choices based on a binary decision tree, which entails sequentially checking for the existence of the corresponding products in the offer set. The population preferences are defined using the {\em decision forest} model, obtained as a distribution over the decision trees. 
The flexibility of the above approaches comes at the cost of interpretability and the risk of overfitting. In contrast, the proposed GSP model aims to strike a balance between expressiveness and structure: it accommodates regularity violations and preference heterogeneity while retaining a simpler representation of consumer types that is easier to estimate. An additional advantage of the GSP model is its interpretability—it extends the standard SP framework by introducing consumer types whose choices remain grounded in product orderings, yet can exhibit systematic regularity violations.



{\bf Choice model estimation}. In this work, we estimate the parameters of the proposed choice model using maximum likelihood estimation (MLE). Our approach builds on the market discovery algorithm of \citet{van2014market}, who developed an iterative column-generation method for MLE of the SP model. In each iteration, their method identifies a new ranking to add to the current support of the ranking distribution. In our setting, the estimation procedure similarly follows a column-generation framework, but instead of adding preference lists, we iteratively introduce consumer types consistent with our proposed model. We note that estimation of the SP model has been an active area of research, see, e.g.,\cite{farias2013nonparametric,mivsic2016data,van2017expectation,jagabathula2019limit,mendez2019analysis,jena2020partially}. For a broader discussion of estimation techniques across various choice models, refer to \citet{berbeglia2022comparative,feng2022consumer}.

Some years after our first version of our paper was made publicly available (March 2018), \citet{jena2022estimation} developed a column-generation-based estimation procedure for the GSP model using a \emph{growing preference tree (GPT)} representation, where nodes in the tree correspond to products and paths represent product orderings. Their empirical evaluation on the previously mentioned IRI Academic Dataset demonstrates that the GSP model yields more accurate demand predictions than the standard SP model and other frameworks that incorporate non-rational customer behavior. For the benefit of the reader, we provide a self-contained summary of these findings in Appendix~\ref{sec:iri}. As discussed earlier, we propose a distinct estimation approach for the GSP model. In addition, we develop an efficient procedure for estimating the GMNL model, a subclass of GSP. Our numerical evaluation in Appendix~\ref{sec:synth_data} compares the GPT-based method of \citet{jena2022estimation} with our approach in terms of both computational efficiency and predictive performance.}
\section{The GSP choice model}\label{sec:model}
Let $[n]=\{1,2,\hdots,n\}$ denote a universe of products or alternatives. We consider a standard setting in which consumers are faced with some choice set or offer set $S \subseteq [n]$ and select one of the products in $S$. \av{For now, we focus on the scenario where consumers always choose one of the offered products---in Section~\ref{sec:assortment}, we consider an extension that incorporates the ``no-choice'' or ``no-purchase'' option where consumers are allowed to not choose anything from the offer set.} 
A discrete choice model (or simply a choice model) is a function $\mathcal{P}$ that given any non-empty choice set $S \subseteq [n]$ and any alternative $i \in S$, specifies $\mathcal{P}(i, S)$ as the probability that a consumer will select alternative $i$ when she is offered the subset of alternatives $S$. Throughout the paper, we use the terms {\em alternative} and {\em product} and similarly, {\em choice set} and {\em offer set}, interchangeably.
%
%


As mentioned earlier, our proposed choice model is defined as a distribution over consumer types. 
Let $\bell = (\ell_1, \ell_2, \ldots, \ell_n)$ denote an ordering of the $n$ products, with $\Lscr$ denoting the set of all such orderings. It is easy to see that $\abs{\Lscr} = n!$ ($n$ factorial). 
Each consumer type in the GSP model is characterized by a pair $(\bell, k)$, where $\bell \in \Lscr$, and $k$ is an integer between $1$ and $n$. We refer to $k$ as the {\em choice index} of the consumer type. 
We denote the set of all consumer types as 
$\Omega = \set{(\bell, k)  \colon \; \bell \in \Lscr, 1 \leq k < n }$ and refer to an arbitrary type as $\omega = (\bell, k)$.\footnote{The reason we do not include $k=n$ will become clear shortly.}

Next, we describe how each consumer type makes choices when presented with an offer set. Given an offer set $S$, consumers of type $\omega = (\bell, k)$
first construct a subsequence, say $s(\bell, S)$, of $\bell$ by removing \emph{all} alternatives that are \emph{not} in $S$. Then, they choose the alternative at the $k^{th}$ position in the subsequence $s(\bell, S)$ if it exists---the $k^{th}$ position does not exist when the length of the subsequence is strictly smaller than $k$---otherwise when $|s(\bell, S)|<k$,  the consumer will select the last alternative in $s(\bell, S)$. More formally, defining $\rank(j, S; \bell)$ as the position of alternative $j \in S$ in the subsequence $s(\bell, S)$, the choice function for any consumer type in the GSP model is given below.
\begin{definition}[choice function of consumer types in gsp model] \label{def:gsp_choice_fn}
Faced with a choice set $S \subseteq [n]$, the binary choice function $C_{\omega}(j, S)$ which states whether consumer type $\omega =(\bell, k)$ chooses alternative $j \in S$ is defined as follows:
\begin{displaymath}
C_{\omega}(j,S) \coloneqq \indicator{\rank(j, S; \bell) = \min(k, \abs{S})} = \left\{ \begin{array}{ll}  \displaystyle 1  & \quad \textrm{if $j$ is at position $\min(k, \abs{S})$ in $s(\bell,S)$} \\[3ex]
0 & \quad \textrm{otherwise}\\
\end{array} \right.
\end{displaymath}
\end{definition}
Given the above definition, for any type $\omega = (\bell, 1)$, it is easy to check that $C_\omega(j, S) = C_{\omega'}(j, S)$ for all $S \subseteq [n]$ and all $j\in S$, where $\omega' = \left(\text{rev}(\bell), n\right)$ and $\text{rev}(\bell)$ is the ordering induced by reversing the positions of all products in $\bell$. This is the reason we do not include choice index $k=n$ in the set of consumer types $\Omega$. With the above notation in place, we define the class of GSP as follows.
\begin{definition}[choice probability under gsp model]\label{def:gsp_prob_fn}
A discrete choice model $\mathcal{P}$ belongs to the class of GSP models if there exists a probability distribution $\lambda(\cdot)$ over the set of consumer types $\Omega$ such that
\begin{eqnarray} \label{stochastic_preference_probabilities}
\mathcal{P}(j, S) = \sum_{\omega \in \Omega} C_{\omega}(j,S) \cdot \lambda(\omega)
\end{eqnarray}
for all $S \subseteq [n]$ and all $j \in S$.
\end{definition}

\av{Having established the mathematical formulation of the GSP model, we now explore behavioral interpretations that capture the proposed choice behavior.
\subsection{Behavioral interpretation of customer types in the GSP model}
\label{sec:gsp_interpretation}
From Definition~\ref{def:gsp_choice_fn}, it follows that any consumer type $(\bell, k)$ with a choice index $k=1$ chooses the product in $S$ which appears {\em first} in the ordering $\bell$. 
Consequently, the choice behavior of such types coincides exactly with that of a consumer type from the SP model, with the ordering $\bell$ representing the preference list or ranking of the $n$ products. For this reason, we term them as {\em standard} types. 


 On the other hand, the choice behavior of consumer types with choice index $k > 1$, which we term {\em non-standard} types, is more complex to characterize. We offer three possible interpretations, grounded in well-documented findings from the behavioral literature. However, we emphasize that we do not claim that customers are necessarily behaving in these specific ways—rather, these interpretations serve as plausible explanatory mechanisms that help us gain insights into the structure and implications of the modeling approach. 
 Our numerical evaluation in Section~\ref{sec:numerics} demonstrates that non-standard types can explain choice data observed in practice as well as generate accurate out-of-sample predictions.
 
 {\bf Context-dependent choice}. The non-standard types in the GSP model are motivated by the comparison-based choice framework proposed by \citet{kleinberg2017comparison}, which provides a formal basis for capturing the classic \emph{compromise effect} first documented by \citet{simonson1989choice}. This effect refers to the tendency of consumers to favor an option that represents a compromise among the available alternatives. For instance, consider a choice set consisting of a mediocre option for \$10, a good option for \$15, and an excellent option for \$20. In this setting, a consumer might choose the good option for \$15. However, if the offer set shifts to include a more expensive alternative—say, good for \$15, excellent for \$20, and outstanding for \$25—then the excellent option becomes the new compromise and the consumer selects it. It is easy to verify that such behavior cannot be rationalized by any standard type. In contrast, it can be captured by a GSP type with $\bell = (\text{outstanding}, \text{excellent}, \text{good}, \text{mediocre})$ and choice index $k = 2$. More generally, $\bell$ can represent a ranking along a one-dimensional Pareto frontier—such as price versus quality or aesthetics versus fuel efficiency—and the choice index $k > 1$ reflects the consumer’s preference for intermediate (compromise) options. See Proposition~\ref{prop:gsp_regularity_violations} for a formal illustration.

Beyond the compromise effect, GSP types with $k > 1$ can capture a broader class of \emph{context-dependent} choice behaviors studied in the behavioral economics literature \citep{tversky1993context},  wherein the consumer's choice is influenced by the specific offer set presented. For example, \citet{kamenica2008contextual} analyze settings where uninformed consumers infer the value of a product from the structure of the product line. Suppose a buyer is uncertain how to evaluate computers based on CPU speed. If the buyer believes that most of the population generally cares more about speed and only a few people value speed less than her, she may infer that selecting the second-slowest computer in the product line aligns with her own preference relative to the population. This is consistent with a GSP type whose ordering $\bell$ ranks computers from slowest to fastest and whose choice index is $k = 2$. Similarly, \citet{sen1993internal} describes a social scenario in which a guest is offered slices of cake. If the guest must choose between slices of size $x$ and $y$ with $x < y$, she may select the smaller slice $x$ to avoid appearing greedy. However, if a larger slice $z > y$ is added to the menu, the guest may now choose $y$, which has become the second-largest slice. This form of a positional choice rule—selecting the second-largest available option—is again incompatible with standard types but naturally arises from a GSP type with $\bell = (z, y, x)$ and $k = 2$.

\paragraph{\bf Lack of attention.} An alternative behavioral rationale for the choice process of a non-standard consumer type is \emph{inattention} (or \emph{limited attention}) on the part of the consumer when making a decision; see, e.g.,~\cite{reutskaja2011search,cattaneo2017random}. Under this interpretation, $\bell$ continues to represent the consumer’s underlying preference list, but due to inattention, the consumer does not always select the most preferred available product. To illustrate, consider a standard RUM model in which consumers choose the utility-maximizing alternative from a given offer set. In contrast, under limited attention, the consumer may instead select the second-highest utility alternative with some non-zero probability. In Section~\ref{sec:general_RUM_models}, we formalize this idea by introducing a generalization of the RUM model in which a consumer chooses the product with the $k^{\text{th}}$ highest utility in the offer set with probability $\lambda_k \geq 0$. We refer to the resulting model as the \emph{Generalized Random Utility Maximization (GRUM)} model, and show that it is strictly contained within the GSP model class.

{\bf Choice under uncertainty}. A final interpretation of non-standard types is that they capture consumer choice behavior under uncertainty. \citet{wernerfelt1995rational} proposed a mathematical model to explain violations of regularity in environments where consumers face product uncertainty. In such settings, customers may lack sufficient information to identify the highest utility product directly and instead rely on relative comparisons or assumptions about the product category to form preferences. The resulting behavior, while not strictly utility-maximizing in the traditional sense, aligns naturally with a GSP customer type having a choice index $k > 1$, as discussed below.

To illustrate this model, we adapt the main example from \citet{wernerfelt1995rational}. A critical attribute in coffee selection is the roasting level. It is typically categorized as (1) light roast which preserves beans' natural acidity, (2) medium roast which balances acidity, sweetness and bitterness, and (3) dark roast which results in a bold flavor without acidity. Consider a consumer whose preferences are such that medium roast is preferred over dark roast, and dark roast over light roast. In utility terms, this can be written as $U(2) > U(3) > U(1)$. Now, imagine the consumer is abroad and visits a coffee shop offering two options $\{A,B\}$: with coffee A having a ``lighter'' roast level compared to coffee B. However, the roasting levels of each individual coffee are not explicitly stated. 
This lack of explicit labeling leaves the consumer uncertain about the exact roasting levels. Although the consumer can visually observe that coffee B is darker than coffee A, she cannot figure out in absolute terms whether the coffees A and B correspond to light roast (1), medium roast (2), or dark roast (3). From their perspective, the assortment $S$ being offered could correspond to one of the following three sets: \{1,2\}, \{1,3\}, or \{2,3\}. One possible heuristic for the consumer in this scenario could be to select the darker coffee among the two being offered. 

To illustrate why this heuristic is reasonable, \citet{wernerfelt1995rational} assumes that the consumer is agnostic about which of the three possible offer sets is more likely and assigns an equal probability of $\frac{1}{3}$ to each. In this case, if the consumer chooses coffee $A$ (which is lighter than B), their (ex-ante) expected utility is $\Exp[U(A)] = \frac{1}{3}U(1) + \frac{1}{3}U(1) + \frac{1}{3}U(2) = \frac{2}{3}U(1) + \frac{1}{3}U(2)$. If the consumer chooses the darker coffee B instead, their expected utility is $\Exp[U(B)] = \frac{1}{3}U(3) + \frac{1}{3}U(3) + \frac{1}{3}U(2) = \frac{2}{3}U(3) + \frac{1}{3}U(2)$. Since $U(3)>U(1)$, coffee $B$ is chosen. This implies that if the offer set was $\set{1, 3}$ or $\set{2,3}$, the consumer would choose 3 (dark roast), whereas if the offer set was $\set{1, 2}$, the consumer would choose 2 (medium roast).
Now, suppose the shop offers three different types of coffee, all with distinct roasting levels. With this additional option, the consumer can observe the relative color differences and correctly infer the roasting levels of each coffee. In this case, they would choose the medium roast coffee (2) consistent with her preference. In summary, the consumer chooses the darker coffee whenever the offer set is of size 2, but chooses the medium roast if all three are offered. 

As this example illustrates, the consumer's choice arises from a utility-based decision made under uncertainty about the available alternatives.
It can be easily verified that this behavior is reflected by a single GSP customer type $(\bell,k)$ where the product ordering ranks the coffees from lighter to darker roast (i.e., $\bell=(1,2,3)$) and the consumer chooses the second one from the list among those offered (i.e., $k=2$). 
We emphasize that the product ordering $\bell$ is different from the consumer's preference list $2 \succ 3 \succ 1$. \citet{wernerfelt1995rational} also showed how the regularity violation from the above example can occur in contexts where there are multiple consumer types (with different preferences) together with non-uniform priors about the possible offer sets.}
\subsection{Representation power of GSP models}
A choice model $\mathcal{P}$ is said to satisfy \textit{regularity} (or weak rationality) if
\begin{equation} \label{regularity_equation}
\mathcal{P}(j, S) \geq \mathcal{P}(j, S\cup\{i\}) \text{ for every } j \in S, i \notin S, S \subseteq [n].
\end{equation}
In words, the probability of choosing an alternative cannot increase when another product is added to the offer set. It is well-known that any model in the SP class must satisfy regularity, see, e.g., \citet{block1960random}. However, many experiments in behavioral economics have revealed that regularity is often violated in practice. The following result illustrates that the GSP model admits regularity violations:
\av{\begin{proposition}[GSP choice model strictly subsumes the SP model class]\label{prop:gsp_regularity_violations}
Every instance of the SP model class is contained inside GSP. Moreover, there exists an instance of the GSP model that admits regularity violations, and, therefore, GSP is strictly larger than SP.
\end{proposition}}
\label{sec:gsp_power}
\proof{Proof.}
The fact that the SP model class is subsumed within GSP follows immediately from the definition of the GSP model. Next, we show that there exists a GSP model instance that can explain the findings of the famous~\cite{simonson1992choice} choice experiment, which exhibited regularity violations. In the experiment, participants were asked to choose amongst three different camera models: (1) Minolta X-370, (2) Minolta MAXXUM 3000i, and (3) Minolta MAXXUM 7000i. Participants were split into two conditions; those under the first condition (condition 1) were asked to choose between alternatives $\{1,2\}$, whereas those under the second condition (condition 2) had to choose a camera among all three. The experiment results are shown below.
$$
\begin{array}{l|l|l|l}
\text{Model} & \text{Price (USD)} & \text{Condition 1}:  \{1,2\} & \text{Condition 2}:  \{1,2,3\} \\
\hline
\text{1: X-370 } & 169.99 & 50\% & 22\%  \\
\text{2: MAXXUM 3000i }& 239.99 & 50\% & 57\% \\
\text{3: MAXXUM 7000i } & 469.99 &  \text{not available} & 21 \%
\end{array}
$$
Since the likelihood of selecting camera 2 increased from 50\% to 57\% when the choice set was enlarged from $\{1,2\}$ to $\{1,2,3\}$, the experiment violates regularity.

Next, we show that these results can be explained \emph{exactly} with a GSP model that has only four consumer types, shown in Table \ref{simonson_tversky_example_cameras_solution}.
\begin{table}[h]
\centering
\tabcolsep=10pt
\begin{tabular}{|c|c|c|c|}\hline
  \multicolumn{3}{|c|} {Consumer type} & Probability \\ \hline
  label & $\bell$ & $k$ & \\
  \hline
  1 & (1,3,2) & 1 & 0.22 \\
  \hline
  2 & (2,3,1) & 1 & 0.29 \\
  \hline
  3 & (3,2,1) & 1 & 0.21 \\
  \hline
  4 & (3,2,1) & 2 & 0.28 \\
  \hline
\end{tabular}
\caption{A GSP model that explains the \citet{simonson1992choice} experiment}
\label{simonson_tversky_example_cameras_solution}
\end{table}
The first three consumer types are standard, whereas the fourth type is non-standard. 
In particular, this type has the ordering $\bell = (3,2,1)$ which corresponds to a ranking of the cameras in decreasing order of prices. 
Consequently, when all cameras are offered, this consumer type will choose the MAXXUM 3000i (camera 2), whereas when only cameras 1 and 2 are offered, the type chooses the X-370 model (camera 1). This behavior precisely captures the compromise effect since the type chooses the camera with the intermediate price when presented with all the three cameras.

It can be shown that the GSP model in Table~\ref{simonson_tversky_example_cameras_solution} generates the same choice probabilities as observed in the experiment. For illustration, we calculate the choice probabilities $\mathcal{P}(2,\{1,2\})$ and $\mathcal{P}(2,\{1,2,3\})$ and show that they coincide with the experimental results. The remaining observations can be verified by following the same procedure. From Definition~\ref{def:gsp_prob_fn}, it follows that
\begin{eqnarray*}
  \mathcal{P}(2,\{1,2\}) &=&  \sum_{\omega=1}^4 C_\omega(2, \set{1,2}) \cdot \lambda(\omega) \\
  &=&
  0.22\cdot C_1(2,\{1,2\}) + 0.29\cdot C_2(2,\{1,2\}) + 0.21\cdot C_3(2,\{1,2\}) + 0.28\cdot C_4(2,\{1,2\})\\
  &=& 0 + 0.29 + 0.21 + 0 = 0.50, \\
  \mathcal{P}(2,\{1,2,3\}) &=&  \sum_{\omega=1}^4 C_\omega(2, \set{1,2,3}) \cdot \lambda(\omega) \\
  &=& 0.22\cdot C_1(2,\{1,2,3\}) + 0.29\cdot C_2(2,\{1,2,3\}) + 0.21\cdot C_3(2,\{1,2,3\}) + 0.28\cdot C_4(2,\{1,2,3\})\\
  &=& 0 + 0.29 + 0 + 0.28 = 0.57
\end{eqnarray*}
Note that by adding the most expensive camera (alternative 3) to the choice set, the probability of choosing alternative 2 increased by 7\%. This is the result of gaining consumer type 4 (with proportion 28\%) and losing consumer type 3 (with proportion 21\%) which now chooses alternative 3.
 \Halmos \endproof
We highlight that the above result differentiates the GSP model from a majority of choice models in extant literature that go beyond the SP class.
In Appendix~\ref{sec:more_regularity_violations}, we provide additional examples of choice experiments exhibiting regularity violations---including the well-documented {\em attraction or decoy effect}~\citep{huber1982adding,simonson1989choice,simonson1992choice}---that can be explained by the GSP model. 

 \av{
The next two results shed further light into the representation power of GSP models.

 \begin{proposition}[Relationship between GSP and regular choice models]
\label{prop:regular_but_not_gsp}
There exist regular choice models that lie outside the SP model class but belong to the GSP class. However, not all regular choice models belong to the GSP family.
\end{proposition}
\begin{proposition}[Sufficient condition for regularity of a GSP model]
\label{prop:when_is_gsp_regular}
    Consider a GSP model with distribution $\lambda(\cdot)$ over consumer types $\Omega$ satisfying the following inequalities:
    \[ \lambda((\bell, k)) \geq \lambda((\bell, k+1)) \text{ for all } \bell \in \Lscr, k \in [n-2] \] 
    Then, the corresponding GSP instance belongs to the class of regular choice models. However, the condition is not necessary to ensure regularity.
\end{proposition}
\begin{figure}[h]
	\centering
	\includegraphics[width=0.8\textwidth]{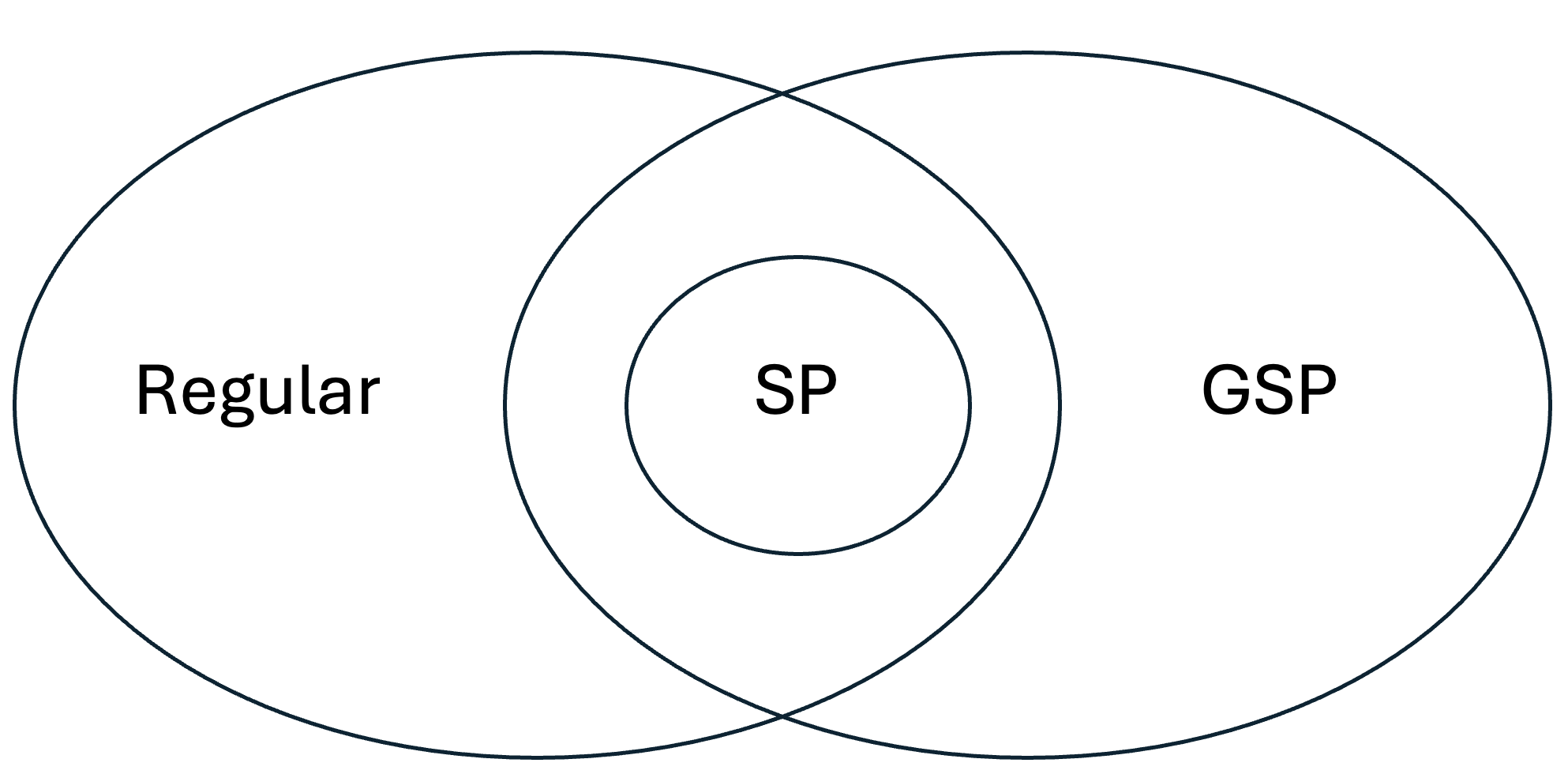}
\caption{\av{Venn diagram illustrating the relationships among the three families of choice models: regular, stochastic preference (SP), and the generalized stochastic preference (GSP).}}\label{fig:GSP_vs_regular}
\end{figure}

Figure~\ref{fig:GSP_vs_regular} summarizes the relationship between SP, GSP, and the class of regular choice models. Proposition~\ref{prop:when_is_gsp_regular} establishes that a GSP model remains regular as long as the proportion of consumers is non-increasing in the choice index. Beyond providing a simple, testable criterion, this result also offers a practical way to control model complexity during estimation by imposing constraints on the type proportions. We elaborate on this point in our discussion on model estimation in Section~\ref{sec:estimation}, where we also explore alternative strategies for regularizing GSP models.
We note that providing a complete characterization of when a given GSP instance satisfies regularity appears to be a challenging problem. To the best of our knowledge, no such characterization exists for any known model in the literature that subsumes the SP model.}
\section{Generalizing the RUM model class}
\label{sec:general_RUM_models}
One limitation of both the SP and GSP models is that they cannot be used to predict consumer choices when a new product is introduced into the market, and/or some features (e.g., price) of existing products are changed. This is because under both models, products are viewed as a fixed collection or bundle of attributes. \av{In this section, we show that our notion of non-standard types can be incorporated into the random utility maximization (RUM)  model class as well, providing the ability to capture dependence on product~features.}

Models in the RUM class assume that each product is associated with a stochastic utility, say $U_i$, and the consumer chooses the product in the offer set with the highest utility. 
Therefore, the choice probability under a general RUM model is given by
\begin{equation}
\label{eq:rum_probs}
    \Pcal(j, S) = \Pr\left(U_j = \max_{i \in S}~U_i\right) \;\;\ \text{for all } S \subseteq [n] \text{ and all } j \in S
\end{equation}

Different specifications on the joint distribution over product utilities $\bm{U} = (U_1, U_2, \ldots, U_n)$ result in different choice models within the RUM class. Prominent examples include the classical MNL model, the nested logit model, the Markov chain choice model and the exponomial choice model; see~\cite{feng2022consumer} for a detailed description of these models.




Our generalization of the RUM model class, which we term the {\em Generalized RUM (GRUM)} model, specifies the choice behavior of consumers as follows: like in the RUM model, consumers associate a random utility $U_i$ with each alternative $i \in \Cscr$. \av{Unlike the RUM model, however, where consumers choose the product with the highest utility, the GRUM model supposes that each customer first samples a choice index $k \in [n]$, and then selects the product with the $k^{th}$ highest utility in the offer set.\footnote{An alternative way is to define a population comprising different customer types, where a customer of type $k$ selects the product with the $k^{\text{th}}$ highest utility in the offer set. We adopt this formulation to align more closely with the representative consumer view of choice models~\citep{anderson1988representative,feng2017relation}, and we thank an anonymous referee for this helpful suggestion.} If $k$ is larger than the number of products offered, the customer chooses the available alternative with the smallest utility.} 


\av{Next, we formally define the parameters of the GRUM model, and the resulting choice probabilities. Because only the relative ordering of the utilities matter as far as the consumer choice is concerned, we can fully specify a general joint distribution over $\bm{U}$ using $n!$ ($n$ factorial) parameters, one for each ordering of the utilities. For each ordering $\bell = (\ell_1, \ell_2, \ldots, \ell_n) \in \Lscr$, define $\alpha_{\bell} = \Pr(U_{\ell_1} > U_{\ell_2} > \ldots > U_{\ell_n})$. Then, the parameters of the GRUM model comprise $\balpha =\left(\alpha_{\bell} \colon \bell \in \Lscr \right)$ and the vector of probabilities governing the sampled choice index $\bm{\lambda} = (\lambda_1, \lambda_2, \ldots, \lambda_n)$.

Given the above, let $\rank(i, S; \bm{U})$ denote the {\em random} position of product $i$'s utility in the relative ordering of the product utilities $\left(U_j \colon j \in S\right)$. That is, if $\rank(i, S; \bm{U}) = 1$, it means that $U_i > U_j$ for all $j \in S \setminus \set{i}$.
With this notation, we first introduce the following definition:
\begin{definition}[choice probability for a fixed choice index $k$ in grum  model]
\label{def:krum_choice_fn}
Faced with a choice set $S \subseteq [n]$, the probability, say $\pi_{k}(j, S; \bm{\alpha})$, that a consumer who samples choice index $k \in [n]$ chooses alternative $j \in S$ is given by:
\begin{displaymath}
\pi_{k}(j,S;\balpha) \coloneqq \Pr\bigg(\rank(j, S; \bm{U}) = \min(k, \abs{S})\bigg) = \sum_{\bell \in \Lscr} \alpha_{\bell} \cdot C_{(\bell, k)}(j, S),
\end{displaymath}
\end{definition}
where the equality follows from the definition of $C_{(\bell, k)}(j, S)$ in Definition~\ref{def:gsp_choice_fn}. 
Using the above, the class of GRUM choice models is defined as follows:
\begin{definition}[choice probability under grum model]
\label{def:krum_prob_fn}
A discrete choice model $\Pcal$ belongs to the GRUM model class if there exists a distribution $\balpha$ over product orderings $\Lscr$, and
a distribution $\bm{\lambda}$ over the choice index $k \in [n]$ such that for all $S \subseteq [n]$ and all $j \in S$,
\[ \Pcal(j, S) \coloneqq \sum_{k=1}^{n} \pi_k(j, S; \balpha) \cdot \lambda_k = \sum_{k=1}^n \left(\sum_{\bell \in \Lscr} \alpha_{\bell} \cdot C_{(\bell, k)}(j, S)\right) \cdot \lambda_k, \]
\end{definition}
where the equality follows from Definition~\ref{def:krum_choice_fn}.}
\av{Now, comparing Definitions~\ref{def:krum_prob_fn} and~\ref{def:gsp_prob_fn}, it is not too hard to see that any GRUM model is an instance of the GSP class; see Proposition~\ref{prop:grum_contained_in_gsp} for the formal result. However, because the GSP model uses significantly more parameters---$(n-1) \times n! - 1$ corresponding to the proportions of each consumer type in $\Omega$---compared to the GRUM model---$(n! - 1) + (n - 1)$ corresponding to the distribution $\balpha$ over product orderings and the distribution $\bm{\lambda}$ over the choice index---the GRUM model is strictly contained within GSP; the proof of Proposition~\ref{prop:grum_contained_in_gsp} also constructs a specific instance of GSP that is not a member of the GRUM class.}

While the GRUM class is more restricted than GSP, it is still rich enough to capture regularity violations. 
In fact, parametric specifications for the joint utility distribution can be exploited to design tractable estimation algorithms, as described in Section~\ref{sec:kmnl_est}. 
\subsection{Generalizing any subclass of RUM models}
\label{sec:generalizing_rum_subfam}
In the same way that the GRUM model was defined based on a general utility distribution, we can incorporate non-rationality into any subclass of RUM models by simply restricting the utility distribution to belong to the specific subclass. This allows us to extend {\em any} class of choice models that belong to the RUM family into one that can accommodate choice behaviours inconsistent with the utility maximization principle. We illustrate this idea by applying it to the most widely used choice model class in the RUM family, the multinomial logit (MNL) model. 

 
\subsubsection{Generalized MNL (GMNL) model.}
We consider here a subclass of the GRUM model, which we call Generalized MNL (GMNL), where the joint distribution on product utilities is consistent with the MNL model. That is, we assume that the utility of product $i \in [n]$ is of the form 
$ U_{j} = v_{j} + \varepsilon_j$, 
where $v_j$ is the deterministic component that is typically a function of product features like price, brand, etc., and $\varepsilon_1, \ldots, \varepsilon_n$ are i.i.d.\ standard Gumbel errors. 
This assumption on product utilities provides a simple closed-form expression for choice probabilities under the MNL model, which is one of the main reasons for its widespread popularity. We show next that by exploiting properties of the Gumbel distribution, choice probabilities under the GMNL model can also be derived in closed form.

For succinct notation, for any offer set $S$ define $w(S) := \sum_{j \in S} e^{v_{j}}$, and given any $j \in S$, denote $\osminus{S}{j} := S \setminus \set{j}$. Further, \av{with a slight abuse of notation}, we denote $\pi_k(i, S; \prefvec)$ as the probability that \av{a consumer with sampled choice index $k$} chooses product $i$ from offer set $S$. When $k=1$, we have the standard MNL choice probability expression:
\[ \pi_1(i, S; \prefvec) = \Pr(\rank(i, S; \prefvec) = 1) = \frac{e^{v_i}}{w(S)} = \frac{e^{v_i}}{\sum_{j \in S} e^{v_j}} \]
For $k > 1$, we can compute the choice probability through an efficient recursion: 
\begin{proposition}[Recurrence for choice probabilities under GMNL model]
\label{prop:kmnl_prob_recurrence}
The following recurrence holds for any choice index $k > 1$ in the GMNL model:
\[ \pi_k(i, S; \prefvec) =
\begin{cases}
\pi_{\abs{S}}(i, S; \prefvec) & \text{if } k > \abs{S} \\
\sum_{j \in \osminus{S}{i}} \pi_1(j, S; \prefvec) \cdot \pi_{k-1}(i, \osminus{S}{j}; \prefvec) & \text{otherwise.}
\end{cases}
\]
\end{proposition}

\av{To conclude this section, we present an instance of the GMNL model that violates regularity and, therefore, shows that the model lies outside the RUM class.
Finally, Proposition~\ref{prop:grum_regular} shows that any GRUM model—and therefore any GMNL model—satisfies regularity if the distribution governing the sampled choice index satisfies $\lambda_1 \geq \lambda_2 \geq \cdots \geq \lambda_n$, by adapting the result from Proposition~\ref{prop:when_is_gsp_regular}.}

\section{Model estimation}\label{sec:estimation}
 In this section, we show how to estimate the parameters of our proposed models given aggregate choice data. 
 Specifically, we assume access to a dataset consisting of choice observations or {\em sales} \footnote{We use the term `sales' since our estimation framework can be readily applied to a retail setting where the firm has access to sales transaction data.} for a collection of $T$ offer sets $S_1, S_2, \ldots, S_T$. We denote by $N_{jt} \geq 0$ the number of sales of product $j \in S_t$, and $M := \sum_{t\in [T]} \abs{S_t}$ as the total number of observations. 
  Such data are typically used for estimating choice models, see, e.g.~\cite{jagabathula2019limit} and \cite{berbeglia2022comparative}.
  We first describe our estimation method for the GSP model which leverages the Frank-Wolfe algorithm \citep{frank1956algorithm}, and then propose an expectation-maximization (EM) algorithm for estimating the GMNL model.

\subsection{Frank-Wolfe algorithm for estimating the GSP model}
\label{sec:gsp_est}
We use maximum likelihood estimation (MLE) to estimate the distribution $\lambda$ over the GSP types $\Omega$:
\begin{align*}\tag{{\sf GSP MLE Problem}}
\label{eq:unconstr_mle}
&\argmax_{\lambda}~\LL(\lambda) := \;\; \sum_{t=1}^T \sum_{j\in S_t} N_{jt} \log \left( \sum_{\omega \in \Omega} C_\omega(j, S_t) \cdot \lambda(\omega)  \right) \\
&\text{s.t.} \;\; \lambda(\omega) \geq 0\;\ \forall~ \omega \in \Omega, \sum_{\omega \in \Omega} \lambda(\omega)=1, \quad \text{and}, \sum_{\omega \in \nonrattypes} \lambda(\omega) \leq \delta, \nonumber
\end{align*}
where $\nonrattypes := \cup_{k=2}^{n-1} \Omega_k$ denotes the set of all non-standard types in the GSP model. We also define $\rattypes := \Omega \setminus \nonrattypes = \Omega_1$ as the set of all standard types. 

%
%
%
%
The~\ref{eq:unconstr_mle} is similar to the problem considered for estimating the SP model in prior work~\citep{farias2013nonparametric,van2014market,van2017expectation,jagabathula2019limit}. For the rank-based model, one typically seeks for a distribution over rankings that is {\em sparse}, i.e., the number of consumer types (or rankings) that have non-zero probability is small, to control the model complexity and prevent overfitting to the observed sales data. For the GSP model, besides finding a sparse distribution, it may also be beneficial to find a distribution where the probability mass assigned to the non-standard consumer types is low. This is because many of the controlled choice experiments that provide evidence for non-rationality in consumer choices show that the rationality violations are typically of a small scale. 
To that end, we limit the total proportion of non-standard types using a hyperparameter $\delta \in [0, 1]$ that
%
%
balances the goodness-of-fit, that is, log-likelihood on the observed data, and out-of-sample prediction error. Consequently, we tune $\delta$ via cross-validation methods in our numerical study in Section~\ref{sec:pred_study}. Note that if we set $\delta = 0$, we only allow standard types in the GSP model and the~\ref{eq:unconstr_mle} reduces to the MLE problem for the SP/rank-based model studied in prior literature. 

It can be verified that the~\ref{eq:unconstr_mle} is a concave maximization problem, see, e.g.~\cite{jagabathula2019limit}. However, solving it is challenging since the feasible region involves $\abs{\Omega} = (n-1) \times n!$ variables, one for each type $\omega \in \Omega$, which quickly becomes intractable even for small values of $n$. To address this issue, we leverage the Frank-Wolfe algorithm~\citep{frank1956algorithm,jaggi2013revisiting}, aka the conditional gradient method, that was employed by~\cite{jagabathula2019limit} to estimate the rank-based model. The Frank-Wolfe (FW) algorithm starts with an initial set of consumer types $\Omega^{(0)}$ and a distribution $\lambdak{0}$ over $\Omega^{(0)}$, and generates a sequence $\left(\lambdak{r} \colon r \geq 1\right)$ of improving solutions to the~\ref{eq:unconstr_mle} by optimizing linear approximations of the negative log-likelihood objective\footnote{The Frank-Wolfe algorithm is used for minimizing convex functions and therefore, we need to consider the negative log-likelihood.} over the feasible region in each iteration. In particular, in each iteration  $r \geq 1$, the FW algorithm solves the following linear program (LP):
\begin{align*} \tag{{\sf Frank-Wolfe Subproblem}}
\label{eq:type_subprob}
&\max_{\lambda} \;\; \sum_{\omega \in \Omega} \left(\sum_{t=1}^T  \sum_{j \in S_t} \mus{r-1}_{jt} \cdot C_\omega(j, S_t)\right) \cdot \lambda(\omega) \\
&\text{s.t.} \;\; \lambda(\omega) \geq 0\;\ \forall~ \omega \in \Omega, \sum_{\omega \in \Omega} \lambda(\omega)=1 \;\;\; \text{and} \;\;\; \sum_{\omega \in \nonrattypes} \lambda(\omega) \leq \delta,
\end{align*}
where $\mus{r-1}_{jt} = \frac{N_{jt}}{\Pcal^{(r-1)}(j, S_t)}$ and $\Pcal^{(r-1)}(j, S_t)$ is the probability of choosing product $j$ from offer set $S_t$ under a GSP model with distribution $\lambdak{r-1}$ over the consumer types. 
Denoting $\tlambdak{r}$ as an optimal solution to the~\ref{eq:type_subprob}, the FW algorithm updates the solution $\lambdak{r}$ as a convex combination of $\lambdak{r-1}$ and $\tlambdak{r}$, which guarantees a feasible solution to the~\ref{eq:unconstr_mle}. The convex combination weight is typically chosen via linesearch~\citep{jaggi2013revisiting} to ensure the maximum improvement in the log-likelihood objective.

It follows from the above description that the performance of the FW algorithm depends on the ability to solve the~\ref{eq:type_subprob} (reasonably) efficiently. The following result characterizes the optimal solution of the~\ref{eq:type_subprob}, 
where for each $\omega \in \Omega$, we denote $\logll(\omega) = \sum_{t=1}^T \sum_{j \in S_t} \mus{r-1}_{jt} \cdot C_\omega(j, S_t)$:
\begin{proposition}[Frank-Wolfe adds at most two consumer types in each iteration]
\label{prop:type_subprob_optimal}
Let $\omega_{\rat} \in \argmax_{\omega \in \rattypes} \logll(\omega)$ and $\omega_{\nonrat} \in \argmax_{\omega \in \nonrattypes} \logll(\omega)$, where recall that $\Omega = \rattypes \cup \nonrattypes$. Then, an optimal solution to the~\ref{eq:type_subprob}, say $\tilde{\lambda}$,  is of the form:
\begin{itemize}
\item If $\logll(\omega_{\rat}) \geq \logll(\omega_{\nonrat})$, then $\tlambda(\omega_{\rat}) = 1$ and $\tlambda(\omega) = 0$ for all $\omega \in \gsptypes \setminus \set{\omega_{\rat}}$.
\item Otherwise, $\tlambda(\omega_{\rat}) = 1 - \delta$, $\tlambda(\omega_{\nonrat}) = \delta$, and $\tlambda(\omega) = 0$ for all $\omega \in \gsptypes \setminus \set{\omega_{\rat}, \omega_{\nonrat}}$.
\end{itemize}
\end{proposition}
The proof is given in Appendix~\ref{app:extreme_point_optimal} and is a consequence of the fact that every LP has an extreme point (of the feasible region) that is an optimal solution.
The above result implies that in each iteration $r \geq 1$ of the FW algorithm, we add either a single standard type, or one standard type {\em and} one non-standard type, to the existing set of consumer types $\Omega^{(r-1)}$. Based on this characterization, we now discuss our solution procedure for the
~\ref{eq:type_subprob}. 


\subsubsection{IP formulation for finding improving solutions.} 
~\cite{jagabathula2019limit} showed that solving the~\ref{eq:type_subprob} (which they termed as the {\sc Rank Aggregation LP}) for the rank-based model, that is, when $\delta = 0$, is NP-hard. Therefore, we resort to integer linear program (ILP) formulations.

\av{For finding the standard type $\omega_{\rat}$, we can leverage the ILP proposed in the {\em market discovery (MD)} algorithm for estimating the rank-based model in~\citet[Section 4.3.2]{van2014market}\footnote{
While Van Ryzin and Vulcano~\citeyear{van2014market} derived their algorithm using duality arguments, it can be shown that the subproblem they solve in each iteration (termed the {\em type discovery subproblem}) is identical to the~\ref{eq:type_subprob}.} In a similar fashion, we show that the non-standard type $\omega_{\nonrat}$ can also be obtained by solving an ILP, which we present in Appendix~\ref{app:single_ilp}. However, this formulation does not scale well because we have an integer-valued variable for the choice index $k$ that can take values in the set $\set{2, 3, \ldots, n-1}$. Aside from computational concerns, allowing the choice index to take large values increases the risk of overfitting and may degrade prediction performance on unseen offer sets. Restricting the choice index to be ``small'' can provide a natural inductive bias that can help mitigate overfitting—particularly when transaction data is limited. A similar insight was leveraged in the GPT-based estimation procedure for the GSP model in \citet{jena2022estimation}. 

Motivated by the empirical findings in~\cite{jena2022estimation}, we restrict the non-standard type to have choice index $k \leq k_{\max}$, where $k_{\max} = O(1)$. This constraint offers the advantage that now we only need to determine the ``best'' ordering for each fixed $k \in \{2, \ldots, k_{\max}\}$, for which we extend the ILP for finding rankings in~\cite{van2014market} to accommodate choice indices $k > 1$.}

Define $\bm{x} = \left(x_{ij} \colon i, j \in [n]\right)$ to be a vector of binary decision variables that encodes the ordering $\bell$, that is, we have a binary variable $x_{ij}$ for all pairs of products $(i, j)$ such that $x_{ij}=1$ if and only if product $i$ appears before $j$ in $\bell$. Further, define $\bm{y} = \left(y_{jt} \colon j \in S_t, t \in [T]\right)$ to be a vector of binary decision variables, with $y_{jt} = C_{\omega}(j, S_t)$ for each $j \in S_t$ and $t \in [T]$. Then, the ILP is as follows:
\begin{subequations}
\label{eq:gsp_mip}
\begin{align}
\max_{\bm{y}, \bm{x}} \;\;\; &\sum_{t=1}^T \sum_{j \in S_t} \mus{r-1}_{jt} \cdot y_{jt} \label{eq:mip_obj}\\
\text{s.t. } \;\;\; &x_{ji} + x_{ij} = 1 \quad \forall ~ i,j \in \Cscr;  i\neq j \label{eq:rank_constr1}\\
&x_{ji} + x_{il} + x_{lj} \leq 2 \quad \forall~i,j,l \in \Cscr; i\neq j \neq l \label{eq:rank_constr2}\\
&y_{jt} \leq x_{ij} \;\;\;\; \forall~j \in S_t, \forall~i \in \osminus{S_t}{j} \;\;\; \forall~t \in [T] \text{ s.t. } \abs{S_t} \leq k \label{eq:feas_constr1}\\
&\av{(k-1) \cdot y_{jt} + \sum_{i \in \osminus{S_t}{j}} x_{ji} \leq \abs{S_t} -1 \;\;\;\; \forall~j\in S_t, \forall~t \in [T] \text{ s.t. } \abs{S_t} > k \label{eq:feas_constr2}}\\
&\av{(k-\abs{S_t}) \cdot y_{jt} + \sum_{i \in \osminus{S_t}{j}} x_{ji} \geq 0 \;\;\; \forall~j\in S_t, \forall~t \in [T] \text{ s.t. } \abs{S_t} > k \label{eq:feas_constr3}}\\
&x_{ij} \in \set{0, 1}  \;\; \forall~i,j \in \Cscr;  \;\; y_{jt} \in \set{0, 1} \;\; \forall~j\in S_t, \forall~t \in [T] \label{eq:var_domains}
\end{align}
\end{subequations}
Constraint~\eqref{eq:rank_constr1} ensures that any two products are ordered in $\bell$, i.e., for all $i \neq j$, either $i$ appears before $j$ or $j$ appears before $i$. 
Constraint~\eqref{eq:rank_constr2} enforces transitivity so that $\bm{x}$ corresponds to a valid permutation of the products. Next, depending on whether the (fixed) choice index $k$ is greater or less than the size of the offer set $S_t$, the set of constraints added to the program are different. If $k \geq \abs{S_t}$, then the {\em last} product in the subsequence $s(\bell, S_t)$ would be chosen. This means that $y_{jt} = 1$ only if product $j$ appears before all other products $i \in \osminus{S_t}{j}$ in the ordering $\bell$, which can be equivalently written as~\eqref{eq:feas_constr1}. Otherwise, if $k < \abs{S_t}$, then $y_{jt} = 1$ implies that product $j$ is at position $k$ in the sequence $s(\bell, S_t)$, which is equivalent to the condition $\sum\limits_{i \in \osminus{S_t}{j}} x_{ji} = \abs{S_t} - k$. The inequality constraints~\eqref{eq:feas_constr2}-\eqref{eq:feas_constr3} enforce this condition.

\av{After solving ILP~\eqref{eq:gsp_mip}, the corresponding ordering, say $\bell^{(k)}$, can be constructed from the vector $\bm{x}$ as follows: the position of product $i \in [n]$ is given by $\sum\limits_{j \in \osminus{[n]}{i}} x_{ji} + 1$. Then, we select the non-standard type  
$(\bell^{(k)}, k)$ that achieves the largest objective~\eqref{eq:mip_obj} among $k \in \set{2, \ldots, k_{\max}}$. 
Having determined the optimal standard and non-standard types to add to the current set of types $\Omega^{(r-1)}$, we update the distribution $\lambdak{r}$ as the optimal solution to the~\ref{eq:unconstr_mle} restricted to the updated set of types $\Omega^{(r)}$. To solve this problem, we leverage the efficient EM algorithm proposed in~\cite{van2017expectation}, which can be easily extended to incorporate the upper bound constraint on the proportion of non-standard types. The entire procedure is summarized in Algorithm~\ref{alg:gsp_fw} in Appendix~\ref{app:single_ilp}.}

\subsection{Expectation-Maximization algorithm for estimating the GMNL model}
\label{sec:kmnl_est}
 Recall from Section~\ref{sec:generalizing_rum_subfam} that the GMNL model is described by the vector of product mean utilities $\prefvec = (v_1, v_2, \ldots, v_n)$ and the vector of type proportions $\blambda = (\lambda_1, \lambda_2, \ldots, \lambda_n)$. Similar to the case of the GSP model above, we estimate the parameters using MLE:
\begin{align}
\argmax_{\prefvec, \blambda} \;\; &\LL(\prefvec, \blambda) := \sum_{t=1}^T \sum_{j \in S_t} N_{jt} \log \left(\sum_{k=1}^n \pi_k(j, S_t; \prefvec) \cdot \lambda_k \right) \label{eq:kmnl_mle} \tag{{\sf GMNL MLE Problem}}\\
&\text{s.t. } \lambda_k \geq 0 \;\; \forall~k \in [n], \sum_{k\in [n]} \lambda_k = 1 \nonumber
 \end{align}
Compared to the standard MNL model, the log-likelihood function for the GMNL is no longer concave. Further, direct maximization is challenging due to the (i) recursive expressions for the choice probability functions $\pi_k(\cdot)$, and (ii) missing information about the consumer type involved in each transaction. To address these issues, we propose an efficient expectation-maximization (EM) algorithm to solve the~\ref{eq:kmnl_mle}. 

While the EM algorithm can be derived for the general case, to keep the exposition simple, we consider the setting where $\lambda_k = 0$ for all $k \geq 3$, i.e., each consumer picks either the product with the highest ($k=1$) or the second-highest ($k=2$) utility. 
\av{Algorithm~\ref{alg:gmnl_em} outlines the procedure for this special case of the GMNL model, which we term GMNL(2); see Appendix~\ref{app:gmnl_em_derivation} for the derivation.
The algorithm for the general case can be obtained in a similar way, although with significantly more complex notation.}
\begin{algorithm}[h]
\caption{\av{EM algorithm for estimating the parameters of GMNL(2) model}}\label{alg:gmnl_em}
\begin{algorithmic}
\STATE \textbf{Initialization:} $\piter{0}_1 \in (0,1)$, $\vveciter{0} \in \Real^n$
\FOR{$r = 1, 2, \ldots$}
\STATE {\bf E-step:} Compute the following quantities
\[ \alpha_{jt}^{(r)} := \frac{ \piter{r-1}_1 \cdot \pi_1(j, S_t; \vveciter{r-1})}{\piter{r-1}_1 \cdot \pi_1(j, S_t; \vveciter{r-1}) + (1-\piter{r-1}_1) \cdot \pi_2(j, S_t; \vveciter{r-1})} \quad \forall~j \in S_t, t \in [T] \]
\[ \beta_{jt}^{(r)}(i) := \frac{\pi_1(i, S_t; \vveciter{r-1})\cdot \pi_1(j, \osminus{S_t}{i}; \vveciter{r-1})}{\sum_{l \in \osminus{S_t}{j}} \pi_1(l, S_t; \vveciter{r-1})\cdot \pi_1(j, \osminus{S_t}{l}; \vveciter{r-1})} \quad \forall~ i \in \osminus{S_t}{j}; \forall~j \in S_t, t \in [T]\]
\STATE {\bf M-step:} Update parameter estimates
\begin{align*} 
\piter{r}_1 &=  \frac{\sum_{t=1}^T\sum_{j\in S_t} N_{jt}\cdot \alpha_{jt}^{(r)}}{\sum_{t=1}^T\sum_{j\in S_t} N_{jt}} \\ \vveciter{r} & \in \argmax_{\prefvec} \sum_{t=1}^T \sum_{j \in S_t} N_{jt} \cdot \set{\alpha_{jt}^{(r)} \cdot \log \left(\frac{e^{v_{j}}}{w(S_t)}\right) + (1 - \alpha_{jt}^{(r)}) \cdot \sum_{i \in \osminus{S_t}{j}} \beta_{jt}^{(r)}(i) \cdot \log\left(\frac{e^{v_i}}{w(S_t)}\cdot \frac{e^{v_{j}}}{w(\osminus{S_t}{i})} \right)}
\end{align*}
\ENDFOR
\end{algorithmic}
\end{algorithm}
We note that the objective function in the M-step update for $\prefvec$ is concave since it is exactly the log-likelihood under an MNL model of the dataset obtained by (i) augmenting and weighting the original sales counts, and (ii) adding dummy (weighted) sales counts. In particular, for each offer set $S_t$, we augment the original sales counts with $N_{jt} \cdot  (1 - \alpha_{jt}^{(r)}) \cdot \beta_{jt}^{(r)}(i)$ sales for each product $i \in \osminus{S_t}{j}$ and all $j \in S_t$. In addition, for each $t \in [T]$, we add $\abs{S_t}$ additional offer sets of the form $\osminus{S_t}{j}$ for all $j \in S_t$, with
$N_{it} \cdot  (1 - \alpha_{it}^{(r)}) \cdot \beta_{it}^{(r)}(j)$ sales for each $i \in \osminus{S_t}{j}$. 
We leverage the MM algorithm~\citep{hunter2004mm} that provides closed-form updates for solving this problem.

The reader familiar with the literature on estimation methods for choice models might have noticed a resemblance between the EM algorithm outlined above and the one used to estimate the latent class MNL (LC-MNL) model, see, e.g.,~\cite{train2008algorithms}. The key distinction between the two approaches is that for the GMNL model, only a {\em single} MNL model needs to be estimated in each iteration, compared to $K$ MNL models for a $K$-class LC-MNL model. Consequently, the EM procedure for the GMNL model converges very fast in practice, as discussed in our numerical results in Section~\ref{sec:pred_study}.

\av{We conclude this section with a discussion on how to modify the estimation procedure to ensure the fitting of a regular GMNL model. For the GMNL(2) model, enforcing the conditions of Proposition~\ref{prop:when_is_gsp_regular} reduces to the inequality $\lambda_1 \geq \lambda_2$, which is equivalent to imposing the constraint $\lambda_1 \geq 0.5$. It is straightforward to verify that this constraint can be easily incorporated into the EM procedure. Specifically, the only change in Algorithm~\ref{alg:gmnl_em} is the update for the proportion $\lambda_1$ in the {\bf M-step}, which now becomes:
\begin{equation}
\label{eq:regular_gmnl}
\lambda_1^{(r)} = \max\left(0.5, \frac{\sum_{t=1}^T\sum_{j\in S_t} N_{jt}\cdot \alpha_{jt}^{(r)}}{\sum_{t=1}^T\sum_{j\in S_t} N_{jt}}\right).
\end{equation}
}
\vspace{-2em}
\av{\section{Assortment optimization}
\label{sec:assortment}
In this brief section, we provide some theoretical results regarding the assortment optimization problem under the GSP and GMNL models. In assortment optimization, it is standard---and often essential---to model consumer choice with the inclusion of a no-purchase alternative, to reflect the possibility that customers may choose not to buy any of the available products. To that end, we begin by describing how the GSP model can be adapted to account for the no-purchase option.

\subsection{GSP model with no-purchase option}
We first introduce some notation. For any $S \subseteq [n]$, we denote by $S^+ =: S \cup \set{0}$, where $0$ represents the no-purchase option. Let $\Gscr$ denote the set of all non-empty sequences comprising entries from the product universe $[n]$ {\em without} repetitions. For instance, when $n=3$, examples of such sequences include $(1,2,3)$, $(2,1)$, $(1)$. 
 Each consumer type in the GSP model with a no-purchase option is characterized by a pair $(\gvec^+, k)$, where $\gvec \in \Gscr$ and $k$ is an integer between $1$ and $\abs{\gvec}$. Here, $\gvec^+$ denotes the sequence $\gvec$ appended by the no-purchase option $0$---for instance, if $\gvec = (2,1)$ then $\gvec^+ := (2, 1, 0)$. 
 As before, we refer to $k$ as the {\em choice index} of the consumer type. 
The set of all consumer types is denoted as 
$\Omega^{\np} = \set{(\gvec^+, k)  \colon \; \gvec \in \Gscr, 1 \leq k \leq \abs{\gvec} }$, with an arbitrary type denoted as $\omega = (\gvec^+, k)$.


Given an offer set $S \subseteq [n]$, the choice behavior of consumers of type $\omega = (\gvec^+, k)$ is identical to that defined in Section~\ref{sec:model}.
Specifically, the consumer first constructs a subsequence $s(\gvec^+, S^+)$ of $\gvec^+$ by removing \emph{all} alternatives that are \emph{not} in $S^+$. Then, they choose the alternative at the $k^{th}$ position in the subsequence $s(\gvec^+, S^+)$ if it exists, otherwise when $|s(\gvec^+, S^+)|< k$, the consumer selects the product in the last position in $s(\gvec^+, S^+)$. Note that by construction, the last alternative in $s(\gvec^+, S^+)$ is always the no-purchase option regardless of the sequence $\gvec$ or offer set $S$. Therefore, it follows that when $|s(\gvec^+, S^+)| \leq k$, the consumer selects the no-purchase option $0$.
Formally, the choice function for any consumer type in the GSP model with a no-purchase option is defined as:
\begin{definition}[choice function of gsp consumer types w/ no-purchase option] \label{def:gsp_choice_fn_with_default}
Faced with a choice set $S \subseteq [n]$, the binary choice function $C_{\omega}(j, S)$ which states whether consumer type $\omega =(\gvec^+, k)$ chooses alternative $j \in S$ is defined as follows:
\begin{displaymath}
C_{\omega}(j,S) \coloneqq \indicator{\rank(j, S^+; \gvec^+) = k} = \left\{ \begin{array}{ll}  \displaystyle 1  & \quad \textrm{if $j$ is at position $k$ in $s(\gvec^+,S^+)$} \\[3ex]
0 & \quad \textrm{otherwise}\\
\end{array} \right.
\end{displaymath}
The choice function for the no-purchase alternative is given by $C_{\omega}(0, S) \coloneqq 1 - \sum_{j \in S} C_{\omega}(j,S)$.
\end{definition}
As before, the class of GSP models in the presence of a no-purchase option is then defined as follows:
\begin{definition}[choice probability under gsp model with no-purchase option]\label{def:gsp_prob_fn_with_default}
A discrete choice model $\mathcal{P}$ belongs to the class of GSP models if there exists a probability distribution $\lambda(\cdot)$ over the set of consumer types $\Omega^\np$ such that
\begin{eqnarray} \label{stochastic_preference_probabilities}
\mathcal{P}(j, S) = \sum_{\omega \in \Omega^\np} C_{\omega}(j,S) \cdot \lambda(\omega)
\end{eqnarray}
for all $S \subseteq [n]$ and all $j \in S \cup \set{0}$. 
\end{definition}

The GRUM model, and by extension the GMNL, can be defined analogously in the presence of a no-purchase option by suitably extending Definitions~\ref{def:krum_choice_fn} and~\ref{def:krum_prob_fn}. Having incorporated the no-purchase option into the customer choice framework, we now proceed to analyze some theoretical and experimental aspects about  assortment optimization for these models.

\subsection{Assortment optimization under the GSP model}
The purpose of this subsection is to derive performance guarantees for the most studied heuristic in assortment optimization, {\em revenue-ordered assortments}~\citep{berbeglia2016assortment}. We show that its worst-case performance is substantially worse under the GSP model compared to its worst-case performance under RUM models.

Given product revenues $r_1, r_2, \ldots r_n$, the assortment optimization problem under the GSP model can be formulated as
\begin{equation}
\label{eq:gsp_assort}
\max_{S \subseteq [n]}~\sum_{j \in S}r_j \cdot \mathcal{P}(j,S),
\end{equation}
where the choice probabilities are given by Definition~\ref{def:gsp_prob_fn_with_default}.

Following Proposition 1, it should not be very surprising that problem~\eqref{eq:gsp_assort} is hard, in general. The strongest negative result about the assortment optimization problem under a general RUM model to date is 
due to \citet{aouad2015approximability}, who established that it is NP-hard to approximate within a factor of $O(n^{1-\epsilon})$ and to within a factor of $O(\log^{1-\epsilon}(r_{\max} / r_{\min}))$, for every $\epsilon > 0$, where $r_{\max} := \max_{i \in [n]} r_i$ and $r_{\min} := \min_{i \in [n]} r_i$ are the largest and smallest product revenues, respectively. Consequently, the analysis of heuristic approaches to solving the problem is of interest.

The heuristic that provides the best revenue guarantees (for a general RUM or a broader class) is \emph{revenue-ordered assortments} proposed by \cite{talluri2004revenue}. 
Suppose $r_{(1)}> r_{(2)} > \ldots > r_{(m)}$ denote the unique values of the product revenues sorted in decreasing order, where $m \leq n$.
Further, define $S_{z} := \set{i \in [n] \colon r_i \geq r_{(z)}}$ for each $z \in [m]$, that is, $S_z$ is the set of products with revenue at least $r_{(z)}$. 
The revenue-ordered assortment heuristic
simply compares the (expected) revenue obtained by the sets $\left(S_z \colon z \in [m]\right)$, and  
chooses the one with the highest revenue. 
\citet{berbeglia2016assortment} showed that, for any regular choice model—including the entire RUM class—the best revenue-ordered assortment guarantees a revenue of at least a $\max\left\{\frac{1}{m},\, \frac{1}{1 + \log\left(r_{\max} / r_{\min}\right)}\right\}$ fraction of the optimal revenue, and that this bound is tight.\footnote{This guarantee cannot be improved even by a factor of $(1 + \epsilon)$ for any $\epsilon > 0$. Restricted to the RUM class, \citet{aouad2015approximability} independently established essentially the same bound (up to a constant factor).} Consequently, revenue-ordered assortments offer the best possible approximation guarantees among all known polynomial-time methods for assortment optimization under regular choice models. This performance bound applies to the subclass of GSP models that satisfy regularity, which can be verified using Proposition~\ref{prop:when_is_gsp_w_nopurch_regular}, that extends the result in Proposition~\ref{prop:when_is_gsp_regular} to account for the no-purchase option.
Unfortunately, the worst-case performance guarantees of the revenue-ordered assortment strategy deteriorate significantly under the GSP class:
\begin{proposition}[Performance of revenue-ordered assortments]
\label{prop:gsp_assort}
Under the GSP model, revenue-ordered assortments guarantee at least an $r_{\min}/ r_{\max}$ fraction of the optimal revenue, and this bound is tight.
\end{proposition}

The proof is provided in Appendix~\ref{app:proof_rev_order_gsp} and builds on Lemma~\ref{lemma_choice_increase_demand}, which establishes that the no-purchase option in the GSP model satisfies the regularity condition stated in~\eqref{regularity_equation}. In fact, the approximation guarantee factor of \( r_{\min} / r_{\max} \) applies to any choice model that exhibits regularity with respect to the no-purchase option such as the RAM~\citep{cattaneo2017random} discussed earlier. Moreover, Proposition \ref{prop:gsp_assort} also applies to the GRUM model class since the GSP instance used to prove the tightness result is also a GRUM instance. Determining whether other efficient algorithms can provide stronger worst-case revenue guarantees for the GSP and GRUM models remains an important open question.


\subsection{Assortment optimization under the GMNL model}
\label{sec:gmnl_assort}
We now turn our attention to the GMNL model. 
We show that the assortment optimization problem with totally unimodular (TU) constraints is NP-hard to approximate.
\footnote{TU constraints can model a wide variety of practical considerations including position dependent choices, cardinality constraints, and discrete pricing.}
By contrast, under the classical MNL model with such constraints, the assortment problem is solvable in polynomial time ~\citep{sumida2021revenue}. 




To establish the hardness of the assortment problem with TU constraints under the GMNL model, we draw a connection to the multi-purchase MNL (MP-MNL) model proposed in~\cite{bai2023assortment}. The MP-MNL model extends the classical MNL framework to allow for multiple purchases within a single shopping visit. 
~\citet[Theorem A.1]{bai2023assortment} proved that it is NP-hard to approximate the assortment optimization problem in the presence of TU constraints under the MP-MNL model within a factor of $O(n^{1-\epsilon})$ for any fixed $\epsilon > 0$. We show that this hardness result also applies to the GMNL model by establishing an equivalence between the subclass of regular GMNL models and the MP-MNL model:
\begin{proposition}[Hardness of assortment optimization under the GMNL model]
\label{prop:gmnl_assortment}
It is NP-hard to approximate TU-constrained assortment optimization under the GMNL model within factor $O(n^{1-\epsilon})$ for any fixed $\epsilon > 0$, even when restricted to the subclass that satisfies regularity.
\end{proposition}

The proof is provided in Appendix~\ref{app:proof_gmnl_assort}. First, we show that the purchase probabilities under the MP-MNL model are equal to the choice probabilities under a specific instance of the GMNL model multiplied by a constant (which, interestingly, equals the expected number of purchases under the MP-MNL model); see Lemma~\ref{lem:mpmnl_subsumed_by_gmnl} for the formal statement. To the best of our knowledge, this is one of the few formal comparisons between traditional single-purchase and the growing literature on multi-purchase models~\citep{tulabandhula2023multi,bai2023assortment,luan2025joint,linexpress}, and we believe it may be of independent interest. In particular, the GMNL instance we construct satisfies $\lambda_1 \geq \lambda_2 \geq \ldots \geq \lambda_n$, which, by Proposition~\ref{prop:grum_regular}, implies that the model is regular. The proposition then follows from the inapproximability result for the MP-MNL model (Theorem~A.1 from~\citealp{bai2023assortment}).



Moreover, in Proposition~\ref{prop:regular_gmnl_subsumed_by_mpmnl}, we show that the choice probabilities under any instance of the GMNL model satisfying $\lambda_1 \geq \lambda_2 \geq \ldots \geq \lambda_n$ are also equivalent to the purchase probabilities under an instance of the MP-MNL model multiplied by $\lambda_1$. Consequently, the two approximation algorithms developed in~\cite{bai2023assortment} for solving the (unconstrained) assortment optimization problem under the MP-MNL can be directly applied to approximate the optimal revenue under those (regular) instances of the GMNL model. Note that, in particular, Proposition~\ref{prop:regular_gmnl_subsumed_by_mpmnl} implies that the MP-MNL model satisfies regularity.

To complement our theoretical analysis, we conduct a numerical study to evaluate the performance of revenue-ordered (RO) assortments under the GMNL model in the unconstrained setting. Specifically, we study the impact of the type distribution $\lambda$ and the maximum choice index of non-standard customers on the revenue performance. Due to space constraints, the results are described in Appendix~\ref{app:gmnl_ro}. Based on our results, in which RO assortments consistently achieved at least half of the optimal expected revenue, establishing whether RO assortments provide a constant-factor revenue guarantee is an interesting open question. In addition, determining the computational complexity of the unconstrained assortment problem under the GMNL model is also an important direction for future study.

}
\section{Numerical evaluation}
\label{sec:numerics}
In this section, we test the efficacy of our proposed models using real choice data. Specifically, we showcase the ability of the GSP model to (i) explain non-rational choice behavior, and (ii) predict out-of-sample~choices. \av{We also conduct a detailed simulation study where we compare the prediction performance of our estimation procedure for the GSP model with the GPT-based procedure of~\cite{jena2022estimation}. Due to space constraints, we discuss the results in Appendix~\ref{sec:synth_data}.}
\subsection{Explaining non-rational time and risk preferences}
\label{sec:explain}
As discussed in Section~\ref{sec:gsp_power}, the GSP model can capture regularity violations in choice data, unlike the SP model. We further demonstrate the existence of such non-rational behavior using data collected from two incentive-aligned experiments in the economics literature. Our results show that the SP model, \av{as well as the Halo-MNL model~\citep{maragheh2018customer,yousefi2020choice}, a parametric extension of the MNL model that can capture regularity violations}, are unable to fully explain the observed choices while the GSP model can perfectly fit the data using a relatively small proportion of non-standard types. \av{In fact, the Halo-MNL model fits worse than the SP model, underscoring the importance of a nonparametric framework like GSP, which generalizes the SP model.} 


\subsubsection{Time preferences via delayed payment plans.} First, we leverage the experiment conducted by~\cite{manzini2006two} to study individuals' time preferences. The experiment involved 102 participants, each of whom were asked to choose from offer sets consisting of distinct plans offering delayed payments. Each plan prescribed a payment schedule in three installments over a duration of 9 months, and the authors set up $n=4$ different payment plans---C (constant), I (increasing), D (decreasing), and J (jump)---shown in Table~\ref{tab:mm_plans} in Appendix~\ref{app:numerics}. Each participant was presented with all possible choice sets consisting of at least two plans and was asked to choose a plan in each instance, resulting in a total of $T=11$ choices per participant. The study was incentivized such that each individual earned a participation fee of \euro{5} and had a $50\%$ chance of receiving additional payment, corresponding to the participant's choice in a randomly selected offer set.


The aggregated choice data, presented in Table~\ref{tab:mm_plans_obs}, indicates the presence of regularity violations. For instance, when payment plan J is added to the offer set \{C, I\}, the proportion of participants choosing plan I increases from $7\%$ to $11\%$. As a result, we know that the SP model will not be able to perfectly match the observed choices as some participants exhibit non-rational behavior. \av{We verify this by fitting a SP model to the choice data using the MD algorithm of~\cite{van2014market} and evaluating the goodness-of-fit using the mean absolute percentage error (MAPE) metric:
\begin{equation}
\label{eq:mape}
\loss = 100 \times \frac{1}{\sum_{t=1}^T \abs{S_t}}\sum_{t=1}^{T} \sum_{j\in S_t} \abs{\frac{f_{jt} - \Pcal(j, S_t)}{f_{jt}}}
\end{equation}
where $f_{jt}$ is the observed choice fraction for product $j \in S_t$, and $\Pcal(j, S_t)$ is the predicted choice probability. We observed that the $\loss$ for the SP model was $\sim 7 $ (or $ 7 \%)$.}
\begin{figure}[t]
\FIGURE
{\includegraphics[width=0.8\textwidth]{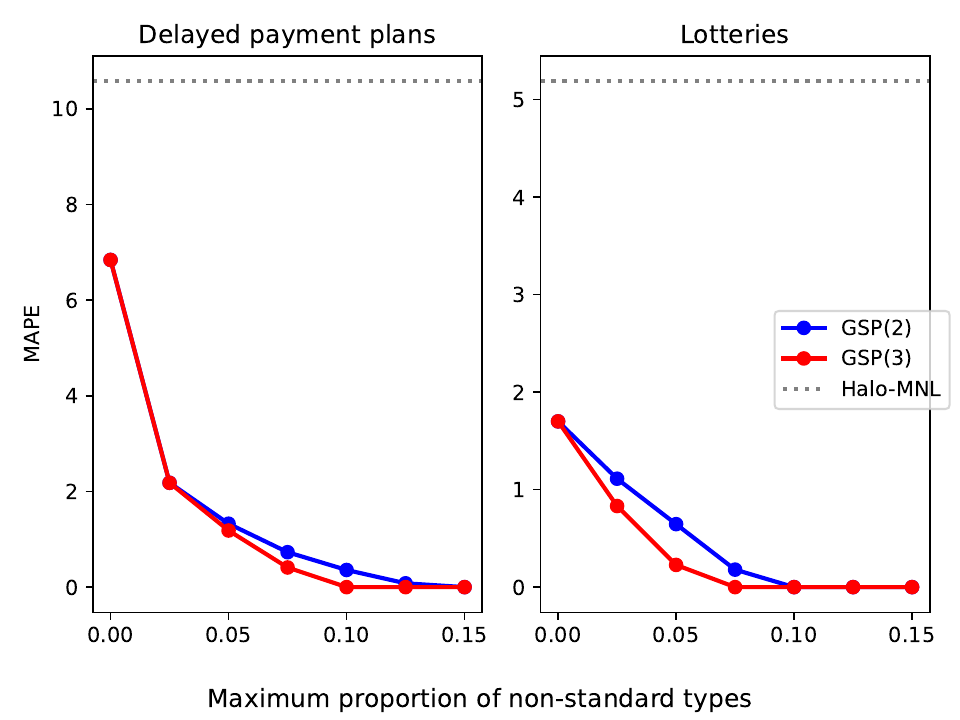}}
{\av{Evidence of non-rational choice behavior in the~\cite{manzini2006two} (left) and~\cite{caliari2020behavioral} (right) choice experiments\label{fig:in-sample}}}
{\av{The figures plot the MAPE~\eqref{eq:mape} obtained when fitting GSP models with increasing proportions of non-standard types. A proportion of $0$ corresponds to fitting a SP model, which leads to a significant MAPE value. Also, shown is the MAPE for the Halo-MNL~\citep{yousefi2020choice} model which is even higher than the SP model.}}
\end{figure}

Next, we fit subclasses of the GSP model by restricting the maximum choice index $k_{\max}=2,3$ of non-standard types. 
We term these models as GSP(2) and GSP(3), analogous to the GMNL models in Section~\ref{sec:gmnl_assort} earlier. For each model class, we vary the maximum proportion of non-standard types ($\delta$) in the~\ref{eq:unconstr_mle} from $2.5\%$ to $15\%$ in increments of $2.5\%$. \av{For comparison, we also fit the Halo-MNL model~\citep{maragheh2018customer,yousefi2020choice}, which introduces pairwise interaction effects into the standard MNL framework and can admit regularity violations. The log-likelihood function under the Halo-MNL model remains concave in the model parameters, facilitating efficient estimation of the $n\times n$ product interaction matrix---refer to Appendix~\ref{sec:halomnl_est} for the details.}

The left panel of Figure~\ref{fig:in-sample} plots the \loss{} as a function of $\delta$ under both model classes, with $\delta=0$ corresponding to the \loss{} value under the SP model. It can be seen that both model classes perfectly fit the observed choices given a large enough $\delta$. 
The minimum proportion of non-standard types needed to explain the choices is $14\%$ and $10\%$ for the GSP(2) and GSP(3) models, respectively, which is reasonable because the GSP(3) class captures a larger array of non-rational behaviors compared to the GSP(2) class. \av{Interestingly, the Halo-MNL fits worse than the SP model, which highlights the potential limitations of using parametric models that incorporate non-rationality.}

\subsubsection{Risk preferences via lotteries.} Next, we consider the experiment reported in~\cite{caliari2020behavioral} to elicit participants' risk preferences by observing their choices amongst different lottery configurations. The study involved 145 subjects who were asked to choose among four lotteries\footnote{The full experiment involved elicitation for other lottery designs as well but we focus on only these 4 alternatives (labeled as ``MAIN'' in~\citealp{caliari2020behavioral}) since participants were asked to choose from all possible choice sets of these alternatives, providing us full visibility into their underlying preferences.}---D (Degenerate), Sa (Safe), 50/50 (Fifty-Fifty), and R(Risky)---that differ in their risk profiles, as shown in Table~\ref{tab:lotteries} in Appendix~\ref{app:numerics}
Table~\ref{tab:lottery_obs} reports the observed choice fractions for this experiment, which also shows the presence of regularity violations.
\av{The SP model achieved a \loss{} of $\sim2\%$, indicating comparatively less non-rationality in choices relative to the experiment in~\cite{manzini2006two}. However, the Halo-MNL model again performs worse than SP, while GSP perfectly explains the observed choices using at most $10\%$ non-standard types; see the right panel of Figure~\ref{fig:in-sample}.} 

\subsection{Prediction performance on real data}
\label{sec:pred_study}
\av{We now proceed with an evaluation of the prediction error of our proposed models.} 
Our results show that the non-rational models, GSP and GMNL, outperform the standard MNL and SP choice models in out-of-sample prediction accuracy. In particular, the GMNL model offers improved predictions compared to the RUM choice models while requiring substantially lower time to fit compared to the GSP models, making it an attractive option for practical use. 

We consider the following datasets for our analysis:
\begin{enumerate}
\item {\tt swissmetro}~\citep{bierlaire2001acceptance}, consisting of 10,719 survey responses collected during 1997-98 to assess the potential demand for Swissmetro, an innovative underground maglev system to connect the major urban centres in Switzerland.
Each respondent was
asked to choose one mode amongst train, Swissmetro, and car for inter-city travel, where each mode was associated with different attributes such as travel time and cost. For individuals without a car, the choice set included only train and Swissmetro.
There were three different travel times for both train and Swissmetro, and following existing work~\citep{osogami2014restricted} we consider these as separate alternatives, resulting in a total of $7$ alternatives: one for car, and three each for train and Swissmetro. 

\item {\tt sfwork} and {\tt sfshop}~\citep{koppelman2006self},  collected using a survey of transportation mode choices around the SF Bay Area. The {\tt sfshop} dataset contains 3,157 observations, each comprising a set of transport modes such as transit, walk, bike, drive, rideshare, etc. available to travel to a shopping center and the corresponding choice.
The {\tt sfwork} dataset contains 5,029 observations consisting of
different commuting options available to work  and the choice made on each visit. 
\end{enumerate}

Table~\ref{tab:real_data} shows the number of alternatives and distinct choice sets for all three datasets.\footnote{Two choice sets received less than $5$ responses, and therefore we dropped them from the analysis.} For each distinct offer set, we compute the sales counts $\left(N_{jt} \colon j \in S_t\right)$ as the number of respondents that chose option $j$ when offered choice set $S_t$.  Then, we evaluate the accuracy in predicting the choice probabilities for an
arbitrary offer set, when the model is trained on the choice data from all the remaining offer sets; referred to as {\em leave-one-out cross validation (loocv)} in the ML literature. 
Next, we describe the different models compared along with details on their estimation procedures.


{\bf Models compared and estimation details}. We compare the predictive performance of the following five models: the standard MNL, the SP, the GMNL(2), the GSP(2), and the GSP(3). 
The MNL model is fit using the MM algorithm~\citep{hunter2004mm} which provides closed-form updates for the mean utilities in each iteration. We fit the GMNL(2) model using the EM algorithm described in Section~\ref{sec:kmnl_est}, with the MNL model in the M-step fit using the MM algorithm. For both models, the mean utilities were initialized to a vector of zeros. For the GMNL(2) model, the proportion of the standard type was initialized to $1/2$. The MD approach of~\cite{van2014market} was used to fit the SP model. We used an initial set of $n$ preference lists $\sigma_1, \sigma_2, \ldots, \sigma_n$ with product $i$ ranked at the top of preference list $\sigma_i$ for each $i \in [n]$.
Finally, the two GSP models were fit using the Frank-Wolfe algorithm described in Section~\ref{sec:gsp_est} starting from the same set of rankings $\sigma_1, \sigma_2, \ldots, \sigma_n$. For both GSP models, we tune the hyperparameter $\delta$ that controls the proportion of non-standard types from the following set $\set{0.05, 0.1, 0.2, 0.3, 0.4}$ using 3-fold cross validation. For each model, the estimation is terminated when the relative change in the train KL divergence loss, defined in~\eqref{eq:lor}, in consecutive iterations is smaller than a threshold ($10^{-8}$):
\begin{equation}
\label{eq:lor}
{\sf KLDiv} = \frac{1}{\sum_{t \in \Tcal_{\text{tr}}} \sum_{j \in S_t} N_{jt}}\sum_{t \in \Tcal_{\text{tr}}} \sum_{j\in S_t} N_{jt} \cdot \log \left(\frac{f_{jt}}{\Pcal(j, S_t)} \right),
\end{equation}
where $\Tcal_{\text{tr}}$ is the collection of training offer sets, and recall that $f_{jt}$ is the observed choice fraction of product $j$ in offer set $S_t$. 
 \begin{table}
\tabcolsep=8.0pt
\TABLE
{Comparison of different models in predicting travel mode choices\label{tab:real_data}}
{\begin{tabular}[t]{*{8}{c}}
\toprule
  Dataset & \# products ($n$) & \# offer sets ($T$) & \multicolumn{5}{c} {Test KL divergence loss ($\times 10^{-2}$)} \\
  \cmidrule(lr){4-8} 
 & &  &  MNL & SP & GMNL(2) & GSP(2) & GSP(3) \\
  \midrule
{\tt swissmetro} & 7 & 18 & 7.2 & 11.1 &  \textbf{4.3} & 10.0 & 13.2
 \\
   {\tt sfshop} & 8 & 8 & 15.3 & 380 & 14.7 & \textbf{11.2} & 191 \\
     {\tt sfwork} & 6 & 12 & 3.20 & 2.99 & \textbf{2.86} & 3.09 & 4.47 \\
     \bottomrule
\end{tabular}}
{Lower values of the KL divergence loss, defined in~\eqref{eq:lor} are preferred. For each dataset, the best performing model(s) is highlighted in bold. \av{Non-rational models outperform their rational counterparts in all datasets. Note that the MNL model outperforms SP in two of the datasets, which suggests that the latter is overfitting. A similar observation can be made for the more complex GSP(3) model, despite the regularization to constrain the proportion of non-standard types.}}
\end{table}

{\bf Results and discussion.} Table~\ref{tab:real_data} reports the KL divergence loss evaluated on the test choice sets for each of the different models. Comparing the performance of the rational and non-rational choice models, we make the following observations from the table:
\begin{enumerate}
    \item {\em Incorporating non-rationality improves prediction accuracy.} \av{For all three  datasets, a non-rational model obtains the lowest test KL divergence loss. 
   Moreover, the improvements can be  substantial: $40\%$ and $27\%$ improvement over the best performing rational model for the {\tt swissmetro} and {\tt sfshop} datasets, respectively.}
    These results show the benefits of incorporating even small amounts of non-rationality---by restricting the maximum choice index and/or the proportion of non-standard types---into the rational choice models. This is appealing since such ``simple'' non-rational models are easier to interpret (as discussed in Section~\ref{sec:gsp_interpretation}) and more tractable to estimate.
    \item {\em GMNL model offers an attractive generalization vs.\ estimation trade-off.} \av{Despite its parametric assumptions, we observe that the GMNL(2) model outperforms both its rational counterpart MNL, as well as the more sophisticated SP choice model across all datasets. This is an impressive feat, since the GMNL(2) model uses only one additional parameter compared to MNL, and therefore fitting the former is typically much faster compared to the SP and GSP models.} Consequently, we believe that the GMNL model is a promising candidate for use in large-scale and real-time applications.
    \end{enumerate}

\section{Conclusions and future research}\label{sec:conclusions}
\av{There has been a growing literature on non-rational choice models that aim to capture regularity violations observed in consumer choices. 
However, most of these models rely on parametric assumptions, and therefore do not subsume the stochastic preference (SP) model, a rational model that contains the classical MNL as well as every choice model based on the random utility maximization framework. As a result, such parametric models may underperform in practice, potentially achieving worse fit than a general SP model when applied to real data. In this paper, we introduce the \emph{generalized stochastic preference (GSP)} model, a nonparametric framework that extends the SP class by explicitly incorporating non-rational behavior into consumer preferences. The flexibility of the GSP model allows it to adapt to a wide range of choice behaviors, depending on the structure of the data. Moreover, we show that our proposed notion of non-rationality can be embedded within any subclass of the RUM model, including the MNL model. We develop tractable algorithms to estimate our proposed models from choice data and demonstrate that they achieve superior predictive performance compared to rational benchmarks. Additionally, we analyze the assortment optimization problem under our proposed models, establishing computational hardness results and deriving worst-case performance guarantees.}

There are a few natural directions for further study of the GSP model. The first is to obtain a characterization for the GSP model. Formally, given a choice model $\mathcal{P}$ as the complete system of probabilities $\mathcal{P}(j,S)$ for each $j \in S$ and $S \subseteq [n]$, is it possible to determine whether $\mathcal{P}$ belongs to the GSP model class? An elegant characterization for the RUM model class was proved by~\citet{falmagne1978representation}, who showed that the Block-Marschak conditions are sufficient (they were known to be a necessary condition).
Obtaining a similar characterization for the GSP model
is an interesting open question.
Second, the ability of the GMNL model to account for product features such as price motivates the study of key operational decisions such as price optimization and joint assortment and pricing optimization under the GMNL model. Finally, applying the non-rationality principle proposed in the paper to other RUM subfamilies, such as the exponomial choice model~\citep{alptekinouglu2016exponomial} which is both easy to estimate and provides good empirical performance~\citep{berbeglia2022comparative,aouad2023exponomial}, is a promising direction for future work.

\ACKNOWLEDGMENT{The first author would like to thank Guiyun Feng and Xiaobo Li for enriching discussions that helped improve the paper.}

\bibliographystyle{informs2014}
\bibliography{references_discrete_choice}

\ECSwitch

\ECDisclaimer

\begin{APPENDICES}
\section{Additional choice experiments that violate regularity}
\label{sec:more_regularity_violations}
In this section, we provide examples of other well-known behavioral experiments that violate regularity and show that the GSP model can explain the findings in each of them.

\begin{example}\label{example_3}
In another experiment reported in \citet{simonson1992choice} to illustrate the compromise effect, participants were split into two groups ($N=60$ and $N=61$). First, all subjects were given pictures and descriptions of three microwave ovens. Second, participants were asked to choose among a subset of the ovens. Those in the first group had to choose among microwaves Emerson and Panasonic I (condition 1), whereas those in the second group had to choose among 3 microwaves: Emerson, Panasonic I and Panasonic II (condition 2). Below we display the results of this experiment:

$$
\begin{array}{l|lll}
\text{Model} & \text{Price (USD)} & \text{Condition 1}:  \{1,2\} & \text{Condition 2}: \{1,2,3\} \\
\hline
\text{1: Emerson} & 109.99 & 57\% & 27\%  \\
\text{2: Panasonic I}& 179.99 & 43\% & 60\% \\
\text{3: Panasonic II} & 199.99 &  \text{not available} & 13 \%
\end{array}
$$

These results cannot be explained using a RUM, as there is a violation of regularity. Specifically, the proportion of participants choosing alternative 2 (Panasonic I) increase considerably when the alternative 3 (Panasonic II) is added to the choice set. However, it is possible to perfectly fit a GSP model with only four consumer types to the results of this experiment, which is shown in \ref{simonson_tversky_example_microwaves_solution}.
\begin{table}[h]\label{simonson_tversky_example_microwaves_solution}
\centering
\caption{A generalized stochastic preference that fits the experiment from Example \ref{example_3}}
\begin{tabular}{|c|c|c|c|}\hline
  \multicolumn{3}{|c|} {Consumer type} & Probability \\ \hline
  label & $\bell$ & $k$ &  \\
  \hline
  1 & (1,2,3) & 1 & 0.27 \\
  \hline
  2 & (2,1,3) & 1 & 0.43 \\
  \hline
  3 & (3,1,2) & 1 & 0.13 \\
  \hline
  4 & (3,2,1) & 2 & 0.17 \\
  \hline
\end{tabular}
\end{table}
Similar to the example in the main body, the first three consumer types are rational and prefer the first alternative in their rankings. The fourth type (which has a probability of 0.17), is equipped with the ordering (3,2,1) and will select the alternative in the second position (if it exists) after removing from the sequence (3,2,1) the alternatives not present in the choice set. If the more expensive Panasonic II microwave is available, this consumer type will select the cheaper Panasonic I model (if available). But if the choice set is composed of only the Emerson (alternative 1) and the Panasonic I (alternative 2) ovens, the consumer will select the Emerson model.

We now calculate $\mathcal{P}(2,\{1,2\})$ and $\mathcal{P}(2,\{1,2,3\})$ and show that it matches with the experiment results. Again, one can calculate the remaining probabilities and recover the different market shares from the experiment under both conditions.
\begin{eqnarray*}
  \mathcal{P}(2,\{1,2\}) &=& 0.27\cdot C_1(2,\{1,2\}) + 0.43\cdot C_2(2,\{1,2\}) + \\
  & & 0.13\cdot C_3(2,\{1,2\}) + 0.17\cdot C_4(2,\{1,2\})\\
  &=& 0 + 0.43 + 0 + 0 = 0.43
\end{eqnarray*}

\begin{eqnarray*}
  \mathcal{P}(2,\{1,2,3\}) &=& 0.27\cdot C_1(2,\{1,2,3\}) + 0.43\cdot C_2(2,\{1,2,3\}) + \\
  & & 0.13\cdot C_3(2,\{1,2,3\}) + 0.17\cdot C_4(2,\{1,2,3\})\\
  &=& 0 + 0.43 + 0 + 0.17 = 0.60
\end{eqnarray*}

\end{example}

\begin{example}\label{the_economist_experiment}
Next, we consider the famous experiment reported in \citet{ariely03} which illustrates the {\em attraction effect}, also known as the {\em decoy effect}. In the experiment, students were asked to choose between three different subscription options for the Economist magazine: (1) Online version only, (2) Print version only and (3) Print and online bundle. Half of the students were shown options $\{1,3\}$ and the other half were asked to choose between $\{1,2,3\}$. The results of the experiment were:

$$
\begin{array}{l|lll}
\text{Subscription} & \text{Price (USD)} & \text{Condition 1}:  \{1,3\} & \text{Condition 2}: \{1,2,3\} \\
\hline
\text{1: Online only} & 50 & 68\% & 16\%  \\
\text{2: Print only} & 125 & \text{not available} & 0\% \\
\text{3: Online and print} & 125 & 32\% & 84 \%
\end{array}
$$
Observe that by adding alternative 2 (print only) to the choice set $\{1,3\}$, the probability of choosing alternative 3 (online and print bundle) increased considerably, which violates regularity. In particular, alternative 2 serves as a ``decoy'' meant to increase the market share for alternative 3, which strictly dominates it. The GSP model comprising the three consumer types shown in Table~\ref{ariely_solution} reproduces the choice probabilities observed in the experiment, with the non-rational type explaining the 52\% jump in market share for alternative 3.
\begin{table}[h]
\centering
\begin{tabular}{|c|c|c|c|}\hline
  \multicolumn{3}{|c|} {Consumer type} & Probability \\ \hline
  label & $\bell$ & $k$ &  \\
  \hline
  1 & (3,1,2) & 1 & 0.32 \\
  \hline
  2 & (1,2,3) & 1 & 0.16 \\
  \hline
  3 & (2,3,1) & 2 & 0.52 \\
  \hline
\end{tabular}
\caption{A generalized stochastic preference that fits the experiment from Example \ref{the_economist_experiment}}\label{ariely_solution}
\end{table}
\end{example}

\begin{example}\label{example_herne1997decoy}
The decoy effect has also been observed when people are asked to make political choices. \citet{herne1997decoy} performed a choice experiment in which participants need to pick one economic union for an imaginary country (initially, this country doesn't belong to any economic union).

Again, participants were split into two conditions. Those under the condition 1 were asked to choose between the unions $\{T,C\}$ whereas those under the second condition (condition 2) had to choose among all three $\{T,C,D\}$. The main features displayed about the unions and the experiment results are summarized below:
$$
\begin{array}{l|llll}
\text{Economic Union} & \text{Inflation(\%)} & \text{Economic growth(\%)} &\text{Condition 1}: \{T, C\} & \text{Condition 2}: \{T, C, D\} \\
\hline
\text{T} & 1 & 2.1 & 53\% & 63\%  \\
\text{C}& 2.2 & 3.9 & 47\% & 37\% \\
\text{D} & 1.5 & 2 & \text{not available} & 0 \%
\end{array}
$$
Observe that the union $D$ acts as a decoy to attract participants to union $T$ which dominates $D$ in both dimensions: inflation and economic growth. The GSP model shown in Table \ref{herne_solution} explains the results observed in this experiment. 
\begin{table}[h]
\centering
\caption{A generalized stochastic preference that explains the experiment reported in \citet{herne1997decoy}.}
\begin{tabular}{|c|c|c|c|}\hline
  \multicolumn{3}{|c|} {Consumer type} & Probability \\ \hline
  label & $\bell$ & $k$ &  \\
  \hline
  1 & $(T,C,D)$ & 1 & 0.53 \\
  \hline
  2 & $(C,T,D)$ & 1 & 0.37 \\
  \hline
  3 & $(D,T,C)$ & 2 & 0.10 \\
  \hline
\end{tabular}
\label{herne_solution}
\end{table}
Again, the non-rational type is responsible for the 10\% jump in market share of union $T$ when union $D$ is added to the choice set.
\end{example}

\newcommand{\bomega}{\bm{\omega}}

\section{Proofs of results in Section~\ref{sec:model}}
\subsection{\av{Proof of Proposition~\ref{prop:regular_but_not_gsp}}}
\av{For the first part, consider the choice model shown in Table~\ref{table_mcfadden1990stochastic}. As can be seen, this choice model satisfies the regularity condition. However, it lies outside the RUM class, see~\citet{berbeglia2016assortment}.}

\begin{table}[h]
\tabcolsep=14.0pt
\TABLE
{Example of a choice model that is not a RUM but satisfies regularity\label{table_mcfadden1990stochastic}}
{\begin{tabular}{|l|c|c|c|c|}\hline
$S$ & $\mathcal{P}(1, S)$ & $\mathcal{P}(2, S)$ & $\mathcal{P}(3, S)$ & $\mathcal{P}(4, S)$ \\
\hline
\hline
$\{1, 2\}$ & 0.5 & 0.5 & - & - \\
$\{1, 3\}$ & 0.5 & - & 0.5 & - \\
$\{1, 4\}$ & 0.5 & - & - & 0.5 \\
$\{2, 3\}$ & - & 0.5 & 0.5 & - \\
$\{2, 4\}$ & - & 0.5 & - & 0.5 \\
$\{3, 4\}$ & - & - & 0.5 & 0.5 \\
\hline
$\{1,2,3\}$ & 0.4 & 0.3 & 0.3 & - \\
$\{1,2,4\}$ & 0.4 & 0.3 & - & 0.3 \\
$\{1,3,4\}$ & 0.4 & - & 0.3 & 0.3 \\
$\{2,3,4\}$ & - & 0.4 & 0.3 & 0.3 \\
\hline
$\{1,2,3,4\}$ & 0.25 & 0.25 & 0.25 & 0.25 \\
\hline
\end{tabular}}
{}
\end{table}
\av{Table~\ref{mcfadden_solution} presents an instance of the GSP model that explains the choice probabilities in Table~\ref{table_mcfadden1990stochastic}.
\begin{table}[h]
\centering
\tabcolsep=10pt
\caption{A GSP model that fits the choice model proposed in Table~\ref{table_mcfadden1990stochastic}}
\begin{tabular}{|c|c|c|c|}\hline
  \multicolumn{3}{|c|} {Consumer type} & Probability \\ \hline
  label & $\bell$ & $k$ &  \\
  \hline
  1 & (1, 2, 3, 4) & 1 & 0.15 \\
2 & (1, 4, 3, 2) & 1 & 0.05 \\
3 & (2, 1, 3, 4) & 2 & 0.05 \\
4 & (2, 4, 3, 1) & 1 & 0.05 \\
5 & (3, 2, 1, 4) & 1 & 0.15 \\
6 & (3, 2, 1, 4) & 2 & 0.10 \\
7 & (3, 4, 1, 2) & 2 & 0.05 \\
8 & (3, 4, 2, 1) & 1 & 0.10 \\
9 & (4, 2, 1, 3) & 1 & 0.15 \\
10 & (4, 2, 1, 3) & 2 & 0.10 \\
11 & (4, 3, 1, 2) & 1 & 0.05 \\
\hline
\end{tabular}
\label{mcfadden_solution}
\end{table}
}

\av{For the second part, suppose $n=4$ and consider the choice model depicted in Table~\ref{tab:regular_but_not_GSP}. It is easy to see that the model does not violate regularity.
\begin{table}[h]
\tabcolsep=14.0pt
\TABLE
{Example of regular choice model outside GSP class \label{tab:regular_but_not_GSP}}
{\begin{tabular}{|l|c|c|c|c|}\hline
  $S$ & $\Pcal(1, S)$ & $\Pcal(2, S)$ & $\Pcal(3, S)$ & $\Pcal(4, S)$ \\
  \hline
  \hline
   \{1, 2\} & 1 & 0 & - & -\\
  \{1, 3\} & 0.5 & - & 0.5 & -\\
  \{1, 4\} & 0.5 & - & - & 0.5\\
  \{2, 3\} & - & 0.5 & 0.5 & -\\
  \{2, 4\} & - & 0 & - & 1\\
 \{3, 4\} & - & - & 0.5 & 0.5\\
  \hline
  \{1, 2, 3\} & 0.5 & 0 & 0.5 & -\\
  \{1, 2, 4\} & 0.5 & 0 & - & 0.5\\
  \{1, 3, 4\} & 0.5 & -& 0 & 0.5 \\
  \{2, 3, 4\} & - & 0 & 0.5 & 0.5 \\
  \hline
  \{1, 2, 3, 4\} & 0.5 & 0 & 0 & 0.5 \\
  \hline
\end{tabular}}
{}
\end{table}
For purposes of contradiction, suppose there exists a GSP model that explains the choice probabilities defined above. Given that $\mathcal{P}(2,\{1,2\}) = 0$, it follows from definitions~\ref{def:gsp_choice_fn} and~\ref{def:gsp_prob_fn} that 
\begin{align} 
&\lambda((\bell, 1)) = 0 \text{ for any } \bell \text{ in which  alternative 2 appears before 1} \label{prod2_1} \\
&\lambda((\bell,2)) = \lambda((\bell,3)) = 0 \text{ for any } \bell \text{ in which  alternative 1 appears before 2} \label{prod2_2}
\end{align}
Similarly, given that $\mathcal{P}(2,\{2,4\}) = 0$ it follows that
\begin{align} 
&\lambda((\bell, 1)) = 0 \text{ for any } \bell \text{ in which alternative 2 appears before 4} \label{prod2_3} \\
&\lambda((\bell,2)) = \lambda((\bell,3)) = 0 \text{ for any } \bell \text{ in which  alternative 4 appears before 2} \label{prod2_4}
\end{align}
Next, since $\mathcal{P}(3,\{1,3,4\}) = 0$ it follows that
\begin{align} 
&\lambda((\bell, 1)) = 0 \text{ for any } \bell \text{ in which alternative 3 appears before both 1 and 4} \label{prod3_1} \\
&\lambda((\bell, 2)) = 0 \text{ for any } \bell \text{ in which  alternative 3 appears between 1 and 4} \label{prod3_2} \\
&\lambda((\bell, 3)) = 0 \text{ for any } \bell \text{ in which  alternative 3 appears after both 1 and 4} \label{prod3_3}
\end{align}
Finally, since $\mathcal{P}(3,\{1,2,3,4\}) = \mathcal{P}(3,\{1,2,3,4\}) = 0$ it follows that
\begin{align} 
&\lambda((\bell, 2)) = 0 \text{ for any } \bell \text{ in which either alternative 2 or 3 appears at position 2} \label{prod3_4} \\
&\lambda((\bell, 3)) = 0 \text{ for any } \bell \text{ in which alternative 2 or 3 appears at position 3} \label{prod3_5}
\end{align}

From the above equations, it follows that only the following 12 GSP types can have non-zero proportions:
\begin{align*}
\lambda((1,3,4,2), 1)) &\quad \lambda((2,4,1,3), 2) &\quad \lambda((2,3,1,4), 3)\\
\lambda((1,4,3,2), 1)) &\quad \lambda((2,1,4,3), 2) &\quad \lambda((2,3,4,1), 3)\\
\lambda((1,4,2,3), 1)) &\quad &\quad \lambda((3,2,1,4), 3)\\
\lambda((4,1,3,2), 1)) &\quad &\quad \lambda((3,2,4,1), 3)\\
\lambda((4,1,2,3), 1)) \\
\lambda((4,3,1,2), 1)
\end{align*}
Now, since $\Pcal(3, \set{3,4}) = 0.5$, it follows that
\begin{equation}
    \lambda((1,3,4,2), 1)) + \lambda((2,4,1,3), 2)) + \lambda((2,1,4,3), 2)) = 0.5   \label{prod_13}
\end{equation}
Next, since $\Pcal(1, \set{1,2,3}) = 0.5$, it follows that
\begin{multline}
    \lambda((1,3,4,2), 1)) + \lambda((1,4,3,2), 1)) + \lambda((4,1,3,2), 1)) + \lambda((1,4,2,3), 1)) \\ +  \lambda((4,1,2,3), 1)) + \lambda((2,4,1,3), 2)) + \lambda((2,1,4,3), 2)) + \lambda((2,3,1,4), 3)) \\ + \lambda((2,3,4,1), 3)) + \lambda((3,2,4,1), 3)) +\lambda((3,2,1,4), 3)) = 0.5   \label{prod_123}
\end{multline}
From~\eqref{prod_13} and~\eqref{prod_123}, it follows that 
only the following 4 GSP types can have non-zero proportions
\[ \lambda((1,3,4,2), 1)) ; \lambda((4,3,1,2), 1) ;  \lambda((2,4,1,3), 2)) ; \lambda((2,1,4,3), 2))  \]
Consider $\Pcal(3, \set{1,2,3}) = 0.5$, this implies that
\begin{equation}
\lambda((4,3,1,2), 1) = 0.5 \label{single_type}
\end{equation}
Now, since $\Pcal(4, \set{2,3,4}) = 0.5$, it follows that
\begin{equation}
    \lambda((4,3,1,2), 1) + \lambda((2,1,4,3), 2) + \lambda((2,4,1,3), 2) = 0.5 \implies \lambda((2,1,4,3), 2) = \lambda((2,4,1,3), 2) = 0, \label{only_rankings}
\end{equation}
where the implication follows from~\eqref{single_type}. Combining~\eqref{prod_13},~\eqref{single_type},~\eqref{only_rankings}, it follows that the GSP model is of the following form
\[  \lambda((4,3,1,2), 1) =  \lambda((1,3,4,2), 1) = 0.5 \]
However, the above GSP model implies that $\Pcal(3, \set{2,3}) = 1$ which contradicts the value in Table~\ref{tab:regular_but_not_GSP}.
\Halmos}

\subsection{\av{Proof of Proposition~\ref{prop:when_is_gsp_regular}}}
\av{We begin with the sufficiency part. Consider any offer-set $S \subseteq [n]$, and alternative $j \in S$. For any $i \notin S$, define $\Lscr_{i < j} = \set{\bell \in \Lscr \colon i \text{ appears before } j \text{ in } \bell}$ and similarly define $\Lscr_{j < i}$. Then, it follows that
\begin{align*}
    \Pcal(j, S) &=  \sum_{\omega \in \Omega} C_{\omega}(j,S) \cdot \lambda(\omega) \\
    &= \sum_{\bell \in \Lscr} \sum_{k \in [n-1]} C_{(\bell, k)}(j, S) \cdot \lambda((\bell, k)) \\
    &= \sum_{\bell \in \Lscr_{i < j}} \sum_{k \in [n-1]} C_{(\bell, k)}(j, S) \cdot \lambda((\bell, k)) + \sum_{\bell \in \Lscr_{j < i}} \sum_{k \in [n-1]} C_{(\bell, k)}(j, S) \cdot \lambda((\bell, k)) \\
    &\stackrel{\rm (a)}{=} \sum_{\bell \in \Lscr_{i < j}} \sum_{k \in [n-1]} C_{(\bell, k+1)}(j, S \cup \set{i}) \cdot \lambda((\bell, k)) + \sum_{\bell \in \Lscr_{j < i}} \sum_{k \in [n-1]} C_{(\bell, k)}(j, S) \cdot \lambda((\bell, k)) \\
    &\stackrel{\rm (b)}{\geq} \sum_{\bell \in \Lscr_{i < j}} \sum_{k \in [n-1]} C_{(\bell, k+1)}(j, S \cup \set{i}) \cdot \lambda((\bell, k+1)) + \sum_{\bell \in \Lscr_{j < i}} \sum_{k \in [n-1]} C_{(\bell, k)}(j, S) \cdot \lambda((\bell, k)) \\
    &\stackrel{\rm (c)}{\geq} \sum_{\bell \in \Lscr_{i < j}} \sum_{k \in [n-1]} C_{(\bell, k+1)}(j, S \cup \set{i}) \cdot \lambda((\bell, k+1)) + \sum_{\bell \in \Lscr_{j < i}} \sum_{k \in [n-1]} C_{(\bell, k)}(j, S \cup \set{i}) \cdot \lambda((\bell, k)) \\
    &= \sum_{\bell \in \Lscr_{i < j}} \sum_{k=2}^{n-1} C_{(\bell, k)}(j, S \cup \set{i}) \cdot \lambda((\bell, k)) + \sum_{\bell \in \Lscr_{i < j}} C_{(\bell, n)}(j, S \cup \set{i}) \cdot \lambda((\bell, n)) \\
    & + \sum_{\bell \in \Lscr_{j < i}} \sum_{k \in [n-1]} C_{(\bell, k)}(j, S \cup \set{i}) \cdot \lambda((\bell, k)) \\
    &\stackrel{\rm (d)}{=} \sum_{\bell \in \Lscr_{i < j}} \sum_{k=1}^{n-1} C_{(\bell, k)}(j, S \cup \set{i}) \cdot \lambda((\bell, k)) + \sum_{\bell \in \Lscr_{i < j}} C_{(\bell, n)}(j, S \cup \set{i}) \cdot \lambda((\bell, n)) \\
    &+ \sum_{\bell \in \Lscr_{j < i}} \sum_{k \in [n-1]} C_{(\bell, k)}(j, S \cup \set{i}) \cdot \lambda((\bell, k)) \\
    &\geq \sum_{\bell \in \Lscr_{i < j}} \sum_{k=1}^{n-1} C_{(\bell, k)}(j, S \cup \set{i}) \cdot \lambda((\bell, k)) + \sum_{\bell \in \Lscr_{j < i}} \sum_{k \in [n-1]} C_{(\bell, k)}(j, S \cup \set{i}) \cdot \lambda((\bell, k)) \\
    &\stackrel{\rm (e)}{=} \Pcal\left(j, S \cup \set{i}\right),
\end{align*}
where the justification for the numbered equations are provided next. For (a), it is clear that $C_{(\bell, k)}(j, S) = C_{(\bell, k+1)}(j, S \cup \set{i})$ for any $\bell \in \Lscr_{i < j}$ and any $k \in [n-1]$. (b) follows from hypothesis. (c) follows $C_{(\bell, k)}(j, S) = 0 \implies 
C_{(\bell, k)}(j, S \cup \set{i}) = 0$ for any $\bell \in \Lscr_{j < i}$ and any $k \in [n-1]$. (d) follows since $C_{(\bell, 1)}(j, S \cup \set{i}) = 0$ for any $\bell \in \Lscr_{i < j}$. (e) follows from~\eqref{stochastic_preference_probabilities}.

Finally, Table~\ref{tab:gsp_model_necessity} presents an example of a GSP instance that does not satisfy the sufficient condition yet is regular, as can be verified from the choice probabilities in Table~\ref{tab:prop3_not_necessary}.
\begin{table}[h]
\centering
\tabcolsep=10pt
\caption{A GSP model that is regular but does not satisfy the sufficient condition in Proposition~\ref{prop:when_is_gsp_regular}\label{tab:gsp_model_necessity}}
\begin{tabular}{|c|c|c|c|}\hline
  \multicolumn{3}{|c|} {Consumer type} & Probability \\ \hline
  label & $\bell$ & $k$ & \\
  \hline
  1 & (1,2,3) & 1 & 0.1 \\
  \hline
  2 & (1,2,3) & 2 & 0.3 \\
  \hline
    3 & (1,3,2) & 1 & 0.2 \\
  \hline
      4 & (1,3,2) & 2 & 0.4 \\
      \hline
\end{tabular}
\end{table}
\begin{table}[h]
\caption{Choice probabilities resulting from GSP model in Table~\ref{tab:gsp_model_necessity}\label{tab:prop3_not_necessary}}
\begin{center}
\begin{tabular}{ c | c c c }
$S$ & $\mathcal{P}(1, S)$ & $\mathcal{P}(2, S)$ & $\mathcal{P}(3, S)$ \\
\hline
$\{1,2\}$ & 0.3 & 0.7 & - \\
$\{1,3\}$ & 0.3 & - & 0.7 \\
$\{2,3\}$ & - & 0.5 & 0.5 \\
$\{1,2,3\}$ & 0.3 & 0.3 & 0.4  \\
\end{tabular}
\end{center}
\end{table} \Halmos
}
\subsection{Analysis of the GRUM model}
\label{app:proof_prop5}
The following proposition characterizes the relation between the GRUM and GSP models:
\begin{proposition}[Relationship between GRUM and GSP models]
\label{prop:grum_contained_in_gsp}
The GRUM model class is strictly contained within the class of GSP models.
\end{proposition}
\proof{Proof.}
\newcommand{\rev}{\text{rev}}
Given an instance of the GRUM model, consider a GSP model with the following distribution $\lambda(\cdot)$ over the consumer types:
\begin{equation}
\label{eq:gsp_instance_equivalent_to_gmnl}    
\lambda\left((\bell, k)\right) = \begin{cases}
\lambda_1 \cdot \alpha_{\bell} + \lambda_n \cdot \alpha_{\rev(\bell)} & \text{if } k=1 \\
\lambda_k \cdot \alpha_{\bell} & \text{if } k > 1
\end{cases},
\end{equation}
where recall that $\rev(\bell)$ is the ordering obtained by reversing product positions in $\bell$.
It can be verified that~\eqref{eq:gsp_instance_equivalent_to_gmnl} results in a valid distribution. Next, consider the choice probabilities under the GSP instance in~\eqref{eq:gsp_instance_equivalent_to_gmnl} according to Definition~\ref{def:gsp_prob_fn}:
\begin{align*}
\Pcal(j, S) &=  \sum_{\omega \in \Omega} C_{\omega}(j,S) \cdot \lambda(\omega) \\
&= \sum_{\bell \in \Lscr} \sum_{k=1}^{n-1}C_{(\bell, k)}(j,S) \cdot \lambda\left((\bell, k)\right) \\
&= \sum_{\bell \in \Lscr} C_{(\bell, 1)}(j,S) \cdot \lambda\left((\bell, 1)\right) + \sum_{\bell \in \Lscr} \sum_{k=2}^{n-1}C_{(\bell, k)}(j,S) \cdot \lambda\left((\bell, k)\right) \\
&= \sum_{\bell \in \Lscr} C_{(\bell, 1)}(j,S) \cdot\left(\lambda_1 \cdot \alpha_{\bell} + \lambda_n \cdot \alpha_{\rev(\bell)} \right)  + \sum_{\bell \in \Lscr} \sum_{k=2}^{n-1}C_{(\bell, k)}(j,S) \cdot (\lambda_k \cdot \alpha_{\bell}) \\
&= \sum_{\bell \in \Lscr} \sum_{k=1}^{n-1}C_{(\bell, k)}(j,S) \cdot (\lambda_k \cdot \alpha_{\bell}) + \sum_{\bell \in \Lscr} C_{(\bell, 1)}(j,S) \cdot (\lambda_n \cdot \alpha_{\rev(\bell)})\\
&= \sum_{\bell \in \Lscr} \sum_{k=1}^{n-1}C_{(\bell, k)}(j,S) \cdot (\lambda_k \cdot \alpha_{\bell}) + \sum_{\bell \in \Lscr} C_{(\rev(\bell), n)}(j,S) \cdot (\lambda_n \cdot \alpha_{\rev(\bell)}) \\
&= \sum_{\bell \in \Lscr} \sum_{k=1}^{n}C_{(\bell, k)}(j,S) \cdot (\lambda_k \cdot \alpha_{\bell}).
\end{align*}
The claim then follows from the expression for the choice probabilities under the GRUM model in Definition~\ref{def:krum_prob_fn}.

Next, we exhibit a GSP instance that lies outside the GRUM model class. Consider the GSP model with three alternatives and three consumer types shown in Table \ref{GSP_but_not_GRUM_example}. For succinct notation, denote $\alpha_{123} = \Pr(U_1 > U_2 > U_3)$ and similarly for the other orderings.
\begin{table}[h]
\centering
\begin{tabular}{|c|c|c|c|}\hline
  \multicolumn{3}{|c|} {Consumer type} & Probability \\ \hline
  label & $\bell$ & $k$ &  \\
  \hline
  1 & (1,2,3) & 1 & 0.1 \\
  \hline
 2 & (1,2,3) & 2 & 0.6 \\
  \hline
  3 & (1,3,2) & 1 & 0.3 \\
  \hline
\end{tabular}
\caption{A GSP model that lies outside the GRUM class}
\label{GSP_but_not_GRUM_example}
\end{table}
Then, if there exists a GRUM model that satisfies the choice probabilities generated by this GSP model, there must exist feasible values of $(\alpha_{\bell} \colon \bell \in \Lscr)$ as well as proportions $\left(\lambda_1, \lambda_2, \lambda_3\right)$ that satisfy the equation in Definition~\ref{def:krum_prob_fn} for all offer-sets $S$ and all alternatives $j \in S$. Expanding the equations for a subset of the offer-sets, we get the following:
\begin{align}
\Pcal(1, \set{1, 3}) &= \lambda_1 \cdot(\alpha_{123} + \alpha_{132} + \alpha_{213}) + \lambda_2 \cdot(\alpha_{312} + \alpha_{321} + \alpha_{231}) + \lambda_3 \cdot(\alpha_{312} + \alpha_{321} + \alpha_{231})  \label{pr1_13} \\
\Pcal(2, \set{2, 3}) &= \lambda_1 \cdot(\alpha_{123} + \alpha_{213} + \alpha_{231}) + \lambda_2 \cdot(\alpha_{132} + \alpha_{312} + \alpha_{321}) + \lambda_3 \cdot(\alpha_{132} + \alpha_{312} + \alpha_{321}) \label{pr2_23}\\
\Pcal(2, \set{1, 2, 3}) &= \lambda_1 \cdot(\alpha_{213} + \alpha_{231}) + \lambda_2 \cdot(\alpha_{123} + \alpha_{321}) + \lambda_3 \cdot(\alpha_{132} + \alpha_{312}) \label{pr2_123}\\
\Pcal(3, \set{1, 2, 3}) &= \lambda_1 \cdot(\alpha_{312} + \alpha_{321}) + \lambda_2 \cdot(\alpha_{132} + \alpha_{231}) + \lambda_3 \cdot(\alpha_{213} + \alpha_{123}) \label{pr3_123}
\end{align}
We consider two different cases:
\begin{enumerate}
\item {\bf Case 1:} $\lambda_1 > 0$. It can be verified that $\Pcal(3, \set{1, 2, 3}) = 0$ under the GSP model in Table~\ref{GSP_but_not_GRUM_example}. Then, it follows from equation~\eqref{pr3_123} that $\alpha_{312} = \alpha_{321} = 0$. Next, we have the following two sub-cases:
\begin{enumerate}
\item {\bf Subcase 1:} $\lambda_1 < 1$. Consider the following possible scenarios:

(i) $\lambda_2 > 0$ and $\lambda_3 > 0$. It follows from equation~\eqref{pr3_123} that $\alpha_{123} = \alpha_{132} = \alpha_{231} = \alpha_{213} = \alpha_{312} = \alpha_{321} = 0$, which is clearly a contradiction.\\
(ii) $\lambda_2 > 0$ but $\lambda_3 = 0$. It follows from equation~\eqref{pr3_123} that $\alpha_{132} = \alpha_{231} = 0$. Then, plugging these into~\eqref{pr1_13} and~\eqref{pr2_23}, it follows
\begin{align*}
\Pcal(1, \set{1,3}) = \lambda_1 \cdot(\alpha_{123} + \alpha_{213}) = 0.4 \\
\Pcal(2, \set{2,3}) = \lambda_1 \cdot(\alpha_{123} + \alpha_{213}) = 0.1
\end{align*}
where the probability values are again computed from Table~\ref{GSP_but_not_GRUM_example}. This results in a contradiction.\\
(iii) $\lambda_2 = 0$ and $\lambda_3 > 0$. It follows from equation~\eqref{pr3_123} that that $\alpha_{213} = \alpha_{123} = 0$. Then, plugging these into~\eqref{pr1_13} and~\eqref{pr2_23}, it follows
\begin{align*}
\Pcal(1, \set{1,3}) = \lambda_1 \cdot \alpha_{132} + \lambda_3 \cdot \alpha_{231} = 0.4 \\
\Pcal(2, \set{2,3}) = \lambda_1 \cdot \alpha_{231} + \lambda_3 \cdot  \alpha_{132} = 0.1
\end{align*}
Adding the above two equations gives $\alpha_{123} + \alpha_{231} = 0.5 < 1$ (since $\lambda_1 + \lambda_3 = 1$), which results in a contradiction since the alphas must sum to one.
\item {\bf Subcase 2:} $\lambda_1 = 1$. Plugging $\lambda_2 = \lambda_3 = 0$ into~\eqref{pr2_23} and~\eqref{pr2_123} gives
\begin{align*}
\Pcal(2, \set{2,3}) &= \alpha_{231} + \alpha_{123} + \alpha_{213} = 0.1 \\
  \Pcal(2, \set{1,2,3})&= \alpha_{213} + \alpha_{231} = 0.6
\end{align*}
Solving the above equations we get $\alpha_{123} = -0.5 < 0$ which leads to a contradiction since all the alphas are non-negative.
\end{enumerate}
\item {\bf Case 2:} $\lambda_1 = 0$. Consider the following three sub-cases:
\begin{enumerate}
\item {\bf Subcase 1:} $\lambda_2 > 0$ and $\lambda_3 > 0$. Since $\Pcal(3, \set{1,2,3}) = 0$, it follows from~\eqref{pr3_123} that $\alpha_{132} = \alpha_{231} = \alpha_{213} = \alpha_{123} = 0$. Then plugging these into~\eqref{pr1_13} and~\eqref{pr2_23} gives
\begin{align*}
\Pcal(1, \set{1,3}) = \alpha_{312} + \alpha_{321} = 0.4 \\
\Pcal(2, \set{2,3}) = \alpha_{312} + \alpha_{321} = 0.1
\end{align*}
which results in a contradiction.
\item {\bf Subcase 2:} $\lambda_2 = 0$ and $\lambda_3 = 1$. Since $\Pcal(3, \set{1,2,3}) = 0$, it follows from~\eqref{pr3_123} that $\alpha_{123} = \alpha_{213} = 0$. Then, plugging these into~\eqref{pr2_23} and~\eqref{pr2_123} gives
\begin{align*}
\Pcal(2, \set{2,3}) &= \alpha_{132} + \alpha_{312} + \alpha_{321} = 0.1 \\
  \Pcal(2, \set{1,2,3}) &= \alpha_{132} + \alpha_{312} = 0.6
\end{align*}
Solving the above equations we get $\alpha_{321} = -0.5 < 0$ which results in a contradiction since all the alphas are non-negative.
\item {\bf Subcase 3:} $\lambda_2 = 1$ and $\lambda_3 = 0$. Since $\Pcal(3, \set{1,2,3}) = 0$, it follows from~\eqref{pr3_123} that $\alpha_{132} = \alpha_{231} = 0$. Then, plugging these into~\eqref{pr1_13} and~\eqref{pr2_23} gives
\begin{align*}
\Pcal(1, \set{1,3}) = \alpha_{312} + \alpha_{321} = 0.4 \\
  \Pcal(2, \set{2,3}) = \alpha_{312} + \alpha_{321} = 0.1
  \end{align*}
which again leads to a contradiction.
\end{enumerate}
\end{enumerate}
The above sequence of arguments show that there exist no set of proportions $(\lambda_1, \lambda_2, \lambda_3)$ that can satisfy the choice probabilities implied by the GSP model in Table~\ref{GSP_but_not_GRUM_example}, and hence the claim follows.
\Halmos \endproof

\av{The following proposition provides a sufficient condition for regularity of the GRUM model, and by extension the GMNL model:
\begin{proposition}[Sufficient condition for regularity of GRUM]
\label{prop:grum_regular}
Suppose the GRUM instance has model parameters of the form $\lambda_1 \geq\lambda_2 \geq \ldots \geq \lambda_n$. Such an instance belongs to the class of regular choice models.    
\end{proposition}
\proof{Proof.}
From the proof of Proposition~\ref{prop:grum_contained_in_gsp}, consider the GSP instance in~\eqref{eq:gsp_instance_equivalent_to_gmnl}. It is easy to see that this GSP instance satisfies the sufficient condition in Proposition~\ref{prop:when_is_gsp_regular} when $\lambda_1 \geq\lambda_2 \geq \ldots \geq \lambda_n$. The claim then follows by combining the results of Proposition~\ref{prop:grum_contained_in_gsp} and Proposition~\ref{prop:when_is_gsp_regular}.
\Halmos \endproof

\subsubsection{Proof of Proposition~\ref{prop:kmnl_prob_recurrence}.}
The case of $k > \abs{S}$ follows directly from Definition~\ref{def:krum_choice_fn}. Therefore, suppose $k \leq \abs{S}$. Then using Definition~\ref{def:krum_choice_fn}, it follows that 
\begin{align*}
\pi_k(i, S; \prefvec) &= \Pr(\rank(i, S; \bm{U}) = k) \\
&= \sum_{j \in \osminus{S}{i}} \Pr(\rank(j, S; \bm{U}) = 1 \text{ and } \rank(i, S; \prefvec) = k) \\
&= \sum_{j \in \osminus{S}{i}} \Pr(\rank(j, S; \bm{U}) = 1) \cdot \Pr(\rank(i, S; \bm{U}) = k \; \vert \; \rank(j, S; \bm{U}) = 1) \\
&= \sum_{j \in \osminus{S}{i}} \Pr(\rank(j, S; \bm{U}) = 1)  \cdot \Pr(\rank(i, \osminus{S}{j}; \bm{U}) = k-1) \\
&= \sum_{j \in \osminus{S}{i}} \pi_1(j, S; \prefvec) \cdot \pi_{k-1}(i, \osminus{S}{j}; \prefvec)
\end{align*}
In the above, the second equality follows since $k > 1$ and the second to last equality follows from the rank-order conditional independence property of Gumbel distributed utilities, see, e.g.,~\citet[Section 3]{bai2023assortment}, which states that conditioned on the top-ranked product(s), the ranking of the remaining products is identical to the unconditional setting.
\Halmos
\subsubsection{Example of regularity violation under GMNL model}

 Suppose $n=3$ and consider the following mean utilities for the products: $v_1 = \log 2, v_2 = \log 1.5, v_3 = 0$. Moreover, let the type distribution $\bm{\lambda}$ be as follows: $\lambda_1=\lambda_3=0$, and $\lambda_2 = 1$. Then, using Definition~\ref{def:krum_prob_fn} and Proposition~\ref{prop:kmnl_prob_recurrence} it follows that:
\begin{eqnarray*}
\Pcal(1, \set{1,3}) &=& \sum_{k=1}^3 \pi_k(1, \set{1,3}; \prefvec) \cdot \lambda_k = \pi_2(1, \set{1,3}; \prefvec) \cdot 1 = \pi_1(3, \set{1,3}; \prefvec) \cdot \pi_1(1, \set{1}; \prefvec)  \\
&=& \frac{e^{v_3}}{e^{v_1} + e^{v_3}} \cdot 1 = \frac{1}{3} \\
\Pcal(1, \set{1,2,3}) &=& \sum_{k=1}^3 \pi_k(1, \set{1,2,3}; \prefvec) \cdot \lambda_k = \pi_2(1, \set{1,2,3}; \prefvec) \cdot 1 \\
&=& \pi_1(2, \set{1,2,3}; \prefvec) \cdot \pi_1(1, \set{1,3}; \prefvec) + \pi_1(3, \set{1,2,3}; \prefvec) \cdot \pi_1(1, \set{1,2}; \prefvec)  \\
&=& \frac{e^{v_2}}{e^{v_1} + e^{v_2} + e^{v_3}} \cdot \frac{e^{v_1}}{e^{v_1} + e^{v_3}} +  \frac{e^{v_3}}{e^{v_1} + e^{v_2} + e^{v_3}} \cdot \frac{e^{v_1}}{e^{v_1} + e^{v_2}} \\
&\simeq& 0.349 > \frac{1}{3}
\end{eqnarray*}}

\section{Model estimation details}
\subsection{Proof of Proposition~\ref{prop:type_subprob_optimal}}
\label{app:extreme_point_optimal}
Define the set of feasible distributions as $\Lambda := \set{\lambda \colon \Omega \to [0,1] \;\; | \;\; \sum_{\omega \in \Omega} \lambda(\omega) = 1, \sum_{\omega \in \Omega_{\nonrat}} \lambda(\omega) \leq \delta}$. We consider the two cases separately:

\underline{{\bf Case 1:} $\logll(\omega_{\rat}) \geq \logll(\omega_{\nonrat})$.} In this case, it follows that for any feasible distribution $\lambda \in \Lambda$:
\begin{align*}
 \sum_{\omega \in \Omega} \logll(\omega) \cdot \lambda(\omega) &=  \sum_{\omega \in \Omega_{\rat}} \logll(\omega) \cdot \lambda(\omega) +  \sum_{\omega \in \Omega_{\nonrat}} \logll(\omega) \cdot \lambda(\omega) \\
 &\leq  \sum_{\omega \in \Omega_{\rat}} \logll(\omega_{\rat}) \cdot \lambda(\omega) +  \sum_{\omega \in \Omega_{\nonrat}} \logll(\omega_{\nonrat}) \cdot \lambda(\omega) \\
  &\leq  \sum_{\omega \in \Omega_{\rat}} \logll(\omega_{\rat}) \cdot \lambda(\omega) +  \sum_{\omega \in \Omega_{\nonrat}} \logll(\omega_{\rat}) \cdot \lambda(\omega) \\
  &= \logll(\omega_{\rat}) \cdot  \sum_{\omega \in \Omega}  \lambda(\omega)  \\
   &= \logll(\omega_{\rat}) \\
   &=  \sum_{\omega \in \Omega} \logll(\omega) \cdot \tlambda(\omega),
\end{align*}
where the first inequality follows from the definitions of $\omega_{\rat}$ and $\omega_{\nonrat}$, the second inequality since $\logll(\omega_{\nonrat}) \leq \logll(\omega_{\rat})$, and the final equality follows from the definition of $\tlambda$ in the statement of the Lemma.

\underline{{\bf Case 2:} $\logll(\omega_{\nonrat}) > \logll(\omega_{\rat})$.} In this case, it follows that for any feasible distribution $\lambda \in \Lambda$:
\begin{align*}
 \sum_{\omega \in \Omega} \logll(\omega) \cdot \lambda(\omega) &=  \sum_{\omega \in \Omega_{\rat}} \logll(\omega) \cdot \lambda(\omega) +  \sum_{\omega \in \Omega_{\nonrat}} \logll(\omega) \cdot \lambda(\omega) \\
 &\leq  \sum_{\omega \in \Omega_{\rat}} \logll(\omega_{\rat}) \cdot \lambda(\omega) +  \sum_{\omega \in \Omega_{\nonrat}} \logll(\omega_{\nonrat}) \cdot \lambda(\omega) \\
  &=  \logll(\omega_{\rat}) \cdot \sum_{\omega \in \Omega_{\rat}}  \lambda(\omega) + \logll(\omega_{\nonrat}) \cdot  \sum_{\omega \in \Omega_{\nonrat}}  \lambda(\omega) \\
  &= \logll(\omega_{\rat}) \cdot  \left( 1- \sum_{\omega \in \Omega_{\nonrat}}  \lambda(\omega) \right) + \logll(\omega_{\nonrat}) \cdot  \sum_{\omega \in \Omega_{\nonrat}}  \lambda(\omega) \\
  &= \logll(\omega_{\rat}) + \left( \logll(\omega_{\nonrat}) -  \logll(\omega_{\rat})\right) \cdot \sum_{\omega \in \Omega_{\nonrat}}  \lambda(\omega) \\
   &\leq \logll(\omega_{\rat}) + \left( \logll(\omega_{\nonrat}) -  \logll(\omega_{\rat})\right) \cdot \delta \\
     &= \logll(\omega_{\rat}) \cdot (1-\delta) + \logll(\omega_{\nonrat}) \cdot \delta \\
   &=  \sum_{\omega \in \Omega} \logll(\omega) \cdot \tlambda(\omega)
\end{align*}
where the first inequality follows from the definitions of $\omega_{\rat}$ and $\omega_{\nonrat}$, the second inequality since $\logll(\omega_{\nonrat}) \leq \logll(\omega_{\rat})$, and the final equality follows from the definition of $\tlambda$ in the statement of the lemma.
\Halmos

\subsection{ILP formulations for finding non-standard type in Proposition~\ref{prop:type_subprob_optimal}}
\label{app:single_ilp}
The decision variables are the same as in~\eqref{eq:gsp_mip}, along with an integer variable $k$ for the choice index:
\begin{subequations}
\label{eq:app_mip}
\begin{align}
\max_{\bm{y}, \bm{x}, k} \;\;\; &\sum_{t=1}^T \sum_{j \in S_t} \mus{r-1}_{jt} \cdot y_{jt} \label{eq:app_mip_obj}\\
\text{s.t. } &x_{ji} + x_{ij} = 1 & \forall ~ i,j \in \Cscr;  i\neq j \label{eq:app_rank_constr1}\\
&x_{ji} + x_{il} + x_{lj} \leq 2 & \forall~i,j,l \in \Cscr; i\neq j \neq l \label{eq:app_rank_constr2}\\
& (\abs{S_t} -1) \cdot \left(y_{jt} -1\right) + \sum_{i \in \osminus{S_t}{j}} x_{ji} \leq \left(\abs{S_t} - k\right)^+ & \forall~j \in S_t,\forall~t \in [T] \label{eq:app_feas_constr} \\
&  (\abs{S_t} -1) \cdot \left(y_{jt} - 1\right) - \sum_{i \in \osminus{S_t}{j}}x_{ji} \leq - \left(\abs{S_t} - k\right)^+  &  \forall~j \in S_t,\forall~t \in [T] \label{eq:app_feas_constr2}\\
&x_{ij} \in \set{0, 1}  \;\; \forall~i,j \in \Cscr;  \;\; y_{jt} \in \set{0, 1} \;\;\forall~j\in S_t, \forall~t \in [T]; \;\;\; k \in \set{2, 3, \ldots, n-1} \label{eq:var_domains}
\end{align}
\end{subequations}
Constraints~\eqref{eq:app_rank_constr1} and~\eqref{eq:app_rank_constr2} ensure that $\bm{x}$ corresponds to a valid ordering of the products. Next, we need to add constraints to ensure validity of the binary variables $\bm{y}$. If $\abs{S_t} \leq k$, then the {\em last} product in the subsequence $s(\bell, S_t)$ would be chosen by the consumer type. This means that for any $j \in S_t$, $y_{jt} = 1$ only if product $j$ appears {\em after} all other products $i \in \osminus{S_t}{j}$ in the ordering $\bell$, which can be equivalently written as $\sum_{i \in \osminus{S_t}{j}}x_{ji} = 0$. Otherwise, if $\abs{S_t} > k$, then $y_{jt} = 1$ implies that product $j$ is at position $k$ in the subsequence $s(\bell, S_t)$, which is equivalent to the condition $\sum_{i \in \osminus{S_t}{j}} x_{ji} = \abs{S_t} - k$.  Both of these conditions can be captured simultaneously using a single equation $\sum_{i \in \osminus{S_t}{j}} x_{ji} = \left(\abs{S_t} -k\right)^+$, where  $\left(\abs{S_t} -k\right)^+ = \max\left(\abs{S_t} -k, 0\right)$ is the positive part function. The inequality constraints~\eqref{eq:app_feas_constr}-\eqref{eq:app_feas_constr2} enforce the condition  that $y_{jt} = 1$ only if $\sum_{i \in \osminus{S_t}{j}} x_{ji} = \left(\abs{S_t} -k\right)^+$ for all $j \in S_t$ and all $t \in [T]$.

We can also reformulate problem~\eqref{eq:app_mip} to linearize the $\max$ term using standard techniques, resulting in an ILP formulation.\footnote{Refer to \url{https://orinanobworld.blogspot.com/2010/12/lps-and-positive-part.html} for an example of how to convert $a \leq \max(b, 0)$ into linear constraints.} Define $z_t = \max(\abs{S_t} - k, 0)$ for each $t \in [T]$. Further, introduce a binary decision variable $\gamma_t$ for each $t \in [T]$. Then, it can be verified that the following formulation is equivalent to~\eqref{eq:app_mip}: 
\begin{align*}
\max_{\bm{y}, \bm{x}, k, \bm{z}, \bm{\gamma}} \;\;\; &\sum_{t=1}^T \sum_{j \in S_t} \mus{r-1}_{jt} \cdot y_{jt} \\
\text{s.t. } &x_{ji} + x_{ij} = 1 & \forall ~ i,j \in \Cscr;  i\neq j \\
&x_{ji} + x_{il} + x_{lj} \leq 2 & \forall~i,j,l \in \Cscr; i\neq j \neq l\\
& (\abs{S_t} -1) \cdot \left(y_{jt} -1\right) + \sum_{i \in \osminus{S_t}{j}} x_{ji} \leq z_t & \forall~j \in S_t,\forall~t \in [T] \\
&  (\abs{S_t} -1) \cdot \left(y_{jt} - 1\right) - \sum_{i \in \osminus{S_t}{j}}x_{ji} \leq -z_t  &  \forall~j \in S_t,\forall~t \in [T]\\
& (\abs{S_t} - n)\cdot \gamma_t \leq \abs{S_t} - k  & \forall~t\in [T]\\
& \abs{S_t} - k \leq (\abs{S_t} - 1)\cdot (1-\gamma_t) & \forall~t\in [T]\\
& z_t + (\abs{S_t} - n)\cdot \gamma_t \leq \abs{S_t} - k & \forall~t\in [T]\\
& z_t  + (\abs{S_t} - n)\cdot \gamma_t \leq \abs{S_t} - 1 & \forall~t\in [T]\\
&z_t \geq 0; \;\; z_t \geq \abs{S_t} - k & \forall~t \in [T] \\
& \gamma_t \in \set{0,1}\;\; \forall~t\in [T]\quad x_{ij} \in \set{0,1}  \;\; \forall~i,j \in \Cscr  \quad y_{jt} \in \set{0, 1} \;\;\forall~j\in S_t, \forall~t \in [T] \\
& k \in \set{2, 3, \ldots, n-1}
\end{align*}

\begin{algorithm}[h]
\caption{\av{Frank-Wolfe procedure for MLE of GSP model}\label{alg:gsp_fw}}
\begin{algorithmic}
\STATE \textbf{Initialization:} consumer types $\Omega^{(0)} := \set{\omega_1, \omega_2, \ldots \omega_{I}}$, distribution $\lambdak{0}$ over $\Omega^{(0)}$, and choice probabilities $\Pcal^{(0)}(j, S_t) = \sum_{\omega \in \itertypes{0}} C_{\omega}(j, S_t) \cdot \lambdak{0}(\omega)$ for all $j \in S_t, t \in [T]$, $k_{\max} \gets$ largest choice index for non-standard types, $\delta \gets$ maximum proportion of non-standard types
\FOR{$r = 1, 2, \ldots$}
\STATE Compute $\mus{r-1}_{jt} := \frac{N_{jt}}{\Pcal^{(0)}(j, S_t)}$ for all $j \in S_t, t \in [T]$.
\STATE Construct consumer types 
\begin{itemize}
    \item
$\omega_{\rat}$ using optimal solution of ILP~\eqref{eq:gsp_mip} for $k=1$.
\item $\omega_{\nonrat}$ as the non-standard type with the largest optimal objective~\eqref{eq:mip_obj} for $2 \leq k \leq k_{\max}$
\end{itemize}
\IF{$\logll(\omega_{\rat}) \geq \logll(\omega_{\nonrat})$}
\STATE $\Omega^{(r)} := \Omega^{(r-1)} \cup \set{\omega_{\rat}}$
\ELSE 
\STATE $\Omega^{(r)} := \Omega^{(r-1)} \cup \set{\omega_{\rat}, \omega_{\nonrat}}$
\ENDIF
\STATE Update distribution $\lambdak{r}$ as optimal solution of the~\ref{eq:unconstr_mle} when $\Omega = \Omega^{(r)}$ 
\STATE Compute choice probabilities $\Pcal^{(r)}(j, S_t) = \sum_{\omega \in \itertypes{r}} C_{\omega}(j, S_t) \cdot \lambdak{r}(\omega)$ for all $j \in S_t, t \in [T]$
\ENDFOR
\end{algorithmic}
\end{algorithm}

\av{{\bf Incorporating the no-purchase option.} 
Here, we present the analogue of~\eqref{eq:gsp_mip} when a no-purchase option is available. In this scenario, the sales data is of the form $\left(N_{jt} \colon j \in S_t^+, t \in [T] \right)$.

Then, for a fixed choice index $k$, we need to find a non-standard type $(\gvec^+, k)$ satisfying the choice behavior defined in Section~\ref{sec:assortment}, with $\abs{\gvec} \geq k > 1$. 
The following ILP can be formulated to determine which non-standard type to add in each iteration of Algorithm~\ref{alg:gsp_fw}:
\begin{subequations}
\label{eq:app_mip_w_nochoice}
\begin{align}
\max_{\bm{y}, \bm{x}} \;\;\; &\sum_{t=1}^T \sum_{j \in \tS_t} \mus{r-1}_{jt} \cdot y_{jt}\label{eq:app_mip_w_nochoice_obj} \\
\text{s.t. } &x_{ji} + x_{ij} = 1 & \forall ~ i,j \in \Cscr \cup \set{0};  i\neq j \label{eq:app_mip_w_nochoice_rankconstr1}\\
&x_{ji} + x_{il} + x_{lj} \leq 2 & \forall~i,j,l \in \Cscr \cup \set{0}; i\neq j \neq l \label{eq:app_mip_w_nochoice_rankconstr2}\\
& \sum_{j=1}^n x_{j0} \geq k \label{eq:app_mip_w_nochoice_rankconstr3}\\
& (k - \abs{S_t}) \cdot y_{jt} + \sum_{i \in \osminus{S_t}{j}} x_{ji} \geq 0 & \forall~j \in S_t,\forall~t\in [T] \label{eq:app_mip_w_nochoice_feasconstr1}\\
&(k-1) \cdot y_{jt} + \sum_{i \in \osminus{S_t}{j}} x_{ji} \leq \abs{S_t} - 1 & \forall~j \in S_t,\forall~t \in [T] \label{eq:app_mip_w_nochoice_feasconstr2}\\
& y_{jt} \leq x_{j0} & \forall~j \in S_t,\forall~t \in [T] \label{eq:app_mip_w_nochoice_feasconstr3}\\
& (k - \abs{S_t} - 1) \cdot y_{0t} + \sum_{i \in S_t} x_{0i} \geq 0 & \forall~t\in [T] \label{eq:app_mip_w_nochoice_feasconstr4} \\
&x_{ij} \in \set{0, 1}  \;\; \forall~i,j \in \Cscr \cup \set{0};  \;\; y_{jt} \in \set{0, 1} \;\;\forall~j\in S_t^+, \forall~t \in [T] \label{eq:app_mip_w_nochoice_var_domains}
\end{align}
\end{subequations}
Constraints~\eqref{eq:app_mip_w_nochoice_rankconstr1}-\eqref{eq:app_mip_w_nochoice_rankconstr2} are the same as~\eqref{eq:rank_constr1}-\eqref{eq:rank_constr2}, except that the no-purchase option $0$ is also included. Constraint~\eqref{eq:app_mip_w_nochoice_rankconstr3} precludes the no-purchase option being among the first $k$ positions in the ordering $\gvec^+$ since otherwise it will always be chosen irrespective of the offer set $S$. Constraints~\eqref{eq:app_mip_w_nochoice_feasconstr1}-\eqref{eq:app_mip_w_nochoice_feasconstr2} are the same as~\eqref{eq:feas_constr1}-\eqref{eq:feas_constr2} and ensure that 
for any $t \in [T]$ and $j \in S_t$, $y_{jt} = 1$ only if product $j$ is at position $k$ in the subsequence $s(\gvec^+, S_t^+)$. 
In addition, constraint~\eqref{eq:app_mip_w_nochoice_feasconstr3} ensures that for any $j \in S_t$, $y_{jt} = 1$ only if product $j$ appears {\em before} product $0$ in $\gvec^+$.
Finally, for the no-purchase option $0$, it follows that for any $t \in [T]$, $y_{0t} = 1$ only if at most $k-1$ products in $S_t$ appear before $0$ in in the subsequence $s(\gvec^+, S_t^+)$, which is equivalent to the condition $\sum_{i \in S_t} x_{i0} \leq (k-1)$. This is captured by constraint~\eqref{eq:app_mip_w_nochoice_feasconstr4}. 

After solving the ILP above, the corresponding sequence $\gvec^+$ can be obtained in the same way as the solution of~\eqref{eq:gsp_mip}, where we ignore products appearing after the no-purchase option $0$.}

\subsection{Deriving the EM algorithm for the GMNL(2) model}
\label{app:gmnl_em_derivation}
\av{In our context, we have two kinds of latent variables: (i) the choice index corresponding to each data point $j \in S_t$ and $t \in [T]$, say $c_{jt} \in \set{1, 2}$, and (ii) the product $z_{jt} \in \osminus{S_t}{j}$ with the highest (realized) utility conditional on the event that a consumer with index $k=2$ generated the sales corresponding to data point $(j,t)$.\footnote{It suffices to consider only these variables and not one variable for each individual transaction since the posterior distribution of the latent variables depend only on the {\em combination} of offer set and chosen product.} The parameters to be estimated are the mean utilities $\prefvec$ and the proportion $\lambda_1$ of standard consumers. Denote the vector of latent variables as $\bm{c} = \left(c_{jt} \colon j \in S_t, t\in [T] \right)$ and $\bm{z} = \left(z_{jt} \colon j \in S_t, t \in [T] \right)$.

Then, the complete data log-likelihood can be written as
\begin{align*}
\LL(\prefvec, \lambda_1 \cond \bm{c}, \bm{z}) = &\sum_{t=1}^T\sum_{j\in S_t} N_{jt} \cdot\bigg\{\indicator{c_{jt} = 1} \cdot \log \left(\pi_1(j, S_t; \prefvec) \cdot \lambda_1 \right)\bigg. \\ 
&\bigg.+ \indicator{c_{jt} = 2} \cdot \sum_{i \in \osminus{S_t}{j}} \indicator{z_{jt} = i} \cdot \log\left(\pi_1(i, S_t; \prefvec) \cdot \pi_1(j, \osminus{S_t}{i}; \prefvec) \cdot (1 - \lambda_1) \right) \bigg\},
\end{align*}
where we substituted $\lambda_2 = 1 - \lambda_1$ and plugged in the choice probability expressions from Proposition~\ref{prop:kmnl_prob_recurrence}. Starting from some initial solution $(\vveciter{0}, \piter{0}_1)$, the EM algorithm performs the following two steps in each iteration $r \geq 1$:

{\em E-step.} Here, we first compute the posterior distribution of the unobserved variables $(\bm{c}, \bm{z})$ in the complete data log-likelihood function, given the current parameter estimates:
\[ \alpha_{jt}^{(r)} := \Pr(c_{jt} = 1 ; \vveciter{r-1}, \piter{r-1}_1) = \frac{ \piter{r-1}_1 \cdot \pi_1(j, S_t; \vveciter{r-1})}{\piter{r-1}_1 \cdot \pi_1(j, S_t; \vveciter{r-1}) + (1-\piter{r-1}_1) \cdot \pi_2(j, S_t; \vveciter{r-1})} \]


\[ \beta_{jt}^{(r)}(i) := \Pr(z_{jt} = i ; \prefvec^{(r-1)}) = \frac{\pi_1(i, S_t; \vveciter{r-1})\cdot \pi_1(j, \osminus{S_t}{i}; \vveciter{r-1})}{\sum_{l \in \osminus{S_t}{j}} \pi_1(l, S_t; \vveciter{r-1})\cdot \pi_1(j, \osminus{S_t}{l}; \vveciter{r-1})} \;\;\; \forall~ i \in \osminus{S_t}{j} \]
Then, the expected complete data log-likelihood takes the form:
\begin{align}
\label{eq:exp_compl_ll}
&\Exp[\LL(\prefvec, \lambda_1 \cond \bm{c}, \bm{z})] \\
&= \sum_{t=1}^T \sum_{j \in S_t} N_{jt} \cdot \set{\alpha_{jt}^{(r)} \cdot \log \left(\frac{e^{v_{j}}}{w(S_t)} \cdot \lambda_1 \right) + (1-\alpha_{jt}^{(r)} ) \cdot \sum_{i \in \osminus{S_t}{j}} \beta_{jt}^{(r)}(i) \cdot \log\left(\frac{e^{v_i}}{w(S_t)}\cdot \frac{e^{v_{j}}}{w(S_t) - e^{v_i}} \cdot (1 - \lambda_1) \right)} \nonumber,
\end{align}
where we substituted the MNL choice probability expression for $\pi_1(\cdot)$ and used the fact that $w(\osminus{S_t}{i}) = \sum\limits_{l \in \osminus{S_t}{i}} e^{v_{l}} = w(S_t) - e^{v_{i}}$.

{\em M-step.} Here, we find the parameter values that maximize the expected complete data log-likelihood~\eqref{eq:exp_compl_ll}. Taking the derivative w.r.t $\lambda_1$ results in the following closed-form update:
\[ \piter{r}_1 =  \frac{\sum_{t=1}^T\sum_{j\in S_t} N_{jt}\cdot \alpha_{jt}^{(r)}}{\sum_{t=1}^T\sum_{j\in S_t} N_{jt}} \]

To update the mean utilities $\prefvec$, we need to solve the following optimization problem:
\[ \vveciter{r} \in \argmax_{\prefvec} \sum_{t=1}^T \sum_{j \in S_t} N_{jt} \cdot \set{\alpha_{jt}^{(r)} \cdot \log \left(\frac{e^{v_{j}}}{w(S_t)}\right) + (1 - \alpha_{jt}^{(r)}) \cdot \sum_{i \in \osminus{S_t}{j}} \beta_{jt}^{(r)}(i) \cdot \log\left(\frac{e^{v_i}}{w(S_t)}\cdot \frac{e^{v_{j}}}{w(S_t) - e^{v_i}} \right)}. \]}

\section{Proof of results in Section~\ref{sec:assortment}}
\label{app:assortment_proofs}
\subsection{Regularity of GSP/GRUM model with no-purchase option.}
\av{The following result extends the sufficient condition in Proposition~\ref{prop:when_is_gsp_regular} and Proposition~\ref{prop:grum_regular} to account for the no-purchase option:
\begin{proposition}[Sufficient condition for regularity w/ no-purchase option]
\label{prop:when_is_gsp_w_nopurch_regular}
    Consider a GSP model with distribution $\lambda(\cdot)$ over consumer types $\Omega^{\np}$ satisfying the following inequalities:
    \[ \lambda((\gvec^+, k)) \geq \lambda((\gvec^+, k+1)) \text{ for all } \gvec \in \Gscr, 1 \leq k \leq \abs{\gvec} - 1 \] 
    Then, the corresponding GSP instance satisfies the regularity condition~\eqref{regularity_equation}. Consequently, a GRUM instance with model parameters of the form $\lambda_1 \geq \lambda_2 \ldots \geq \lambda_n$ also satisfies~\eqref{regularity_equation}.
    \end{proposition}
\proof{Proof.}
Consider any offer-set $S \subseteq [n]$, and alternative $j \in S$. For any $i \notin S$ and any $\gvec \in \Gscr$, note there are four possible scenarios: (1) both $i, j \in \gvec$, (2) $i \in \gvec$ but $j \notin \gvec$, (3) $i \notin \gvec$ but $j \in \gvec$, and (4) both $i, j \notin \gvec$. Under scenarios (2) and (4), it can be verified from Definition~\ref{def:gsp_choice_fn_with_default} that $C_{(\gvec^+, k)}(j, S) = C_{(\gvec^+, k)}(j, S \cup \set{i}) = 0$ for any $1 \leq k \leq \abs{\gvec}$. Further, under scenario (3), it is also the case that $C_{(\gvec^+, k)}(j, S) = C_{(\gvec^+, k)}(j, S \cup \set{i})$ for any $1 \leq k \leq \abs{\gvec}$. From Definition~\ref{def:gsp_prob_fn_with_default}, this implies that we can restrict attention to only those $\gvec$ that contain both $i$ and $j$. 

Define $\Gscr_{i < j} := \set{\gvec \in \Gscr \colon i \text{ appears before } j \text{ in } \gvec}$ and similarly define $\Gscr_{j < i}$. Then, it follows that
\begin{align*}
    &\sum_{\gvec \in \Gscr_{i < j}} \sum_{k=1}^{\abs{\gvec}} C_{(\gvec^+, k)}(j, S) \cdot \lambda((\gvec^+, k)) + 
     \sum_{\gvec \in \Gscr_{j < i}} \sum_{k=1}^{\abs{\gvec}} C_{(\gvec^+, k)}(j, S) \cdot \lambda((\gvec^+, k)) \\
    &\stackrel{\rm (a)}{=} \sum_{\gvec \in \Gscr_{i < j}} \sum_{k=1}^{\abs{\gvec}-1} C_{(\gvec^+, k+1)}(j, S \cup \set{i}) \cdot \lambda((\gvec^+, k)) +   \sum_{\gvec \in \Gscr_{j < i}} \sum_{k=1}^{\abs{\gvec}} C_{(\gvec^+, k)}(j, S) \cdot \lambda((\gvec^+, k))  \\
    &\stackrel{\rm (b)}{\geq} \sum_{\gvec \in \Gscr_{i < j}}\sum_{k=1}^{\abs{\gvec}-1} C_{(\gvec^+, k+1)}(j, S \cup \set{i}) \cdot \lambda((\gvec^+, k+1)) +   \sum_{\gvec \in \Gscr_{j < i}} \sum_{k=1}^{\abs{\gvec}} C_{(\gvec^+, k)}(j, S) \cdot \lambda((\gvec^+, k)) \\
    &\stackrel{\rm (c)}{\geq} \sum_{\gvec \in \Gscr_{i < j}}\sum_{k=1}^{\abs{\gvec}-1} C_{(\gvec^+, k+1)}(j, S \cup \set{i}) \cdot \lambda((\gvec^+, k+1)) + \sum_{\gvec \in \Gscr_{j < i}} \sum_{k=1}^{\abs{\gvec}} C_{(\gvec^+, k)}(j, S \cup \set{i}) \cdot \lambda((\gvec^+, k)) \\
    &= \sum_{\gvec \in \Gscr_{i < j}} \sum_{k=2}^{\abs{\gvec}} C_{(\gvec^+, k)}(j, S \cup \set{i}) \cdot \lambda((\gvec^+, k)) +  \sum_{\gvec \in \Gscr_{j < i}} \sum_{k=1}^{\abs{\gvec}} C_{(\gvec^+, k)}(j, S \cup \set{i}) \cdot \lambda((\gvec^+, k)) \\
    &\stackrel{\rm (d)}{=} \sum_{\gvec \in \Gscr_{i < j}} \sum_{k=1}^{\abs{\gvec}} C_{(\gvec^+, k)}(j, S \cup \set{i}) \cdot \lambda((\gvec^+, k)) +  \sum_{\gvec \in \Gscr_{j < i}} \sum_{k=1}^{\abs{\gvec}} C_{(\gvec^+, k)}(j, S \cup \set{i}) \cdot \lambda((\gvec^+, k)),
\end{align*}
where the justification for the numbered equations are provided next. For (a), it is clear that for any $\gvec \in \Gscr_{i < j}$,  $C_{(\gvec^+, \abs{\gvec})}(j, S) = 0$ and $C_{(\gvec^+, k)}(j, S) = C_{(\gvec^+, k+1)}(j, S \cup \set{i})$ whenever $1 \leq k \leq \abs{\gvec}-1$, according to Definition~\ref{def:gsp_choice_fn_with_default}. (b) follows from hypothesis. (c) follows $C_{(\gvec^+, k)}(j, S) = 0 \implies 
C_{(\gvec^+, k)}(j, S \cup \set{i}) = 0$ for any $\gvec \in \Gscr_{j < i}$ and any $1 \leq k \leq \abs{\gvec}$. (d) follows since $C_{(\gvec^+, 1)}(j, S \cup \set{i}) = 0$ for any $\gvec \in \Gscr_{i < j}$. The result then follows.
\Halmos \endproof}

\subsection{Proof of Proposition~\ref{prop:gsp_assort}.}
\label{app:proof_rev_order_gsp}
\av{We first establish the following lemma which shows that regularity, as stated in~\eqref{regularity_equation}, is satisfied for the no-purchase option in the GSP model:
\begin{lemma}[Regularity of no-purchase option under GSP]
\label{lemma_choice_increase_demand}
For the GSP model with a no-purchase option, it must be the case that  $\Pcal(0, S) \geq \Pcal(0, S')$
for every $ S\subseteq S'\subseteq [n]$. Equivalently, $\sum_{j \in S} \mathcal{P}(j, S) \leq \sum_{j \in S'}\mathcal{P}(j, S')$ for every $ S\subseteq S'\subseteq [n]$.

\end{lemma}
\proof{Proof.}
From Definition~\ref{def:gsp_prob_fn_with_default}, it is enough to show that for every consumer type $\omega =(\gvec^+, k)$, whenever $C_{\omega}(0, S') = 1$, it holds that $C_{\omega}(0, S) = 1$. 
If $C_{\omega}(0, S') = 1$, it must imply that $\abs{s\left(\gvec^+, (S')^+\right)} \leq k$, since otherwise one of the offered products in $S'$ will be chosen according to Definition~\ref{def:gsp_choice_fn_with_default}. 
Since $S \subseteq S'$, it follows that $k \geq \abs{s\left(\gvec^+, (S')^+\right)} \geq \abs{s(\gvec^+, S^+)}$, and therefore $C_{\omega}(0, S) = 1$ as well. The second part of the lemma follows since $\Pcal(0, S) = 1 - \sum_{j \in S} \Pcal(j, S)$ for any $S \subseteq [n]$.
\Halmos \endproof

We now proceed to establish the approximate guarantee for revenue-ordered assortments.

First, observe that the highest revenue obtained using revenue-ordered assortments is never less than what can be achieved by showing all alternatives. Denoting $S^*$ as an optimal assortment and $R^*$ the optimal revenue, it follows that:
\begin{eqnarray}
R^* &=& \sum_{i \in S^*}r_i \cdot \mathcal{P}(i,S^*) \nonumber \\
& \leq&  \sum_{i \in S^*}r_{\max} \cdot \mathcal{P}(i,S^*) \nonumber \\
& \leq &  r_{\max} \cdot \sum_{i \in [n]}\mathcal{P}(i, [n]) \nonumber \\
&= & \frac{r_{\max}}{r_{\min}} \sum_{i \in [n]} r_{\min} \cdot \mathcal{P}(i, [n])\nonumber \\
& \leq &  \frac{r_{\max}}{r_{\min}}  \sum_{i \in [n]} r_i \cdot \mathcal{P}(i, [n]) \nonumber
\end{eqnarray}
where the second inequality is due to Lemma \ref{lemma_choice_increase_demand}.

Next, we show that the bound $\frac{r_{\min}}{r_{\max}}$ is tight. Consider a GSP choice model with $n=3$ alternatives and a single consumer type $\omega= \left((1,2,3,0),2\right) \in \Omega^{\np}$. Let the product revenues be as follows: $r_1 = r_2 < r_3$. Then, it can be verified than an optimal solution for problem~\eqref{eq:gsp_assort} is $S^*=\{1,3\}$, which provides a revenue of $R^* = r_3 = r_{\max}$. The best revenue that can be obtained using revenue-ordered assortments, on the other hand, equals $r_2 = r_1 = r_{\min}$.
\Halmos}

\subsection{Proof of Proposition~\ref{prop:gmnl_assortment}.}
\label{app:proof_gmnl_assort}
\av{We first define the recurrence for choice probabilities under the GMNL model in the presence of a no-purchase option, analogous to Proposition~\ref{prop:kmnl_prob_recurrence}:
\begin{proposition}[Recurrence for choice probabilities under GMNL w/ no-purchase option]
\label{prop:gmnl_nopurch_recurrence}
Given any offer set $S \subseteq [n]$ and any product $i \in S$, the following recurrence holds for the choice probabilities in the GMNL model with a no-purchase option:
\[ \pi_k(i, S; \prefvec) =
\begin{cases}
\frac{v_i}{1 + w(S)} & \text{if } k = 1 \\
\sum_{j \in \osminus{S}{i}} \pi_1(j, S; \prefvec) \cdot \pi_{k-1}(i, \osminus{S}{j}; \prefvec) & \text{if } 1 < k \leq \abs{S} \\
0 & \text{otherwise.}
\end{cases}
\]
\end{proposition}
\proof{Proof.}
When $k=1$, $\pi_1(i, S; \prefvec)$ is the standard MNL probability, where we normalize the attraction for the no-purchase option to be $1$ without loss of generality. When $k > \abs{S}$, the customer chooses the no-purchase option by definition and therefore, $\pi_k(i, S; \prefvec) = 0$. Finally, when $1 < k \leq \abs{S}$, the expression is identical to that in Proposition~\ref{prop:kmnl_prob_recurrence}. In particular, in this scenario, we assume that product $i$ can be purchased only if its utility $U_i$ is higher than the utility of the no-purchase option $U_0$.
\Halmos \endproof
In the following, given any choice model instance $\Mcal$, we denote by $\Pcal_{\Mcal}(i, S)$, the choice probability of product $i$ in offer set $S$ for any $S \subseteq [n]$ and all $i \in S$. As mentioned in the main body, we begin by establishing the following equivalence between GMNL and MP-MNL:
\begin{lemma}[GMNL can recover MP-MNL choice probabilities]
\label{lem:mpmnl_subsumed_by_gmnl}
Given any instance $\Mcal$ of the MP-MNL model, there exists an instance $\Mcal'$ of the GMNL model such that 
\[ \Pcal_{\Mcal}(i, S) = C \cdot \Pcal_{\Mcal'}(i, S) \quad \text{for all } i \in S \text{ and } S \subseteq [n], \]
where $C > 0$ is a constant independent of $i$ and $S$. Moreover, the instance $\Mcal'$ satisfies regularity.
\end{lemma}
\proof{Proof.}
The MP-MNL model parameters comprise a vector of product attractions $\bm{v} = (v_1, v_2, \ldots, v_n)$, similar to the MNL and GMNL models, and a random variable $\bm{M}$ specifying the maximum number of products the customer is interested in purchasing. Given the model parameters, the choice probabilities under an instance $\Mcal$ of the MP-MNL model are of the following form:
\begin{equation}
\label{eq:mp-mnl-choiceprob}
\Pcal_{\Mcal}(i, S) = \sum_{m\in[n]} \Pr[\bm{M}=m]\cdot \pi_m^{\mpmnl}(i, S; \prefvec), 
\end{equation}
where $\pi_m^{\mpmnl}(i, S; \prefvec)$ is the probability of purchasing product $i$ from offer set $S$ given that the customer is willing to make at most $m$ purchases. When $m=1$, $\pi_1^{\mpmnl}(i, S; \prefvec) = \frac{v_i}{1 + w(S)}$ which is nothing but the probability of selecting product $i$ from $S$ under the standard MNL model.  For $m > 1$, these probabilities satisfy the recurrence (Lemma 1 in~\citealp{bai2023assortment}):
\begin{equation} 
\pi_m^{\mpmnl}(i, S; \prefvec) = \pi_1^{\mpmnl}(i, S; \prefvec) + \sum_{j \in \osminus{S}{i}} \pi_1^{\mpmnl}(j, S; \prefvec)\cdot \pi_{m-1}^{\mpmnl}(i, \osminus{S}{j}; \prefvec)
\label{eq:mpmnl-recurrence}
\end{equation}

We first establish a relation between $\pi_m^{\mpmnl}(i, S)$ and the GMNL probabilities $\pi_k(i, S)$, where we drop the explicit dependence on the product attractions $\prefvec$ to simplify the notation.

\underline{\bf Claim}: For any $1 \leq m \leq n$, $\pi_m^{\mpmnl}(i, S) = \sum_{k=1}^m \pi_k(i, S)$ for all $S \subseteq [n]$ and all $i \in S$.

We prove the claim via induction. When $m=1$, the LHS and RHS both reduce to the MNL probability. Suppose the claim is true for $1 < m < \abs{S}$ and consider the following
\begin{align*}
    \pi_{m+1}^{\mpmnl}(i, S) &\stackrel{\rm (a)}{=} \pi_1^{\mpmnl}(i, S) +  \sum_{j \in \osminus{S}{i}} \pi_1^{\mpmnl}(j, S)\cdot \pi_{m}^{\mpmnl}(i, \osminus{S}{j}) \\
    &\stackrel{\rm (b)}{=} \pi_1^{\mpmnl}(i, S) +  \sum_{j \in \osminus{S}{i}} \pi_1^{\mpmnl}(j, S)\cdot \left(\sum_{k=1}^m \pi_k(i, \osminus{S}{j}) \right) \\
    &\stackrel{\rm (c)}{=}  \pi_1(i, S) + \sum_{k=1}^m \left(\sum_{j \in \osminus{S}{i}} \pi_1(j, S) \cdot \pi_k(i, \osminus{S}{j})\right) \\
    &\stackrel{\rm (d)}{=} \pi_1(i, S) + \sum_{k=1}^m \pi_{k+1}(i, S) \\
    &= \sum_{k=1}^{m+1} \pi_k(i, S),
\end{align*}
where (a) follows from~\eqref{eq:mpmnl-recurrence}, (b) follows from the induction hypothesis, (c) follows since $\pi_1^\mpmnl(i, S) = \pi_1(i, S)$, and (d) follows from Proposition~\ref{prop:gmnl_nopurch_recurrence} whenever $k + 1 \leq m + 1 \leq \abs{S}$.

Finally,~\cite{bai2023assortment} showed that $\pi_m^\mpmnl(i, S) = \pi_{\abs{S}}^\mpmnl(i, S)$ for all $m > \abs{S}$. Since $\pi_k(i, S) = 0$ when $k > \abs{S}$ by Proposition~\ref{prop:gmnl_nopurch_recurrence}, the claim continues to hold even when $m > \abs{S}$.

With the claim established, given any instance $\Mcal$ of the MP-MNL model, consider a GMNL model with the same product attractions and the distribution $\bm{\lambda}$ defined as follows
\begin{equation} 
\label{eq:gmnl-lambdak}
\lambda_k := \frac{\sum_{m=k}^n \Pr[\bm{M}=m]}{\sum_{m=1}^n \Pr[\bm{M}=m]\cdot m} \;\;\ \forall~ k \in [n] \end{equation}
Clearly, $\lambda_k \geq 0$ and it can be verified that $\sum_{k=1}^n \lambda_k = 1$. Moreover, it is easy to see that $\lambda_1 \geq \lambda_2 \geq \ldots \geq \lambda_n$ which establishes that the above GMNL instance satisfies regularity according to Proposition~\ref{prop:when_is_gsp_w_nopurch_regular} and Lemma~\ref{lemma_choice_increase_demand}.

Then, it follows that the choice probability under the constructed GMNL model, say $\Mcal'$, is of the form:
\begin{align*}
\Pcal_{\Mcal'}(i, S) &= \sum_{k=1}^n \pi_k(i, S) \cdot \lambda_k \\
&\stackrel{\rm (a)}{=} \sum_{k=1}^n \pi_k(i, S) \cdot  \frac{\sum_{m=k}^n \Pr[\bM=m]}{\sum_{m=1}^n \Pr[\bM=m]\cdot m} \\
&= \frac{\sum_{m=1}^n \Pr[\bM=m] \cdot \left(\sum_{k=1}^m \pi_k(i, S)\right)}{\sum_{m=1}^n \Pr[\bM=m]\cdot m} \\
&\stackrel{\rm (b)}{=} \frac{\sum_{m=1}^n \Pr[\bM=m] \cdot \pi_m^\mpmnl(i, S)}{\sum_{m=1}^n \Pr[M=m]\cdot m} \\
&\stackrel{\rm (c)}{=} \frac{\Pcal_{\Mcal}(i,S)}{\sum_{m=1}^n \Pr[\bM=m]\cdot m},
\end{align*}
where (a) follows from equation~\eqref{eq:gmnl-lambdak}, (b) follows from the claim above, and (c) follows from equation~\eqref{eq:mp-mnl-choiceprob}.
\Halmos \endproof
The result establishes that the choice probabilities under the MP-MNL model can be recovered by the GMNL model, up to a constant factor $C > 0$ that is independent of the product or offer set. The constant $C = \sum_{m=1}^n \Pr[\bM = m] \cdot m = \Exp[\bM]$, where $\Exp[\bM]$ is the average number of purchases a random customer is interested in making, can be interpreted as a scaling parameter. In traditional single-purchase choice models such as GMNL, it always holds that $\sum_{i \in S} \Pcal(i, S) \leq 1$ for any offer set $S$. However, this constraint does not apply in multi-purchase models, where the total probability across all products in an offer set may exceed one.

Lemma~\ref{lem:mpmnl_subsumed_by_gmnl} directly implies the hardness of approximating the TU-constrained assortment problem under the subclass of regular GMNL models, using the inapproximability result in \citet[Theorem A.1]{bai2023assortment}.
 This is because the expected revenue under the MP-MNL model is simply a scaled version of the expected revenue under the corresponding GMNL instance.}
\subsection{Performance of revenue-ordered assortments for GMNL model}
\label{app:gmnl_ro}
\av{We focus on the GMNL($k_{\max}$) model for $k_{\max}=2,3,4,5$ and consider the following different distributions for consumer types: (1) {\tt standard majority}, where $\lambda_1 = 0.8$ and $\lambda_k = 0.2 / (k_{\max} - 1)$ for all $1< k \leq k_{\max}$; (2) {\tt equal}, where $\lambda_k = 1/k_{\max}$ for all $1 \leq k\leq k_{\max}$; (3) {\tt non-standard majority}, where $\lambda_1 = 0.2$ and $\lambda_k = 0.8 / (k_{\max} - 1)$ for all $1< k \leq k_{\max}$ (4) {\tt pure non-standard}, where $\lambda_{k_{\max}} = 1$; and (5) {\tt random}, where we sample $\lambda_{k} \sim U(0,1)$ independently for each $k \in \set{1, \hdots, k_{\max}}$ and then normalize so that $\sum_{k \in [k_{\max}]} \lambda_k = 1$.


For each scenario and for each value of $n \in \{3, \ldots, 10\}$, we generate 5,000 random problem instances as follows. Product utilities are sampled uniformly from the range $(-4, 4)$, and product revenues are sampled uniformly from the range $(1, 100)$. For each instance, we compute the revenue ratio of RO assortments, defined as the fraction of the optimal revenue achieved by the best revenue-ordered assortment. Table~\ref{fig:gmnl_ro} reports the worst-case and average revenue ratio for each of the five scenarios described above when $n=10$; the insights are similar for smaller values of $n$.

In general, we see that the relative performance under GMNL($k_{\max}$) models with $k_{\max} \in \{2,3, 4, 5\}$ is quite similar. In particular, we observe that RO assortments perform extremely well under the {\tt standard majority} scenario, with the revenue ratio exceeding $94\%$ across all instances. This is promising as it is usually the most realistic model in practice. In addition, RO assortments also exhibit robust performance under the {\tt equal} scenario. As the proportion of non-standard types increases, performance gradually declines. However, even in the {\tt pure non-standard} case (i.e., $\lambda_{k_{\max}} = 1$), we find that RO assortments still achieve a worst-case revenue ratio of $62\%$. In addition, the average-case revenue ratio of RO assortments is significantly better, with near-optimal performance across all scenarios. We also note that the performance was robust to the price range---we obtained similar numbers when experimenting using $U(1, p_{\max})$ with $p_{\max} \in \set{10, 25, 50, 75}$.}

\begin{table}
\tabcolsep=10.0pt
\TABLE
{\av{Performance of revenue-ordered assortments for GMNL model with $n=10$ products\label{fig:gmnl_ro}}}
{\begin{tabular}[t]{*{9}{c}}
\toprule
  Type distribution &  \multicolumn{2}{c} {GMNL(2)} &  \multicolumn{2}{c} {GMNL(3)} &  \multicolumn{2}{c} {GMNL(4)} &  \multicolumn{2}{c} {GMNL(5)} \\
  \cmidrule(lr){2-3} \cmidrule(lr){4-5} \cmidrule(lr){6-7} \cmidrule(lr){8-9}
 & Min & Avg  & Min & Avg & Min & Avg & Min & Avg \\
  \midrule
{\tt standard majority} & 0.94 & 1.00 & 0.95 & 1.00& 0.94 & 1.00& 0.97 & 1.00
 \\
   {\tt equal} & 0.94 & 1.00 & 0.94 & 1.00& 0.95 & 1.00& 0.97 & 1.00 \\
     {\tt non-standard majority} & 0.74 & 0.99 & 0.88 & 1.00& 0.94 & 1.00& 0.97 & 1.00
 \\
     {\tt pure non-standard} & 0.62 & 0.98 &  0.62 & 0.97& 0.63 & 0.97& 0.63 & 0.98
 \\
     {\tt random} & 0.76 & 1.00& 0.80 & 1.00& 0.78 & 1.00& 0.83 & 1.00
 \\
     \bottomrule
\end{tabular}}
{``Min'' and ``Avg'' report the worst-case and average {\em revenue ratio}, that is, fraction of the optimal revenue achieved by the best revenue-ordered assortment.}
\end{table}

\section{Comparison of proposed models with existing models}\label{sec:properties}
\subsection{MP-MNL model}
\av{As before, we denote by $\Pcal_{\Mcal}(i, S)$ the choice probability of product $i$ in offer set $S$ for any $S \subseteq [n]$ and all $i \in S$ for any choice model instance $\Mcal$. The following result establishes the converse of Lemma~\ref{lem:mpmnl_subsumed_by_gmnl} for a subclass of the GMNL model:
\begin{proposition}[MP-MNL can recover choice probabilities under a subclass of GMNL]
\label{prop:regular_gmnl_subsumed_by_mpmnl}
Consider an instance $\Mcal$ of the GMNL model that satisfies $\lambda_1 \geq \lambda_2 \ldots \geq \lambda_n$. There exists an instance $\Mcal'$ of the MP-MNL model such that 
\[ \Pcal_{\Mcal}(i, S) = \lambda_1 \cdot \Pcal_{\Mcal'}(i, S) \quad \text{for all } i \in S \text{ and } S \subseteq [n]. \]
\end{proposition}
\proof{Proof.}
We construct the MP-MNL instance $\Mcal'$ as follows. The product attractions are the same as in the GMNL instance $\Mcal$. The random variable $\bM$ is assumed to have the following distribution (see the proof of Lemma~\ref{lem:mpmnl_subsumed_by_gmnl} for the description of the model parameters under the MP-MNL model):
\begin{equation}
\label{eq:mpmnl-purch-distr}
    \Pr[\bM = m] = \frac{\lambda_{m} - \lambda_{m+1}}{\lambda_1} \quad \forall~m \in [n-1]; \quad \Pr[\bM = n] = \frac{\lambda_n}{\lambda_1}
\end{equation}
Because $\lambda_1 \geq \ldots \geq \lambda_n$, it follows that $\Pr[\bM = m] \geq 0$ for all $m \in [n]$. Moreover, it is easy to see that $\sum_{m \in [n]} \Pr[\bM = m] = 1$.

Then, it follows that the choice probability under the constructed MP-MNL instance is of the form:
\begin{align*}
    \Pcal_{\Mcal'}(i, S) &\stackrel{\rm (a)}{=} \sum_{m=1}^n \Pr[\bM = m] \cdot \pi_m^{\mpmnl}(i, S; \prefvec) \\
    &\stackrel{\rm (b)}{=}  \sum_{m=1}^{n-1} \frac{\lambda_{m} - \lambda_{m+1}}{\lambda_1} \cdot \pi_m^{\mpmnl}(i, S; \prefvec) + \frac{\lambda_n}{\lambda_1}\cdot \pi_n^{\mpmnl}(i, S; \prefvec) \\
    &\stackrel{\rm (c)}{=}  \sum_{m=1}^{n-1} \frac{\lambda_{m} - \lambda_{m+1}}{\lambda_1} \cdot \left(\sum_{k=1}^m \pi_k(i, S; \prefvec) \right) + \frac{\lambda_n}{\lambda_1}\cdot \left(\sum_{k=1}^n \pi_k(i, S; \prefvec)\right) \\
    & \stackrel{\rm (d)}{=} \frac{1}{\lambda_1} \sum_{k=1}^{n} \lambda_k \cdot \pi_k(i, S; \prefvec) \\
    &= \frac{\Pcal_{\Mcal}(i, S)}{\lambda_1},
\end{align*}
where (a) follows from~\eqref{eq:mp-mnl-choiceprob}, (b) follows from~\eqref{eq:mpmnl-purch-distr}, (c) follows from the Claim in the proof of Lemma~\ref{lem:mpmnl_subsumed_by_gmnl}, and (d) follows from the telescoping sum.}
\Halmos \endproof

\subsection{Random attention model (RAM)}
\label{app:ram_and_gsp}
The RAM consists of a product ranking $\sigma \in \Lscr$, and a function $\mu(T \cond S)$, with $T \subseteq S$, which is interpreted as the probability that an individual would consider alternatives in the subset $T$ when presented with offer-set $S$. Therefore, we have the constraint $\sum_{\emptyset \neq T \subseteq S} \mu(T \cond S) = 1$ for any $S \subseteq [n]$. In any RAM model, this function must satisfy $\mu(T \cond S) \geq \mu(T \cond S')$ for all $T \subseteq S \subset S'$. Intuitively, the inequality says that the probability of considering a subset $T$ of alternatives cannot {\em increase} if the offer-set that is presented to consumers is enlarged.
Equipped with $\mu$ and $\sigma$, the probability of choosing alternative $j \in S$ is given by
\begin{equation}
\mathcal{P}(j,S)=\sum_{\emptyset \neq T \subseteq S} \mathds{1}[\rank(j, T; \sigma) = 1] \cdot \mu(T \cond S)
\end{equation}


The next result shows that neither GSP nor RAM is strictly subsumed by the other:
\begin{proposition}[Relationship between RAM and GSP models]
    The class of GSP models is not contained in the RAM class and vice-versa.
\end{proposition} 
\proof{Proof.}
We consider the GSP model in Table \ref{GSP_but_not_RAM_example} consisting of $n=5$ alternatives. 
\begin{table}[h]
\centering
\caption{A GSP model that does not belong to the RAM class}
\begin{tabular}{|c|c|c|c|}\hline
  \multicolumn{3}{|c|} {Consumer type} & Probability \\ \hline
  label & $\bell$ & $k$ &  \\
  \hline
  1 & (2,3,1,4,5) & 1 & 0.41 \\
  \hline
  2 & (2,4,1,3,5) & 1 & 0.09 \\
  \hline
  3 & (2,1,3,4,5) & 2 & 0.10 \\
  \hline
  4 & (3,1,2,4,5) & 2 & 0.01 \\
  \hline
  5 & (1,3,2,4,5) & 2 & 0.09 \\
  \hline
  6 & (5,1,2,3,4) & 1 & 0.30 \\
  \hline
\end{tabular}
\label{GSP_but_not_RAM_example}
\end{table}
For this GSP model, we can compute the following choice  probabilities:
\begin{eqnarray}
\mathcal{P}(1,\{1,2,3,4,5\}) &=& \lambda(((2,1,3,4,5),2)) + \lambda(((3,1,2,4,5),2)) = 0.1 + 0.01 = 0.11 \nonumber \\
\mathcal{P}(1,\{1,3,4,5\}) &=& \lambda(((3,1,2,4,5),2)) = 0.01 \nonumber \\
\mathcal{P}(2,\{2,3,4,5\})&=& \lambda(((2,3,1,4,5),1)) + \lambda(((2,4,1,3,5),1)) + \lambda(((3,1,2,4,5),2)) + \lambda(((1,3,2,4,5),2)) \nonumber \\
&=&0.41 + 0.09 + 0.01 + 0.09 = 0.6 \nonumber \\
\mathcal{P}(2,\{2,4,5\})&=& \lambda(((2,3,1,4,5),1)) + \lambda(((2,4,1,3,5),1))  = 0.41 + 0.09 = 0.5 \nonumber \\
\mathcal{P}(3,\{1,3,5\})&=& \lambda(((2,3,1,4,5),1)) + \lambda(((2,4,1,3,5),1)) + \lambda(((2,1,3,4,5),2)) + \lambda(((1,3,2,4,5),2)) \nonumber \\
&=& 0.41 + 0.09 + 0.1 + 0.09 = 0.69 \nonumber \\
\mathcal{P}(3,\{3,5\})&=&  \lambda(((2,3,1,4,5),1)) + \lambda(((2,4,1,3,5),1)) = 0.41 + 0.09 = 0.5 \nonumber
\end{eqnarray}

\citet{cattaneo2017random} consider the following binary relation $\sim$ among the elements in $[n]$: $i \sim j$ if and only if there exists $S \subseteq [n]$ such that $i,j \in S$ and $\mathcal{P}(i,\osminus{S}{j})< \mathcal{P}(i,S)$, where recall that $\osminus{S}{j} = S \setminus \set{j}$. They showed that a discrete choice model can be represented by a RAM model if and only if $\sim$ does not induce a preference cycle. Observe now that the probabilities computed above show that the pairs $(1,2),(2,3),(3,1)$ all belong to $\sim$ which leads to a preference cycle. Therefore, it follows that this choice model doesn't belong to the RAM class.

For the other direction, consider the choice model outlined in Table \ref{table_counter_example}. We will show that this model cannot be explained by the GSP class.

\begin{table}[h]
\caption{Example of a choice model that is not in the GSP class.}
\begin{center}
\begin{tabular}{ c | c c c }
$S$ & $\mathcal{P}(1, S)$ & $\mathcal{P}(2, S)$ & $\mathcal{P}(3, S)$ \\
\hline
$\{1,2\}$ & 1 & 0 & - \\
$\{1,3\}$ &  0 & - & 1 \\
$\{2,3\}$ &  - & 1 & 0 \\
$\{1,2,3\}$  & 1 & 0 & 0 \\
\end{tabular}
\end{center}
\label{table_counter_example}
\end{table}

For purposes of contradiction, suppose there exists a GSP model that explains the choice probabilities defined above. Given that $\mathcal{P}(1,\{1,2\})=1$, it follows from definitions~\ref{def:gsp_choice_fn} and~\ref{def:gsp_prob_fn} that
\begin{equation} \label{eq_1}
\lambda((1,2,3),1) + \lambda((1,3,2),1) + \lambda((3,1,2),1) + \lambda((2,1,3),2) + \lambda((2,3,1),2) + \lambda((3,2,1),2) = 1
\end{equation}
Above, the first three terms account for all consumer types $(\bell, k)$ with choice index $k=1$ and where alternative 1 appears before 2. The remaining three terms are from the consumer types who rank alternative 2 higher than 1 and choose the alternative in the second position ($k=2$).

Similarly, using the fact that $\mathcal{P}(2,\{2,3\})=1$, and $\mathcal{P}(3,\{1,3\})=1$, we can derive the following two equations:
\begin{eqnarray}
\lambda((2,3,1),1) + \lambda((2,1,3),1) + \lambda((1,2,3),1) + \lambda((3,2,1),2) + \lambda((3,1,2),2) + \lambda((1,3,2),2) = 1 \label{eq_2} \\
\lambda((3,1,2),1) + \lambda((3,2,1),1) + \lambda((2,3,1),1) + \lambda((1,3,2),2) + \lambda((1,2,3),2) + \lambda((2,1,3),2) = 1 \label{eq_3}
\end{eqnarray}
Based on equations \eqref{eq_1}, \eqref{eq_2}, and \eqref{eq_3}, the existence of a distribution $\lambda$ implies that the following matrix equation has a solution,
\[
\begin{bmatrix}
1 & 1 & 1 & 1 & 1 & 1 & 1 & 1 & 1 & 1 & 1 & 1\\
1 & 1 & 1 & 0 & 0 & 0 & 1 & 1 & 1 & 0 & 0 & 0\\
1 & 0 & 0 & 1 & 1 & 0 & 0 & 0 & 1 & 1 & 1 & 0\\
0 & 0 & 1 & 1 & 0 & 1 & 1 & 0 & 0 & 0 & 1 & 1
\end{bmatrix}
\cdot
\begin{bmatrix}
\lambda((1,2,3),1)\\
\lambda((1,3,2),1)\\
\lambda((3,1,2),1)\\
\lambda((2,3,1),1) \\
\lambda((2,1,3),1)\\
\lambda((3,2,1),1)\\
\lambda((2,1,3),2)\\
\lambda((2,3,1),2)\\
\lambda((3,2,1),2)\\
\lambda((3,1,2),2)\\
\lambda((1,3,2),2)\\
\lambda((1,2,3),2)
\end{bmatrix}
=
\begin{bmatrix}
1 \\
1 \\
1 \\
1
\end{bmatrix};
\]
where all the variables are non-negative.\footnote{The equation induced by the first row of the matrix must hold because $\lambda$ is a probability distribution.}

If there exists a solution, by Farkas' Lemma \citep{dinh2014farkas}, there is no vector $(y_1,y_2,y_3,y_4) \in \mathbb{R}^4$ such that {\em both} of the following conditions are satisfied:
\[
\begin{bmatrix}
 1 & 1 & 1 & 0\\
 1 & 1 & 0 & 0\\
1 & 1 & 0 & 1 \\
1 & 0 & 1 & 1 \\
1 & 0 & 1 & 0 \\
1 & 0 & 0 & 1 \\
1 & 1 & 0 & 1 \\
1 & 1 & 0 & 0 \\
1 & 1 & 1 & 0 \\
1 & 0 & 1 & 0 \\
1 & 0 & 1 & 1 \\
1 & 0 & 0 & 1
\end{bmatrix}
\cdot
\begin{bmatrix}
y_1 \\
y_2 \\
y_3 \\
y_4 \\
\end{bmatrix}
\geq
\begin{bmatrix}
0 \\
0 \\
0 \\
0
\end{bmatrix} \quad \text{and} \quad \begin{bmatrix}
1 & 1 & 1 & 1
\end{bmatrix}
\cdot
\begin{bmatrix}
y_1 \\
y_2 \\
y_3 \\
y_4 \\
\end{bmatrix}
<
\begin{bmatrix}
0
\end{bmatrix};
\]

However, the vector $(y_1,y_2,y_3,y_4)=(2,-1,-1,-1)$ satisfies both these inequalities. This is a contradiction, and therefore it follows that no GSP model can explain the system of choice probabilities proposed in Table~\ref{table_counter_example}.

In contrast, based on the characterization in \citealt{cattaneo2017random}[Theorem 2], one can deduce that this model does belong to the RAM class as there are no cycles in the relation $\sim$ among the alternatives.
\Halmos \endproof

\section{Additional numerical results and details}
\subsection{Aggregated choice data for experiments reported in Section~\ref{sec:explain}}
\label{app:numerics}
Table~\ref{tab:mm_plans} shows the different payment plans offered to participants, and Table~\ref{tab:mm_plans_obs} reports the aggregated choice fractions across all possible choice sets. Similarly, tables~\ref{tab:lotteries} and~\ref{tab:lottery_obs} show the offered lotteries and aggregated choices. 

\begin{table}[h]
\tabcolsep=20pt
\TABLE
{The four different payment plans in the time preferences~\citep{manzini2006two} experiment\label{tab:mm_plans}}
{\begin{tabular}{*{5}{c}}
\toprule
  Delay & C (Constant) & I (Increasing) & D (Decreasing) & J (Jump) \\
  \midrule
   3 months & \euro{16} & \euro{8} & \euro{24} & \euro{8} \\
  6 months & \euro{16} & \euro{16} & \euro{16} & \euro{8}\\
 9 months &  \euro{16} & \euro{24} & \euro{8} & \euro{32}\\
  \bottomrule
  \end{tabular}}
  {}
\end{table}
\begin{table}[h]
\tabcolsep=22.0pt
\TABLE
{Observed choice fractions in the delayed payment plans experiment\label{tab:mm_plans_obs}}
{
\begin{tabular}{|l|c|c|c|c|}\hline
  $\Scal$ & $\hPcal(C, \Scal)$ & $\hPcal(I, \Scal)$ & $\hPcal(D, \Scal)$ & $\hPcal(J, \Scal)$ \\
  \hline
  \hline
   \{C, I\} & 0.93 & 0.07 & - & -\\
  \{C, D\} & 0.35 & - & 0.65 & -\\
  \{C, J\} & 0.91 & - & - & 0.09\\
  \{I, D\} & - & 0.19 & 0.81 & -\\
  \{I, J\} & - & 0.91 & - & 0.09\\
 \{D, J\} & - & - & 0.84 & 0.16\\
  \hline
  \{C, I, D\} & 0.32 & 0.08 & 0.60 & -\\
  \{C, I, J\} & 0.86 & 0.11 & - & 0.03\\
  \{C, D, J\} & 0.29 & -& 0.65 & 0.06 \\
  \{I, D, J\} & - & 0.15 & 0.80 & 0.05 \\
  \hline
  \{C, I, D, J\} & 0.34 & 0.05 & 0.56 & 0.05 \\
  \hline
\end{tabular}}
{The first column lists all the possible choice sets $\Scal$ consisting of at least 2 alternatives. In each of the remaining columns, $\hat{\Pcal}(j, \Scal)$ shows the fraction of participants choosing alternative $j$ from choice set $\Scal$.}
\end{table}

\begin{table}[h]
\tabcolsep10.0pt
\TABLE
{The four different lotteries used in the risk preferences~\citep{caliari2020behavioral} experiment\label{tab:lotteries}}
{\begin{tabular}{l|l|r|r}
\toprule
  Lottery & Odds & Mean win  & Stdev in win \\
  \midrule
   D (Degenerate) & Win 50 tokens w.p. 1 & 50.0 & 0.0 \\
  Sa (Safe) & Win 65 tokens w.p. 0.8 or 25 tokens w.p. 0.2 & 57.0 & 16.0 \\
 50/50 (Fifty-Fifty) & Win 95 tokens w.p. 0.5 or 20 tokens w.p. 0.5 & 57.5 & 37.5 \\
 R (Risky) & Win 300 tokens w.p. 0.2 or 5 tokens w.p. 0.8 & 64.0 & 118.0 \\
  \bottomrule
  \end{tabular}}
  {``w.p.'' denotes with probability. ``Mean win'' denotes the expected number of tokens won and ``Stdev in win'' denotes the standard deviation in the number of tokens won. The exchange rate was fixed to 10:1 pounds in the experiment.}
\end{table}

\begin{table}[h]
\tabcolsep=18.0pt
\TABLE
{Observed choice fractions in the lotteries experiment\label{tab:lottery_obs}}
{\begin{tabular}{|l|c|c|c|c|}\hline
  $\Scal$ & $\hPcal(D, \Scal)$ & $\hPcal(Sa, \Scal)$ & $\hPcal(50/50, \Scal)$ & $\hPcal(R, \Scal)$ \\
  \hline
  \hline
   \{D, Sa\} & 0.61 & 0.39 & - & -\\
  \{D, 50/50\} & 0.47 & - & 0.53 & -\\
  \{D, R\} & 0.64 & - & - & 0.36\\
  \{Sa, 50/50\} & - & 0.48 & 0.52 & -\\
  \{Sa, R\} & - & 0.65 & - & 0.35\\
 \{50/50, R\} & - & - & 0.59 & 0.41\\
  \hline
  \{D, Sa, 50/50\} & 0.41 & 0.26 & 0.33 & -\\
  \{D, Sa, R\} & 0.39 & 0.36 & - & 0.25\\
  \{D, 50/50, R\} & 0.41 & -& 0.32 & 0.27 \\
  \{Sa, 50/50, R\} & - & 0.35 & 0.39 & 0.26 \\
  \hline
  \{D, Sa, 50/50, R\} & 0.31 & 0.23 & 0.18 & 0.28 \\
  \hline
\end{tabular}}
{}
\end{table}

\subsection{Estimation procedure for the Halo-MNL model}
\label{sec:halomnl_est}
\av{We estimate the following form of the Halo-MNL which was proposed in~\cite{yousefi2020choice}:
\begin{equation}
\label{eq:halo-mnl-prob}
\Pcal(j, S) = \frac{\exp\left(\alpha_{jj} + \sum_{i \in \osminus{S}{j}} \alpha_{ij}\right)}{\sum_{l \in S}\exp\left(\alpha_{ll} + \sum_{i \in \osminus{S}{l}} \alpha_{il}\right)}.
\end{equation}
We note that this is a different formulation from the one proposed in the original Halo-MNL model paper~\citep{maragheh2018customer}, although the two forms are mathematically equivalent. It is easy to check that the log-likelihood function for the Halo-MNL model is concave w.r.t.\ to the $n \times n$ product interaction matrix $\bm{A} = \left(\alpha_{ij}\right)_{i \in [n], j \in [n]}$. We use Python's {\tt scipy.optimize.minimize} function to determine the MLE of the matrix $\bm{A}$.}

\section{GPT procedure of~\cite{jena2022estimation}}
\subsection{Empirical findings on the IRI Academic Dataset}
\label{sec:iri}
\av{In this section, we briefly discuss the empirical findings in~\citet[Section 5.2]{jena2022estimation}, who leveraged the IRI Academic Dataset~\citep{bronnenberg2008database} to evaluate the predictive performance of the GSP model in comparison with several rational and non-rational choice models.
To enable an scalable estimation of GSP, \citet{jena2022estimation} adopted the framework of partially ranked preference sequences from~\citet{jena2020partially}. Specifically, they introduced the notion of a {\em partially ranked preference with irrationality}, where instead of a complete linear ordering, a consumer type is characterized by a partial ranking over a subset of alternatives, with the consumer being indifferent among the remaining products.
Given an offer set, the consumer either selects an available product from the partially ranked list (analogous to Definition~\ref{def:gsp_choice_fn}), chooses uniformly at random from the products comprising her indifference list, or opts for the no-purchase alternative—depending on the type’s choice index. The key intuition behind partially ranked preferences is that low-ranked alternatives tend to have limited influence on observed choices, and modeling them precisely adds unnecessary complexity and can lead to overfitting. We refer the reader to Section 3 of~\citet{jena2022estimation} for a detailed exposition of this representation.

The authors implement a column generation procedure to estimate the GSP model with partially ranked preferences given access to sales transaction data. Customer types are modeled as paths in a rooted tree, with products represented as nodes. In each iteration, the algorithm identifies which path(s) to ``grow'' by appending products that have not yet appeared in the path. As a result, the procedure is referred to as the {\em growing preference tree (GPT)} algorithm. The authors proposed two variants of the algorithm: GPT-I, which adds new types based solely on reduced costs, and GPT-IC, which incorporates a dominance rule designed to prioritize customer behaviors with small consideration sets---that is, those that strictly rank only a few products, thereby promoting sparse, low-order interactions among the products.

The empirical results clearly demonstrate the benefits of explicitly modeling irrational behavior. In particular, the GSP-based representations—GPT-I and GPT-IC—consistently achieved lower prediction errors compared to both rational (SP model-based) benchmarks~\citep{jagabathula2019limit,jena2020partially} as well as other non-RUM approaches such as the Halo-MNL~\citep{maragheh2018customer} and the pairwise choice Markov Chain (PCMC)~\citep{ragain2016pairwise} model---a context-sensitive model that represents consumer choices as arising from the stationary distribution of a continuous time Markov chain over the set of available alternatives---where the performance was measured using 5-fold cross validated mean absolute error (MAE) between the predicted and observed sales. Among all models evaluated, GSP emerged as the strongest overall performer, achieving the lowest prediction errors in 17 out of 29 product categories evaluated, with GPT-IC performing best in 11 of them. On average, GSP yielded a 12.4\% reduction in MAE relative to the best SP model baseline, outperforming it in 21 of the 29 categories. The improvements were often substantial: 48\% for the {\em cigarettes} category, and around 40\% for the {\em coffee} and {\em beer} categories. 
The detailed results can be found in Table 8 and Figure 5 in~\cite{jena2022estimation}. These findings show the importance and practical value of modeling systematic violations of regularity through the GSP framework.}
\subsection{Prediction performance on synthetic data}
\label{sec:synth_data}
\av{In this section, we evaluate the prediction accuracy of the GMNL and GSP models using synthetic transaction data. Our primary goal is to test the performance of these models in predicting non-rational customer choices. To benchmark our approach, we also compare our estimation method for the GSP model with the GPT procedure proposed in~\citet{jena2022estimation}.

\paragraph{Setup.} We leverage the synthetic instances used in the simulation framework from Section~5.1 of~\citet{jena2022estimation}.\footnote{These instances can be downloaded from \url{https://github.com/ds4dm/IrratDCM/} and we thank~\cite{jena2022estimation} for making them available.} Their setup generates multiple synthetic datasets, each with $n=10$ products (including the no-purchase option), under a variety of ground-truth choice models and transaction volumes. Two classes of non-rational ground-truth models were considered: Halo-MNL and GSP.

For the Halo-MNL model, the authors simulated different degrees of context effects by varying the proportion of pairwise interactions in the utility specification~\eqref{eq:halo-mnl-prob} to $10\%$ and $25\%$. Both symmetric interactions (mutual enhancement of utility) and asymmetric interactions (decoy effects) were considered, for a total of 120 instances each under both the $10\%$ and $25\%$ configurations. Additionally, more complex ground-truths were generated using a mixture of 10 Halo-MNL models, again with 120 instances each under both configurations.

For the GSP model, two ground-truth classes were considered with 10 and 100 customer types, respectively. The proportion of non-standard types was varied across $10\%$, $20\%$, and $50\%$. Choice indices for each type were sampled uniformly at random from $[k_{\max}]$, with $k_{\max} \in \{2, 6, 10\}$ to simulate varying levels of non-rationality.
In total, there were 180 instances for each of the proportions $10\%$, $20\%$, and $50\%$, under both GSP ground-truths.

As a robustness check, the authors also considered rational counterparts of the above models as ground-truths: the MNL and 10-class LC-MNL, and the SP model with 10 and 100 rankings, respectively. 

\paragraph{Models compared.} In addition to the GPT-I, GPT-IC, Halo-MNL, and PCMC models described in Section~\ref{sec:iri}, the authors implement two rational benchmarks: (i) a full enumeration SP model, denoted RB-R (rank-based rational), and (ii) a partially ranked SP model estimated using the GPT procedure from~\citet{jena2020partially}. To assess the impact of our proposed regularization mechanisms (limiting the maximum choice index and proportion of non-standard types) on the predictive performance, we implement two structured variants: (i) a regular GMNL(2) model using Algorithm~\ref{alg:gmnl_em} with the update rule in~\eqref{eq:regular_gmnl}, and (ii) the GSP(3) model using Algorithm~\ref{alg:gsp_fw} with $k_{\max}=3$ and $\delta=0.5$.

\paragraph{Results and Discussion.} All models were evaluated using the mean absolute error (MAE) metric:
\[
\text{MAE} = \frac{1}{|\Scal_{\text{test}}|} \sum_{S \in \Scal_{\text{test}}} \sum_{j \in S} \left| \Pcal_{\text{pred}}(j, S) - \Pcal_{\text{true}}(j, S) \right|,
\]
where $\Scal_{\text{test}}$ is the collection of test assortments, and $\Pcal_{\text{pred}}(j, S)$ and $\Pcal_{\text{true}}(j, S)$ denote the predicted and true choice probabilities, respectively.

Table~\ref{tab:jena2022-table7} summarizes the results, grouped by ground-truth instance. We note that all model performances, except for GMNL(2) and GSP(3), are taken directly from Table~7 in~\citet{jena2022estimation}. Their findings show that the GSP model outperform both rational and other non-rational benchmarks across most irrational ground-truth settings, and the GPT-based procedures are highly scalable with an average runtime of less than 2 seconds per instance.

We now focus on the performance of GMNL(2) and GSP(3). First, consistent with our results in Section~\ref{sec:pred_study}, we find that GMNL(2) achieves excellent predictive performance—even outperforming the GPT-based estimators under two ground-truth settings. Moreover, it is computationally efficient, with an average runtime of just 5 seconds per instance.

Second, GMNL(2) performs worse under the GSP ground-truth with 10 types, where the level of non-rationality—as measured by the LoR metric—is higher (see Section~5.1.2 and Figure~2 in~\citealp{jena2022estimation}). In these cases, the performance of the GSP(3) model is significantly better, especially when the proportion of non-rational types is high. In particular, the average MAE obtained by GSP(3) is lower than all baselines, including the GPT-based procedures, thereby highlighting the strength of our proposed Frank-Wolfe estimator. It is worth noting that estimation of the GSP(3) model was significantly more complex compared to GMNL(2), with a median runtime of approximately 3 minutes per instance.

Third, under the mixture of Halo-MNL and GSP with 100 types ground-truth settings, \citet{jena2022estimation} observed that the degree of irrationality (again measured via LoR) is substantially lower than in the other two settings.\footnote{A similar observation was made in~\citet{berbeglia2022comparative}.} This likely explains why GSP(3) underperforms in these instances, and why the GPT-based procedures with partially ranked lists generalize better. Nonetheless, the GMNL(2) model remains competitive and robust across most settings.}

\begin{table}[htbp]
\TABLE
{\av{Test MAE for rational and non-rational approaches on synthetic instances from~\cite{jena2022estimation}}\label{tab:jena2022-table7}}
{\begin{tabular}{lccccccccc}
\toprule
Instances & \% Irrat.\ & GMNL(2) & GSP(3) & RB-R&GPT-R & GPT-I & GPT-IC & PCMC & Halo-MNL \\
\midrule
\multicolumn{9}{l}{Irrational instances} \\
\qquad Halo-MNL & 10 & 0.1835 &  0.2221 & 0.2176 & 0.2305 & 0.1693 & \textbf{0.1648} & 0.2606 & 0.2028 \\
            & 25 & 0.3091 & 0.3124  & 0.3229 & 0.3280 & 0.2843 & 0.2801 & 0.3804 & \textbf{0.2370} \\
            & (all) Mean & 0.2463 &  0.2672 & 0.2702& 0.2792 & 0.2268 & 0.2224 & 0.3205 & \textbf{0.2199} \\
            \midrule
\qquad Mixture of Halo-MNL & 10 &\textbf{0.0943} & 0.1883  & 0.1462 & 0.1128 & 0.0995 & 0.0979 & 0.2196 & 0.1917 \\
 \qquad (10 class)        & 25 & 0.1578 & 0.2122 & 0.1689 &  0.1574 & 0.1421 & \textbf{0.1378} & 0.2518 & 0.2038 \\
             & (all) Mean & 0.1261 & 0.2002 & 0.1575 & 0.1351 & 0.1208 & \textbf{0.1179} & 0.2357 & 0.1977 \\
\midrule
\midrule
\qquad GSP w/ 10 types & 10 & 0.3567 & 0.2306 & 0.2441 & 0.2208 & \textbf{0.2171} & 0.2337 & 0.4271 & 0.6258 \\
        & 20 & 0.3795 & 0.2544 & 0.2805 & 0.2625 & \textbf{0.2538} & 0.2675 & 0.4624 & 0.6868 \\
        & 50 & 0.4145 &  \textbf{0.3203} & 0.3735 & 0.3651 & 0.3388 & 0.3563 & 0.5314 & 0.8100 \\
        & (all) Mean & 0.3836 & \textbf{0.2684} & 0.2994 & 0.2828 & 0.2699 & 0.2858 & 0.4736 & 0.7075 \\
        \midrule
\qquad GSP w/ 100 types & 10 & 0.1528 &  0.2054 & 0.1679 &  \textbf{0.1409} & 0.1440 & 0.1438 & 0.2890 & 0.2702 \\
         & 20 & 0.1651 & 0.2089 & 0.1801 & \textbf{0.1573} & 0.1617 & 0.1587 & 0.3009 & 0.2737 \\
         & 50 & \textbf{0.1986} & 0.2288  & 0.2160 &  0.2044 & 0.2046 & 0.2030 & 0.3394 & 0.3154 \\
         & (all) Mean & 0.1721 & 0.2144 & 0.1880 & \textbf{0.1675} & 0.1701 & 0.1685 & 0.3098 & 0.2864 \\
         \midrule
         \midrule
\qquad (all) Mean &  & 0.2495 &  0.2390 & 0.2345 & 0.2196 & \textbf{0.2058 }& 0.2096 & 0.3567 & 0.4083 \\
\qquad (all) Median  & & 0.2045 & 0.2227 & 0.2083 & 0.1882 & 0.1772 &\textbf{ 0.1734} & 0.3279 & 0.3078 \\
\qquad (all) Max    &  & \textbf{0.7248} & 0.7352 & 0.9454 &  0.7877 & 0.8087 &  0.7835 & 1.0035 & 1.5872 \\
\midrule
\midrule
\multicolumn{9}{l}{Rational instances} \\
\qquad MNL    &-- & \textbf{0.0288} & 0.2086  & 0.1227 &  0.0916 & 0.0933 & 0.0918 & 0.1404 & 0.1772 \\
\qquad LC-MNL (10 class) &--   & \textbf{0.0345} & 0.2178 & 0.1294 & 0.0724 & 0.0762 & 0.0760 & 0.1529 & 0.1823 \\
\qquad SP w/ 10 types &-- & 0.3531 & 0.2357 & 0.1622 & \textbf{0.1468} & 0.1636 & 0.1812 & 0.3841 & 0.5539 \\
\qquad SP w/ 100 types &--& 0.1447 & 0.2567 & 0.1559 & \textbf{0.1251} & 0.1317 & 0.1282 & 0.2793 & 0.2498 \\
\qquad (all) Mean &  &0.1403  & 0.2297 & 0.1425 &  \textbf{0.1090} & 0.1162 & 0.1193 & 0.2392 & 0.2908 \\
\qquad (all) Median & &  \textbf{0.0993} & 0.2220 & 0.1361 &  0.1012 & 0.1097 & 0.1088 & 0.2310 & 0.2291 \\
\qquad (all) Max & & 0.4713 & 0.4712  & 0.4697 & 0.4117 & 0.4211 & \textbf{0.4075} & 0.6795 & 1.2308 \\
\bottomrule
\end{tabular}}
{The best performing model (lowest MAE) for each row is highlighted in bold. Under the Halo-MNL ground-truths, the first two rows report the MAE averaged over 120 instances, while the last row reports the average over all 240 instances. Similarly, for the GSP ground-truths, the first two rows are averaged over 180 instances, while the last row is average over all 540 instances. For each rational ground-truth, the MAE is averaged over 60 instances.}
\end{table}
\end{APPENDICES}
%
%
%
%
%






\end{document}